\documentclass[a4paper,11pt]{article}
\pdfoutput=1
\makeatletter
\newcommand\notsotiny{\@setfontsize\notsotiny\@viipt\@viiipt}
\makeatother
\usepackage{jheppub}
\usepackage[utf8]{inputenc}
\usepackage[T1]{fontenc}
\usepackage{lmodern}
\usepackage{float}
\usepackage[usenames,dvipsnames,svgnames,table]{xcolor}
\usepackage{rotating}
\usepackage{lscape} 
\usepackage{multirow}
\usepackage{colortbl}
\usepackage{placeins}
\usepackage{adjustbox}
\usepackage{cancel}
\usepackage{xspace}
\usepackage[zerostyle=d]{newtxtt}
\usepackage{booktabs}
\usepackage{subcaption}
\usepackage{tabu}
\usepackage{hhline}
\usepackage{makecell}
\usepackage[normalem]{ulem}

\makeatletter\g@addto@macro\bfseries{\boldmath}\makeatother
\hypersetup{
    unicode=false,          % non-Latin characters in Acrobat's bookmarks
    pdftoolbar=true,        % show Acrobat's toolbar?
    pdfmenubar=true,        % show Acrobat's menu?
    pdffitwindow=false,     % window fit to page when opened
    pdfstartview={FitH},    % fits the width of the page to the window
    pdftitle={On the future of Higgs, electroweak and diboson measurements at lepton colliders},    % title
    pdfauthor={Jorge deBlas, Gauthier Durieux, Christophe Grojean, Jiayin Gu and Ayan Paul},     % author
    pdfnewwindow=true,      % links in new window
    colorlinks=true,       % false: boxed links; true: colored links
    linkcolor=blue,          % color of internal links
    citecolor=blue,%violet,        % color of links to bibliography
    filecolor=blue,%magenta,      % color of file links
    urlcolor=blue,%cyan,           % color of external links
    linktocpage=true
}
\definecolor{Gray}{gray}{0.95}

\def\equationautorefname~#1\null{equation\,(#1)\null}
\newcommand{\appendixref}[1]{\hyperref[#1]{appendix~\ref{#1}}}

\newcommand{\CG}{\cellcolor{Gray}}
\newcommand{\CW}{\cellcolor{white}}
\newcommand{\as}{{\color{Mahogany}*}}
\newcommand{\da}{{\color{Mahogany}$^\dagger$}}
\definecolor{dGray}{gray}{0.95}
\definecolor{lGray}{gray}{0.55}
\newcolumntype{a}{>{\columncolor{dGray}}c}
\newcolumntype{b}{>{\columncolor{lGray}}c}
\newcommand{\PreserveBackslash}[1]{\let\temp=\\#1\let\\=\temp}
\newcolumntype{C}[1]{>{\PreserveBackslash\centering}p{#1}}
\newcolumntype{R}[1]{>{\PreserveBackslash\raggedleft}p{#1}}
\newcolumntype{L}[1]{>{\PreserveBackslash\raggedright}p{#1}}
\newcommand{\rb}[1]{\rotatebox{90}{#1}}
\newcommand{\tp}{\tiny(\%)}
\newcommand{\tpm}{\tiny(\textperthousand)}

\newcommand{\inab}{\,{\rm ab}^{-1}}

\newcommand{\eehz}{e^+e^- \to hZ}

\newcommand{\eeww}{e^+e^- \to WW}

\DeclareMathOperator{\BR}{BR}

\newcommand{\HEPfit}{\texttt{HEPfit}\xspace}

\newcommand{\invisible}[1]{}
\title{On the future of Higgs, electroweak and diboson measurements at lepton colliders}

\author[a,b]{Jorge de~Blas,}
\author[c,d]{Gauthier~Durieux,}
\author[c,e]{Christophe~Grojean,}
\author[f]{Jiayin~Gu,}
\author[c,e]{and Ayan~Paul}

\affiliation[a]{Dipartimento di Fisica e Astronomia ``Galileo Galilei'', Universit\`a di Padova, Via Marzolo 8, I-35131 Padova, Italy}
\affiliation[b]{INFN, Sezione di Padova, Via Marzolo 8, I-35131 Padova, Italy}
\affiliation[c]{DESY, Notkestra{\ss}e 85, D-22607 Hamburg, Germany}
\affiliation[d]{Physics Department, Technion -- Israel Institute of Technology, Haifa 32000, Israel}
\affiliation[e]{Institut f\"ur Physik, Humboldt-Universit\"at zu Berlin, D-12489 Berlin, Germany}
\affiliation[f]{PRISMA$^+$ Cluster of Excellence, Institut f\"ur Physik,\\
Johannes Gutenberg-Universit\"at, 55099 Mainz, Germany}

\emailAdd{Jorge.DeBlasMateo@pd.infn.it}
\emailAdd{durieux@campus.technion.ac.il}
\emailAdd{christophe.grojean@desy.de}
\emailAdd{jiagu@uni-mainz.de}
\emailAdd{ayan.paul@desy.de}

\newcommand*{\mg}{\texttt{MG5\_aMC@NLO}}
\newcommand*{\gev}{\text{GeV}}

\DeclareMathOperator*{\cov}{cov}

\abstract{%
LEP precision on electroweak measurements was sufficient not to hamper the extraction of Higgs couplings at the LHC.
But the foreseen permille-level Higgs measurements at future lepton colliders might suffer from parametric electroweak uncertainties in the absence of a dedicated electroweak program.
We perform a joint, complete and consistent effective-field-theory analysis of Higgs and electroweak processes.
The full electroweak-sector dependence of the $\eeww$ production process is notably accounted for, using statistically optimal observables.
Up-to-date HL-LHC projections are combined with CEPC, FCC-ee, ILC and CLIC ones.
For circular colliders, our results demonstrate the importance of a new $Z$-pole program for the robust extraction of Higgs couplings.
At linear colliders, we show how exploiting multiple polarizations and centre-of-mass energies is crucial to mitigate contaminations from electroweak parameter uncertainties on the Higgs physics program.
We also investigate the potential of alternative electroweak measurements to compensate for the lack of direct $Z$-pole run, considering for instance radiative return to these energies.
Conversely, we find that Higgs measurements at linear colliders could improve our knowledge of the $Z$ couplings to electrons.
}

\begin{document} 

\begin{flushright}
DESY 19-077\\
HU-EP-19/33\\
MITP/19-028
\end{flushright}

\maketitle

%%%%%%%%%%%%%%%%%%%%%%%%%%%
\section{Introduction}
\label{sec:intro}
%%%%%%%%%%%%%%%%%%%%%%%%%%%
The precision electroweak (EW) program carried out at electron-positron colliders had a long-lasting impact on particle physics and definitively contributed to establishing the standard model (SM) at the quantum level~\cite{Taylor:2017edx}.
It even gave the first concrete evidences for the top quark and the Higgs boson that were out of direct reach.
Since its discovery in 2012, the Higgs boson has actually become the central figure in the field and it very much sets the agenda for the next future colliders~\cite{deBlas:2019rxi}.
While there is a strong consensus towards a $e^+e^-$ Higgs factory, its actual design, either circular or linear, is still under fierce debate.
Both options have their own pros and cons: the linear colliders are extendable in energy and there is neither technical nor unbearable economical impediments to polarize their beams.
On the other hand circular colliders can reach higher luminosity, especially at low energies.
Indeed, from centre-of-mass energies relevant for a Higgs program around 250\,GeV, circular colliders are able to deliver integrated luminosities growing roughly like $1/\sqrt{s}^3$ towards lower $\sqrt{s}$ (power losses from synchrotron radiation scale like $\sqrt{s}^4$), while linear collider luminosities rising approximately linearly towards higher $\sqrt{s}$ can be achieved.
That is why a circular Higgs factory would already start as a $Z$ and $W$ pair factory.

Bringing Higgs coupling determinations to the level of precision achieved by the previous generation of lepton colliders on electroweak parameters naturally raises the question of the interplay between those two sectors.
They have remained mostly decoupled so far due to the large hierarchy in the constraints they are subject to~\cite{Pomarol:2013zra, Gupta:2014rxa}.
Higgs coupling measurements did not have sufficient accuracy either to be affected by electroweak parameter uncertainties nor to reduce them, except in the well-known case of anomalous triple gauge couplings (aTGCs)~\cite{Falkowski:2015jaa}.
Our knowledge of Higgs couplings is nevertheless sufficient for their quantum corrections to electroweak processes not to significantly entangle the two sectors.
On the contrary, it was recently shown that the much more uncertain couplings of the top quark to gauge bosons have sizeable loop-level impact on Higgs coupling determinations, especially at future lepton colliders~\cite{Boselli:2018zxr, Durieux:2018ggn, Vryonidou:2018eyv}.
So precision Higgs physics requires the accurate determination of top-quark couplings.
Leaving this issue aside, i.e.\ assuming that top-quark couplings are well constrained by dedicated measurements like that of $e^+ e^- \to t\,\bar{t}$~\cite{Durieux:2018tev}, in this work we address the cross-talk between the Higgs and electroweak sectors at future lepton colliders.

All future lepton colliders target integrated luminosities in the few inverse attobarn range at centre-of-mass energies between $240$ and $380\,\gev$ where the Higgsstrahlung cross section peaks.
Updated electroweak measurements at the $Z$ pole and the $W$ pair production threshold, with astonishingly large statistics, are also part of the program of future circular collider projects.
Collecting large amounts of luminosity at the corresponding energies is, on the contrary, costly and time-consuming for linear machines, as already explained.
Therefore, the interplay between Higgs and electroweak coupling determinations is qualitatively different for the two types of colliders.
These specificities naturally lead us to explore the following questions:
\begin{itemize}
\item What is the deterioration in Higgs couplings determination incurred from electroweak uncertainties?
\item How important are the $Z$-pole and $WW$-threshold runs for Higgs physics?
\item Are measurements at higher energies and with polarized beams sufficient to mitigate their possible absence at linear colliders?
\item Conversely, could Higgs measurements help constraining electroweak parameters?
\end{itemize}
In particular, we will compare the Higgs coupling sensitivity obtained under three different scenarios for the EW measurements: \textit{1.\ LEP/SLD EW} when the experimental uncertainties in the EW sector do not improve over the measurements done at LEP/SLD, \textit{2.\ real EW} when the actual EW projections at the respective lepton collider are used, \textit{3.\ perfect EW} when an infinite statistical and systematic precision in the EW observables is assumed.

In this paper, we shall try to address the questions listed above in the effective field theory (EFT) framework, assuming a mass gap between the weak scale and the scale of new physics such that all the effects of the new degrees of freedom and the new interactions can be captured by local interactions among the SM particles at the energies probed by the colliders. 
We present detailed studies of HL-LHC and future lepton collider prospects based on the most up-to-date experimental studies carefully implemented with available correlations and complemented ---where needed--- with reasonable estimates.
Two different implementations and fitting procedures are used for cross-validation.
They respectively employ the so-called Higgs~\cite{Falkowski:2001958} and Warsaw~\cite{Grzadkowski:2010es} bases of dimension-six operators.
Many of our results are, however, presented in terms of deviations to decay rates in specific channels, to avoid the arbitrariness of the choice of effective operator basis and to provide a more intuitive understanding of prospective limits.
Our study extends and complements that of refs.~\cite{Ellis:2015sca, Ellis:2017kfi, Durieux:2017rsg, Barklow:2017suo, Barklow:2017awn, deBlas:2016nqo, deBlas:2017wmn, DiVita:2017vrr, Chiu:2017yrx, Ellis:2018gqa, Almeida:2018cld, Biekotter:2018rhp,deBlas:2019rxi}.

The figure of merit of the various colliders should not be asserted by looking at single Higgs couplings or electroweak parameters alone, irrespective of their physical relevance.
High-energy machines also give access to new processes, like multiple Higgs production, that can inform us directly about new couplings.
It is instructive to compare, for instance, how much the Higgs self-coupling will be known indirectly from the corrections it induces quantum mechanically on processes with single Higgs~\cite{McCullough:2013rea, Degrassi:2016wml, DiVita:2017eyz, Maltoni:2017ims} and the direct bounds obtained from the study of double Higgs production above 500\,GeV.
The impact of the EW uncertainties in this comparison (so far done only in the perfect EW limit~\cite{DiVita:2017vrr}) is left for future work.

The rest of this paper is organised as follows.
We first review various aspects of the framework employed in \autoref{sec:framework} including a description of our EFT parametrization, of the collider run scenarios and measurements used as input, as well as of our fitting methodologies.
Our results are presented in \autoref{sec:results} with an overview of the Higgs coupling reaches at various colliders followed by a detailed investigation of the effects of EW measurements.
Conclusions are drawn in \autoref{sec:summary}.
A comparison between our results and that of the Higgs@Future Colliders ECFA Working Group report of ref.~\cite{deBlas:2019rxi} is provided in \autoref{sec:ecfa}.
Our results are also expressed in standard effective-field-theory bases in \autoref{app:d6basis}.
Further details about the correlations between Higgs and electroweak parameters are provided in \autoref{app:z-pole-run}.
A brief review of the method of optimal observables is provided in \autoref{sec:oo}.
The measurement inputs used in our global analysis are further detailed in \autoref{app:inputs}. 

%%%%%%%%%%%%%%%%%%%%%%%%%%%
\section{Framework}
\label{sec:framework}
%%%%%%%%%%%%%%%%%%%%%%%%%%%

Introducing our framework, our parametrization of the effective-field-theory space is first presented in \autoref{sec:eft}.
We then provide a brief summary of the future collider run scenarios considered as well as the prospects and measurements used as input to our analysis in \autoref{sec:scenarios}.
The information contained in angular distributions of both diboson and Higgsstrahlung production is extracted using statistically optimal observables (see \autoref{sec:oo} for a brief review), as explained in \autoref{sec:wwinput} and~\ref{sec:hzlloo}.
The assumptions made for scaling statistical and systematic uncertainties with beam polarization are explained in \autoref{sec:ILCPol}.
Our statistical methods for the global fit is described in \autoref{sec:fit}.

%%%%%%%%%%%%%%%%%%%%%%%%%%%
\subsection{Effective field theory}
\label{sec:eft}
%%%%%%%%%%%%%%%%%%%%%%%%%%%

Despite its success on the phenomenological side, the standard model is nowadays viewed as the low-energy limit to a more fundamental ultraviolet (UV) theory, i.e.\ as an EFT. This UV completion is supposed to replace the SM at a given energy scale $\Lambda$, which acts as the cut off of the EFT. For energies  $E \ll \Lambda$, however, the observed phenomena can be well described by an effective Lagrangian expanded in inverse powers of $\Lambda$, 
\begin{equation}
\label{eq:EFT_Lops}
{\cal L}_{\mathrm{Eff}}={\cal L}_{\rm SM}
+  \sum_{d>4} \frac{1}{\Lambda^{d-4}}{\cal L}_{d}
,~~~~~~~
{\cal L}_{d}= \sum_{i}  c_{i}^{(d)} {\cal O}_{i}^{(d)}\, . 
\end{equation}
In the so-called standard-model effective field theory (SMEFT), each operator ${\cal O}_i^{(d)}$ is a local analytical $SU(3)_c \times SU(2)_L \times U(1)_Y$-invariant operator of canonical mass dimension $d$, built from the fields describing particles with mass $m< \Lambda$. 
Such particles are assumed to be the SM ones, with the Higgs field belonging to an $SU(2)_L$ doublet.\footnote{The alternative formulation of the non-linear Higgs effective theory~\cite{Buchalla:2013rka, Brivio:2013pma, deFlorian:2016spz}
is obtained removing the condition of analyticity from the EFT assumptions~\cite{Falkowski:2019tft}.
This leads to a more general structure of interactions but is also characterized by a relatively lower range of applicability to new physics scenarios, for in this case the cut-off of the EFT is necessarily connected with the electroweak vacuum expectation value, $\Lambda\lesssim 4\pi v$.}
The operator coefficients $c_i^{(d)}$ encode the indirect effects of heavy new physics and can be computed in terms of the parameters of particular UV completions~\cite{delAguila:2000rc,delAguila:2008pw,delAguila:2010mx,Henning:2014wua,deBlas:2014mba,Drozd:2015rsp,delAguila:2016zcb,Ellis:2016enq,Fuentes-Martin:2016uol,Zhang:2016pja,Ellis:2017jns,deBlas:2017xtg,Ellis:2019zex}.
The success of the SM and usefulness of this EFT expansion are then justified from the fact that observable effects from an operator of dimension $d$ are suppressed by $(q/\Lambda)^{d-4}$, where $q$ is the relevant energy scale involved in the observable or the electroweak vacuum expectation value $v\approx 246$\,GeV.
Assuming $B$ and $L$ number conservation, the leading order new physics corrections to SM physics are given by the dimension-six terms.
Under all the assumptions above, a complete basis of physically independent dimension-six interactions contains a total of 59 types of operators~\cite{Grzadkowski:2010es} (2499 taking into account flavour indices, still under the assumption of $B$ and $L$ number conservation~\cite{Alonso:2013hga}).

The SMEFT framework provides a convenient and consistent framework to describe indirect effects of new physics and to parametrize possible deformations with respect to the SM predictions.
Since the correlations imposed by gauge invariance and/or the assumption of linearly realized electroweak symmetry breaking (EWSB) are manifest in this framework, the EFT formalism allows to easily study and exploit the complementarities between different types of processes to constrain new physics effects.
Preserving its systematic coverage of heavy new physics scenarios in general also requires one to consider simultaneously all contributions to studied observables up to a given order.
New physics would generate several operators at a high scale.
Renormalization-group running to the scale of measurements could provide non-vanishing coefficients to an even larger set of operators.
The projection of those onto a basis, using equation of motion, integration by part, Fierz identity, etc.\ would have the same effect.
Focusing on arbitrary subset of operators is also technically inconsistent at the quantum level where counterterms from discarded operators may be needed.
This naturally leads to the necessity to consider {\it global} studies of indirect effects, where one includes all possible types of physical observables to probe simultaneously all SMEFT directions open at a given order.
In this work, where we are mainly interested in the study of possible deformations in Higgs processes.
This will necessarily lead us to include in the discussion the observables typically considered in electroweak precision tests and diboson production.
In what follows, we describe the different operators that enter in all these processes in the dimension-six SMEFT formalism.
For that purpose, we will use the convenient parametrization of ref.~\cite{Falkowski:2001958,Falkowski:2015fla}, the so-called {\it Higgs basis}, where the leading order effects of new physics are presented in the unitary gauge.

For simplicity, we will restrict our study to CP-conserving interactions.
CP-violating interactions can be constrained separately with specifically designed CP-odd observables that are insensitive to CP-even effects.
The two sectors can thus be decoupled.
The new physics contributions to the couplings of the Higgs boson to vector bosons can then be written in terms of the following interactions in the physical basis:
\begin{eqnarray}
\label{eq:Lhvv}
\Delta {\cal L}^{hVV}_{6}   &= & {\frac{h}{v}} \left [ 
2 \delta c_w   m_W^2 W_\mu^+ W_\mu^- +   \delta c_z   m_Z^2 Z_\mu Z_\mu
\right . \nonumber\\ & & \left . 
+ c_{ww}  \frac{g^2}{2} W_{\mu \nu}^+  W_{\mu\nu}^-  
+ c_{w \Box} g^2 \left (W_\mu^- \partial_\nu W_{\mu \nu}^+ + {\rm h.c.} \right )  
\right . \nonumber\\ & & \left . 
+  c_{gg} \frac{g_s^2}{4} G_{\mu \nu}^a G_{\mu \nu}^a   + c_{\gamma \gamma} \frac{e^2}{4} A_{\mu \nu} A_{\mu \nu} 
+ c_{z \gamma} {\frac{e \sqrt{g^2+g^{\prime~\!2}}}{2}} Z_{\mu \nu} A_{\mu\nu} + c_{zz} {\frac{g^2+g^{\prime~\!2}}{4}} Z_{\mu \nu} Z_{\mu\nu}
\right . \nonumber\\ & & \left .
+ c_{z \Box} g^2 Z_\mu \partial_\nu Z_{\mu \nu} + c_{\gamma \Box} g g^\prime Z_\mu \partial_\nu A_{\mu \nu}
\right ].
\nonumber\\  \label{eq:L6hVV}
 \end{eqnarray}
In the previous Lagrangian only $c_{gg}, \  \delta c_z,  \ c_{\gamma \gamma}, \ c_{z \gamma},  \ c_{zz},  \ c_{z \Box}$ are independent parameters, with the others being related by gauge invariance:
\begin{eqnarray}
\label{eq:dep_pars}
\delta  c_{w} &=&  \delta c_z + 4 \delta m , 
\nonumber\\
c_{ww} &=&  c_{zz} + 2 \sin^2{\theta_w} c_{z \gamma} + \sin^4{\theta_w} c_{\gamma \gamma}, 
\nonumber\\
c_{w \Box}  &= & {\frac{1}{g^2 - g^{\prime~\!2}}} \left [ 
g^2 c_{z \Box} + g^{\prime~\!2} c_{zz}  - e^2 \sin^2{\theta_w}   c_{\gamma \gamma}  -(g^2 - g^{\prime~\!2}) \sin^2{\theta_w}  c_{z \gamma} 
\right ],  
\nonumber\\
  c_{\gamma \Box}  &= &  
  {\frac{1}{g^2 - g^{\prime~\!2}}} \left [ 
2 g^2 c_{z \Box} + (g^2+ g^{\prime~\!2}) c_{zz}  - e^2  c_{\gamma \gamma}  -(g^2 - g^{\prime~\!2})   c_{z \gamma} 
\right ].
 \end{eqnarray}
In the previous equations $g_s$, $g$ and $g^\prime$ denote the $SU(3)_c \times SU(2)_L \times U(1)_Y$ gauge coupling constants, $e$ is the electric charge, $\theta_w$ denotes the weak mixing angle and $m_{Z,W}$ are the electroweak vector boson masses.
The parameter $\delta m$ describes new physics contributions to the $W$ mass, and it is the only source of custodial symmetry breaking in Higgs couplings to dimension six. The same dynamics generating some of the couplings in \autoref{eq:Lhvv} also induces modifications in the so-called anomalous triple-gauge couplings (aTGC). These anomalous interactions are parametrized by means of three couplings: $\delta g_{1,z}$, $\delta \kappa_\gamma$, $\lambda_z$~\cite{Hagiwara:1986vm}, 
\begin{eqnarray}
   \label{eq:LaTGC}
\Delta {\cal L}^{\mathrm{aTGC}}\!\! &=\! & i e \delta \kappa_\gamma\, A^{\mu\nu} W_\mu^+ W_\nu^- +
i g  \cos{\theta_w} \bigg[ \delta g_{1Z}\, (W_{\mu\nu}^+ W^{-\mu} - W_{\mu\nu}^- W^{+\mu})Z^\nu +\\
&&+\! (\delta g_{1Z}\!-\!\frac{g^{\prime~\!2}}{g^2}\delta \kappa_\gamma)\,\! Z^{\mu\nu} W_\mu^+ W_\nu^-\! \bigg] 
\!+\!\frac{ig \lambda_z}{m_W^2}\!\left( \sin{\theta_w} W_{\mu}^{+\nu} W_{\nu}^{-\rho} A_{\rho}^{\mu}\! +\! \cos{\theta_w} W_{\mu}^{+\nu} W_{\nu}^{-\rho} Z_{\rho}^{\mu} \right)\!,\nonumber
\end{eqnarray}
where the first two can be written as:
\begin{eqnarray}
 \delta  g_{1,z} &=& 
\frac{1}{2 (g^2 - g^{\prime~\!2})} \left [   c_{\gamma\gamma} e^2 g^{\prime~\!2} + c_{z \gamma} (g^2 - g^{\prime~\!2}) g^{\prime~\!2}  - c_{zz} (g^2 + g^{\prime~\!2}) g^{\prime~\!2}  - c_{z\Box} (g^2 + g^{\prime~\!2}) g^2 \right ], 
 \nonumber\\
 \delta \kappa_\gamma  &=& - \frac{g^2}{2} \left ( c_{\gamma\gamma} \frac{e^2}{ g^2 + g^{\prime~\!2}}   + c_{z\gamma}\frac{g^2  - g^{\prime~\!2}}{g^2 + g^{\prime~\!2}} - c_{zz} \right ).
 \end{eqnarray}  
Because of this, and as it was pointed out in ref.~\cite{Falkowski:2015jaa}, the study of aTGC measured, e.g.\ in $e^+ e^- \to W^+ W^-$, provides complementary information that can be used to constrain new physics in the Higgs couplings.
To make use of this complementarity in a general way, however, one also needs to consider the third aTGC, $\lambda_z$,
which parametrizes dimension-six interactions of the form $\varepsilon_{abc}W_{\mu}^{a~\nu}W_\nu^{b~\rho}W_\rho^{c~\mu}$, into the analysis.

The Higgs boson interactions with fermions are given by,
\begin{equation}
\label{eq:hff}
\Delta {\cal L}^{hff}_{6}  =  - \frac{h}{v} \sum_{f \in u,d,e}    (\delta y_f)_{ij} \, (m_f)_{jj}  \bar{f}_i f_j  + {\rm h.c.},
\end{equation}
where, again, all possible CP-violating phases have been set to zero.
As we are interested in exploring all possible deformations that could be tested at future lepton colliders, we will assume that $\delta y_{t}\equiv(\delta y_u)_{33}$, $\delta y_{c}\equiv(\delta y_u)_{22}$, $\delta y_b\equiv(\delta y_d)_{33}$, $\delta y_{\tau}\equiv(\delta y_e)_{33}$ and $\delta y_{\mu}\equiv(\delta y_e)_{22}$ are independent parameters.
Note, however, that this will typically induce flavour changing neutral currents (FCNCs), requiring a fine tuning in the new dynamics generating $\delta y_{f}$.
Off-diagonal flavour structures will not be considered.
In models generating flavour-changing interactions, flavour observables could however impose stronger constraints.

So far all the terms we have written are directly connected to pure Higgs interactions (with the exception of $\lambda_z$). Higgs and diboson production at hadron or lepton colliders are however also sensitive to modifications of the neutral and charged current couplings, $Vf\bar{f}$. Moreover, in the SMEFT framework and to dimension six, such modifications are directly connected to contact interactions of the form $hVf\bar{f}$, which also modify the EW production of the Higgs. These are particularly relevant for associated production of the Higgs with a vector boson, as their effects in the amplitudes grow with the energy relative to the SM contributions. These modifications, either at hadron or lepton colliders, are described in the Higgs basis by,
\begin{eqnarray} 
\label{eq:vff}
\Delta {\cal L}^{(h)Vff}_{6}\!\!\!&=\!& \frac{g}{\sqrt{2}} \left(\!1+2 \frac{h}{v}\right)\! W_\mu^+ \!  \left (
  (\delta g^{\ell}_{W})_{ij}~\bar{\nu}_L^i \gamma^\mu  \ell_L^j
\! +\!   (\delta g^{q}_{W,L})_{ij}~\bar{u}_L^i \gamma^\mu  d_L ^j
\!+\!  (\delta g^{q}_{W,R})_{ij}~\bar{u}_R^i  \gamma^\mu   d_R^j
\! +\! \mathrm{h.c.}\!  \right )
\nonumber\\ 
&+\!& \sqrt{g^2 + g^{\prime~\!2}}  \left(\!1+2 \frac{h}{v} \right)  Z_\mu \!
\left [ \sum_{f = u,d,e,\nu} \!\!\!   (\delta g^{f}_{Z,L})_{ij}  \bar{f}_L^i \gamma^\mu f_L^j  + \!\!\! 
\sum_{f = u,d,e}\!\!  (\delta g^{f}_{Z,R})_{ij}  \bar{f}_R ^i \gamma^\mu f_R^j  \right ]\! ,
\end{eqnarray}
where, again, not all terms are independent.
At the dimension-six level one has:
\begin{equation} 
\delta g^{\ell}_{W} = \delta g^{\nu}_{Z,L} - \delta g^{\ell}_{Z,L},~~~~~~~\delta g^{q}_{W,L} = \delta g^{u}_{Z,L} V_{CKM} - V_{CKM} \delta g^{d}_{Z,L},
\end{equation} 
where $V_{CKM}$ is the {\it Cabibbo--Kowayashi--Maskawa} matrix.
The electroweak precision measurements are most sensitive to flavour-preserving interactions in \autoref{eq:vff}, which are also the relevant ones for Higgs production.
They can also probe most of the couplings to the different light fermion families independently.
We will therefore restrict our analysis to the case where the $(h)Vff$ couplings ---like $hff$ ones--- are diagonal in the fermion-family space.
Considering independent diagonal entries, however, leads again to the problem of FCNCs.
These are especially severe for the light quark families and we impose that the couplings to the first two families of quarks are related via an $U(2)$ flavour symmetry.
This also forbids right-handed charged currents and dipole interactions for the light quarks which do not generate amplitudes interfering with SM ones in the limit $m_q \to 0$.
For the Higgs measurements, the impact of this $U(2)$ flavour symmetry is mostly on the $Vh$ processes at LHC, which has quarks as the initial states.
On the other hand, we keep all lepton couplings independent given that Higgs and diboson production processes single out the $Zee$ coupling.
Finally, as mentioned in the introduction, the electroweak and four-fermion interactions of the top quark are assumed to be well constrained by other measurements ($e^+e^-\to t\,\bar{t}$ in particular).
At tree level, they could mostly impact the precision extracted on the top-quark Yukawa coupling from $e^+e^-\to t\bar{t}h$ measurements.

The above-mentioned set of operators is sufficient to describe $Z$-pole, diboson and single Higgs processes at future lepton colliders.
It counts 
1 ($\delta m$) + 6 ($hVV$/aTGC) + 1 (aTGC) + 5 ($hff$) + 6 ($Z\ell\ell$)  + 3 ($W\ell\nu$) + 2 ($Zuu$) + 4 ($Zdd$)  = 28 new physics parameters.
Note that di-Higgs production would provide further constraints on these operators  in addition to offer a bound on the Higgs trilinear self-coupling~\cite{Azatov:2015oxa, Barklow:2017awn}.
Four-fermion operators that would be relevant for Drell--Yan production at high energies~\cite{Farina:2016rws, Dawson:2019xfp} are omitted from our analysis.

One of our main goals is to study the impact of EW uncertainties on the Higgs coupling determinations.
To serve as reference, we consider an artificial benchmark scenario where infinitely precise EW measurements take exactly SM values.
As {\it perfect EW measurements} we will assume, in particular, $Z$-pole observables as well as the $W$ mass, width and branching ratios.
In the Higgs basis, this corresponds to 1 ($\delta m$) + 6 ($Z\ell\ell$)  + 3 ($W\ell\nu$) + 2 ($Zuu$) + 4 ($Zdd$) = 16 deviations from the SM being set to zero, leaving a total number of 12 Higgs and aTGCs parameters, as considered in ref.~\cite{Durieux:2017rsg}.

As motivated in the introduction, using more physical observables has advantages for the presentation and comparison of measurement prospects.
As replacement for the $\delta m$, $\delta c_z$, $c_{zz}$, $c_{z\Box}$, $c_{z\gamma}$, $c_{\gamma\gamma}$, and $c_{gg}$ degrees of freedom, we therefore  keep $\delta  g_{1,z}$, $\delta\kappa_\gamma$ anomalous triple gauge couplings and define
\begin{align}
\delta g_H^{x}&\equiv \sqrt{\frac{\Gamma(h\to x)}{\Gamma(h\to x)^\text{SM}}} -1,
\end{align}
for $x=WW^*, ZZ^*, Z\gamma, \gamma\gamma, gg$.
The effective couplings $\delta g_{H}^{WW}$ and $\delta g_{H}^{ZZ}$ include the corrections to the full decay chain $H\to VV^* \to 4f$, where all four-fermion final states are summed over.
In addition to modifications of the $HVV$ couplings, this includes corrections to the electroweak vertices $Vff$, contact terms of the $HVff$ form, and modifications of the vector boson propagators (see, e.g.\ section~7.2 of ref.~\cite{Barklow:2017awn}).
A similar approach for Yukawa coupling modifications only fails for the top-quark.
In this case, we simply fix $\delta g_H^{tt}=\delta y_t$ as would have been obtained, to the linear level, if the corresponding decay was kinematically allowed.

In summary, we will present our main results in terms of following 28 parameters:
\begin{equation}
\begin{gathered}
\begin{array}{*{5}{c@{,\quad}}}
 \delta g^{\mu\mu}_H
&\delta g^{\tau\tau}_H
&\delta g^{cc}_H
&\delta g^{tt}_H
&\delta g^{bb}_H
\end{array}
\\
\begin{array}{*{5}{c@{,\quad}}}
 \delta g^{ZZ}_H
&\delta g^{WW}_H
&\delta g^{\gamma\gamma}_H
&\delta g^{Z\gamma}_H
&\delta g^{gg}_H
\end{array}
\\
\begin{array}{*{6}{c@{,\quad}}}
 \delta g_{1,Z}
&\delta \kappa_\gamma
&\lambda_Z
\end{array}
\\
\begin{array}{*{3}{c@{,\quad}}}
 \delta g^{ee}_{Z,L}\equiv (\delta g^{\ell}_{Z,L})_{11}
&\delta g^{\mu\mu}_{Z,L}\equiv (\delta g^{\ell}_{Z,L})_{22}
&\delta g^{\tau\tau}_{Z,L}\equiv (\delta g^{\ell}_{Z,L})_{33}
\end{array}
\\
\begin{array}{*{3}{c@{,\quad}}}
 \delta g^{ee}_{Z,R}\equiv (\delta g^{\ell}_{Z,R})_{11}
&\delta g^{\mu\mu}_{Z,R}\equiv (\delta g^{\ell}_{Z,R})_{22}
&\delta g^{\tau\tau}_{Z,R}\equiv (\delta g^{\ell}_{Z,R})_{33}
\end{array}
\\
\begin{array}{*{3}{c@{,\quad}}}
 \delta g_W^{e\nu}\equiv (\delta g^{\ell}_{W})_{11}
&\delta g_W^{\mu\nu}\equiv (\delta g^{\ell}_{W})_{22}
&\delta g_W^{\tau\nu}\equiv (\delta g^{\ell}_{W})_{33}
\end{array}
\\
\begin{array}{*{6}{c@{,\quad}}}
 \delta g^{uu}_{Z,L}\equiv (\delta g^{u}_{Z,L})_{11}=(\delta g^{u}_{Z,L})_{22}
&\delta g^{dd}_{Z,L}\equiv (\delta g^{d}_{Z,L})_{11}=(\delta g^{d}_{Z,L})_{22}
&\delta g^{bb}_{Z,L}\equiv (\delta g^{d}_{Z,L})_{33}
\end{array}
\\
\begin{array}{*{3}{c@{,\quad}}}
 \delta g^{uu}_{Z,R}\equiv (\delta g^{u}_{Z,R})_{11}=(\delta g^{u}_{Z,R})_{22}
&\delta g^{dd}_{Z,R}\equiv (\delta g^{d}_{Z,R})_{11}=(\delta g^{d}_{Z,R})_{22}
&\omit $\delta g^{bb}_{Z,R}\equiv (\delta g^{d}_{Z,R})_{33}$.
\end{array}
\end{gathered}
\label{eq:parameters}
\end{equation}
Note that the normalization of the fermion couplings to gauge bosons is such that $g^{\rm SM}_W =1$ and $g^{\rm SM}_Z = T^3 - Q\, s^2_w $.
The $\delta g^{ff}_V$ parameters quantify absolute departures from those values.
They are all taken to vanish, when perfect EW measurements are assumed: $\delta g^{ff}_V\equiv 0$.
Furthermore, from the point of view of the Higgs couplings, making the measurement of the $W$ mass ``perfect'', i.e.\ $\delta m\equiv0$, causes $\delta g^{ZZ}_H$ and $\delta g^{WW}_H$ to no longer be independent.

Finally, let us briefly comment on the impact of operators that enter the Higgs and electroweak processes at the loop level.
A general expectation is that these impacts are under control as long as the corresponding operators are also probed at tree level by other measurements.
Possible exceptions are the operators that modify the Higgs self-coupling or the top couplings, as mentioned earlier.
It was shown in ref.~\cite{DiVita:2017vrr} that, with a single 240\,GeV run at a lepton collider alone, the inclusion of the triple Higgs coupling at one-loop order could worsen the reach of $\delta c_Z$ by a factor of two.
However, once the current projections for the triple Higgs coupling measurement at the HL-LHC are included (with a expected precision of around $50\%$~\cite{Cepeda:2019klc}), we find this impact to be at most around $20\%$ for $\delta g^{ZZ}_H$ and $\delta g^{WW}_H$, and much smaller for other parameters at future lepton colliders.
Reference~\cite{Durieux:2018ggn} found the impact of several top-quark operators on Higgs couplings, in particular the top dipole operators, to be potentially sizeable.%
\footnote{References~\cite{Zhang:2012cd, deBlas:2015aea,Feruglio:2017rjo}, on the other hand, discuss some partial results on the sensitivity to top operators via electroweak precision observables. References~\cite{Brod:2014hsa, Feruglio:2016gvd,Feruglio:2017rjo} also discuss similar effect in the flavour sector.
}
Measurements of top-quark pair production at future lepton colliders~\cite{Durieux:2018tev} are thus important for constraining them.
Needless to say, the top loop contribution to the Higgs vertices also receives corrections from a modification of the top Yukawa coupling, $\delta y_t$.
This contribution is particularly important for the $hgg$ vertex given that the corresponding decay can be measured very well at future lepton colliders \cite{Durieux:2017rsg}.  However, note that our definition of $\delta g^{gg}_H$ (and other parameters) absorbs all contributions to the vertex, so the contribution of $\delta y_t$ is only relevant when mapping $\delta g^{gg}_H$ to the $c_{gg}$ in \autoref{eq:L6hVV}.   
The loop contributions to the $Z$ and $W$ pole observables and the diboson processes in the SMEFT framework have been also discussed in other recent studies, such as refs.~\cite{Baglio:2017bfe, Baglio:2018bkm, Dawson:2019clf}, leading to effects larger than expected. It would therefore be interesting to quantity the impact of these EW loops on our global fit, but such an analysis is certainly beyond the scope of the paper.

%%%%%%%%%%%%%%%%%%%%%%%%%%%
\subsection{Run scenarios and input measurements}
\label{sec:scenarios}
%%%%%%%%%%%%%%%%%%%%%%%%%%%
\begin{figure}
\centering
\includegraphics[width=\textwidth]{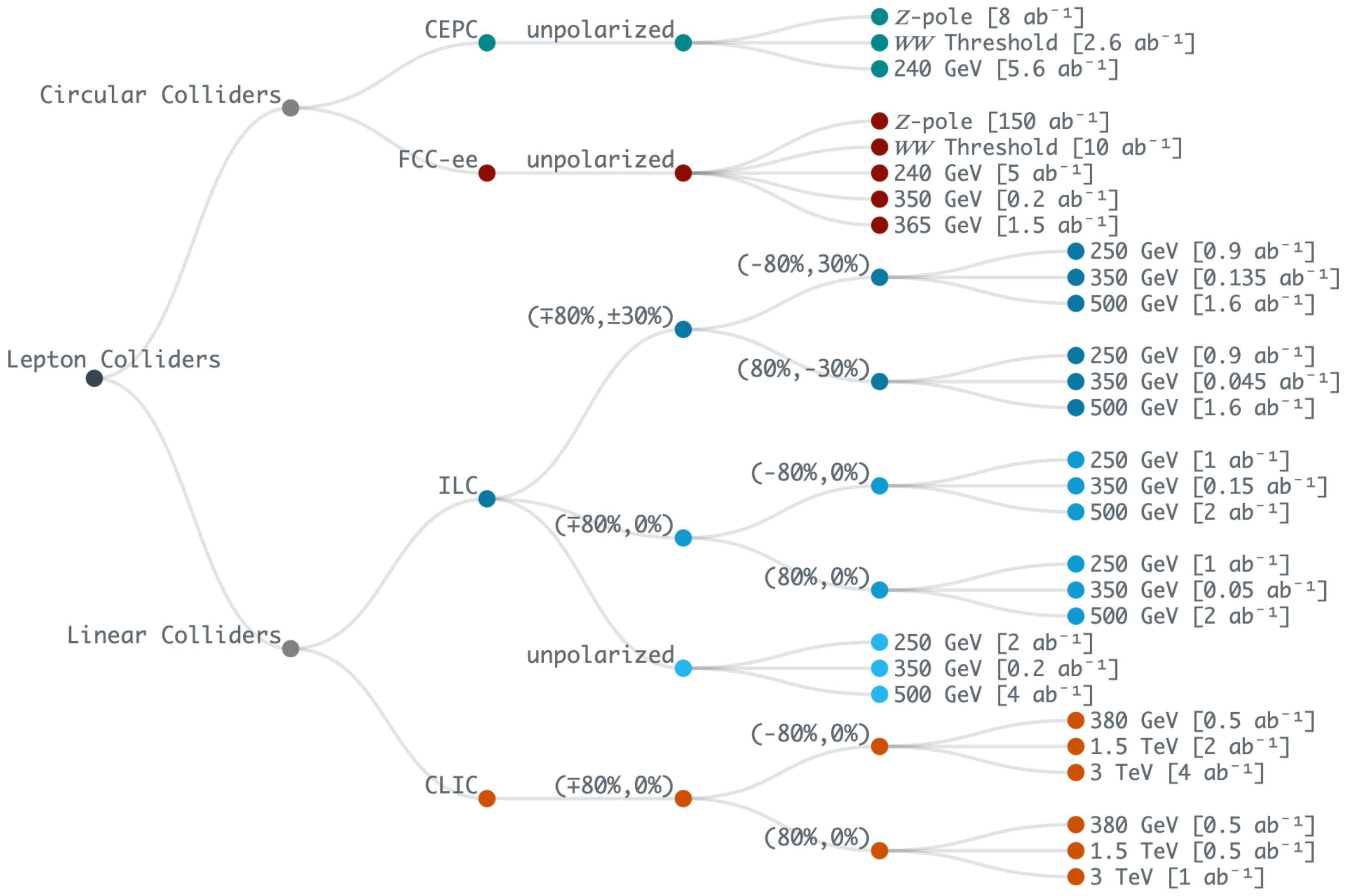}
\caption{\it A summary of run scenarios for CEPC, FCC-ee, ILC and CLIC considered in our analysis, with the corresponding integrated luminosities.
The impact of beam polarization at the ILC is examined by considering $P(e^-,e^+)=(\mp80\%,\pm30\%)$, $(\pm80\%,0\%)$ and unpolarized configurations.
}
\label{fig:scenarios}
\label{tab:scenarios}
\end{figure}

We detail below the future lepton collider run scenarios we assume, together with the sources of measurement precision estimates adopted as input, also for the HL-LHC.
The EW precision measurements of LEP and SLD that we use are listed too.
More details are provided in \autoref{app:inputs}.
{\setlength{\leftmargini}{5mm}
\begin{itemize}

    \item {\bf LEP and SLD:}  The current constraints on EW precision observables from $Z$-pole measurements at LEP and SLD are taken from ref.~\cite{ALEPH:2005ab}.
For the $W$ mass and width, we take the values from PDG~\cite{Tanabashi:2018oca} which also includes measurements from the Tevatron and LHC.
We include the constraints on leptonic branching ratios of $W$ from ref.~\cite{Schael:2013ita}.
Diboson measurements from LEP~II would be completely surpassed by future lepton colliders ones.

    \item {\bf HL-LHC:} For Higgs measurements with $3\inab$ of integrated luminosity, we use the projections provided in the HL-LHC/HE-LHC working group report of ref.~\cite{Cepeda:2019klc}, under the so-called S2 assumption on systematic uncertainties.
We also combine the measurements from ATLAS and CMS to obtain the ultimate reach of the entire HL-LHC program.
We use only the inclusive Higgs measurements.
A differential analysis is left for future work.
Improvements can be expected from the high $p_T$ regions of the Higgsstrahlung ($pp\to Vh$) processes which have enhanced sensitivities~\cite{Franceschini:2017xkh, Banerjee:2018bio} although the validity of EFT becomes a potential issue there~\cite{Contino:2016jqw}.
For the measurements of the diboson $pp\to WW,WZ$ production, we use the results of ref.~\cite{Grojean:2018dqj} which implements the full EFT parametrization with CP-even dimension-six operators including both the aTGCs and the modifications of the quark couplings to gauge bosons.
$W$-mass measurement improvements, to a precision of $7\,$MeV, are also included.
Other electroweak measurements expected at the HL-LHC are not considered.

    \item {\bf CEPC:} We use the official run scenario from the CEPC CDR~\cite{CEPCStudyGroup:2018ghi} which includes runs at the $Z$ pole ($8\inab$), $WW$ threshold ($2.6\inab$) and at $\sqrt{s}=240\,\gev$ ($5.6\inab$) without beam polarization.  A potential upgrade to run at the top-quark pair production threshold, while plausible, is not considered in the current CEPC design.
To palliate the lack of official projections, we make several estimates for $Z$-pole measurement reaches (see~\appendixref{app:Zpole}).

    \item {\bf FCC-ee:} We also use the full FCC-ee run scenario established in its CDR~\cite{Abada:2019lih, Abada:2019zxq}, with runs at the $Z$ pole ($150~\inab$), $WW$ threshold ($10~\inab$), 240\,GeV ($5~\inab$), 350\,GeV ($0.2~\inab$) and 365\,GeV ($1.5~\inab$) without beam polarization.\footnote{Both $10$ and $12\inab$ are considered as benchmark integrated luminosity at the $WW$ threshold.
The difference between them has a negligible impact on our study.
The details of how the 350\,GeV and 365\,GeV runs are combined can be found in~\autoref{app:inputs}.
In the rest of the text, we will refer to this combination of runs as `365\,GeV' unless explicitly specified otherwise.}

    \item {\bf ILC:}  We follow the most recent ILC document~\cite{Bambade:2019fyw} and consider runs at centre-of-mass energies of 250, 350 and 500\,GeV, with total integrated luminosities of $2$, $0.2$ and $4\inab$, respectively.
An upgrade to 1\,TeV, which we do not include, has also been considered in previous documents~\cite{Asner:2013psa}.
It is assumed a longitudinal polarization of $80\%$ ($30\%$) can be achieved for the $e^-$ ($e^+$) beam.
A small fraction of the luminosity planned to be collected with same-helicity beams is mostly useful for controlling systematic uncertainties.  
For Higgs and diboson measurement prospects, we only consider the opposite-helicity runs.
To assess the impact of beam polarization on the Higgs coupling reach, we also investigate scenarios featuring only electron beam polarization or unpolarized beams.
The integrated luminosities considered for each beam polarization run are detailed in~\autoref{tab:scenarios}.
The impact of beam polarizations is further discussed in \autoref{sec:ILCPol}.

    \item {\bf CLIC:}  Following the recent CLIC report~\cite{deBlas:2018mhx}, we use projections for runs at 380\,GeV, 1.5\,TeV and 3\,TeV centre-of-mass energies with $80\%$ longitudinal polarization for the $e^-$ beam and total integrated luminosities of $1$, $2.5$ and $5\,\inab$, respectively.
Their share-out between $P(e^-,\,e^+)=(- 0.8,\, 0) \,/\, (+ 0.8,\, 0)$ configurations are reminded in~\autoref{tab:scenarios}.
Higgs production at centre-of-mass energies beyond about $500\,\gev$ is dominated by the $WW$-fusion mode which involves left-handed electrons. A larger share of luminosity is therefore collected with $(-0.8,0)$ polarization configuration.

\end{itemize}}

For both ILC and CLIC, we will consider EW measurements performed using radiative return to the $Z$ pole under two different assumptions of the systematic uncertainties (see~\autoref{app:Zpole} for detail). But we will not include in our analyses potential EW measurements at the $Z$ pole in dedicated GigaZ runs.\footnote{We refer the reader to the last version of ref.~\cite{deBlas:2019rxi}, where results including such a run at the $Z$ pole at linear colliders are also discussed.}

%%%%%%%%%%%%%%%%%%%%%%%%%%%
\subsection{Diboson production analysis}
\label{sec:wwinput}
%%%%%%%%%%%%%%%%%%%%%%%%%%%

The diboson process, $\eeww$, can be measured both in dedicated runs at the $WW$ threshold at circular machines (CEPC and FCC-ee), and at higher centre-of-mass energies for which the primary targets are Higgs and top-quark measurements.
A line-shape measurement of the diboson pair production at threshold provides precise determinations of the $W$-boson mass and width.
At higher energies, $m_W$ can be measured by the reconstruction of the $W$ from its decay products.
The projected measurement precision from various colliders are summarized in
\autoref{tab:Z-poleInput} of \appendixref{app:Zpole}.

The measurements of diboson production are also essential for the determination of the triple-gauge and $W$-to-fermions couplings.
Various analyses of this process have been performed by future collider collaborations.  
They often adopt the so-called TGC dominance assumption, where any modifications other than those coming from $\delta g_{1,z}$, $\delta \kappa_\gamma$ and $\lambda_z$ in 
\autoref{eq:LaTGC} are neglected.
This was appropriate for LEP~II analyses but it is no longer justified at the LHC already~\cite{Zhang:2016zsp}.
This approximation is also not guaranteed to be valid at future lepton colliders, unless more precise measurements of the other electroweak parameters entering in the process are also available.
For our prospects at the different future colliders, we use an analysis method where we consistently implement 
a full EFT parametrization of the new physics effects entering in $\eeww$ production.
The information contained in the normalized fully differential distribution ---including the production polar angle and the $W$ decay angles--- is maximally exploited using the method of optimal observables.
We briefly review the method in~\autoref{sec:oo}.
Following the theoretical study of ref.~\cite{Diehl:1993br}, LEP~II analyses were already using this technique but in an anomalous coupling approach~\cite{Abbiendi:2003mk, Achard:2004ji, Schael:2004tq, Abdallah:2010zj}.
For simplicity, the narrow-width approximation is employed.
The rate information in all the distinguishable channels is treated separately.
This also allows to easily assess the respective constraining power of rate and differential measurements, each subject to different systematic uncertainties.
Assuming the $W$ has no exotic decays, the inclusive $WW$ cross section as well as $\BR(W\to e\nu \,/\, \mu\nu \,/\, \tau\nu)$ and $\BR(W\to jj)$ branching ratios can all be extracted with a simultaneous fit to all $WW$ channels.
The reach we obtain from this rate analysis, for the various colliders, is summarized in~\autoref{tab:WInput}.
A likelihood is constructed from the $WW$ rate and differential measurements at all available energies, considering only statistical uncertainties, under the assumptions of negligible backgrounds and perfect reconstruction. 
A conservative overall efficiency of $50\%$ is however applied at all centre-of-mass energies and for all beam polarization configurations.\footnote{A selection efficiency of about $70\%$ was found in the ILC diboson analysis performed at $500\,$GeV in the semileptonic channel~\cite{Marchesini:2011aka}.
Background yields were much smaller than signal ones after all selection cuts.
This partially justifies the simple assumptions made here.}

For the differential analysis, we focus on the semileptonic final state with one $W$ decaying to either $e$ or $\mu$ and the other decaying hadronically.
With a sizeable branching fraction of about $29\%$, this channel is well reconstructed and allows for a reliable identification of the $W$ charges.
Some further constraining power could nevertheless be extracted from the fully leptonic and hadronic channels.
Ambiguities in the decay angles of the hadronic $W$ (due to the difficulty to identify quark charges) are taken into account.
Different values of the effective efficiency are also explored in~\autoref{sec:wwscale} to further investigate the impact of the $WW$ measurements on the global fit.

For both rate and differential analyses, we assume all the threshold-scan luminosity is collected at a single centre-of-mass energy of 161\,GeV, although it would be distributed in the 157-172\,GeV range in practice.
To avoid double counting the constraints on the $W$ mass ---note that $m_W$ is also measured using the data at the $WW$ threshold--- we ignore its modifications in our rate analyses (which moreover neglect systematic uncertainties).
Ultimately, experimental collaborations should simultaneously include mass and coupling dependences (the width being function of these parameters in the EFT) at the threshold and above.
Suppressed sensitivities however limit the reach of threshold runs on aTGCs, despite the sizeable luminosities collected.
The $W$ branching ratios are also best measured with the $240$\,GeV run at the circular colliders, which provides the largest $WW$ sample.  
Overall, the $WW$ threshold run is thus mostly relevant for the $W$ mass and width measurements.

In~\autoref{sec:oo} we compare the results of the study using statistically optimal observables with the aTGC reach from a simpler analysis using binned angular distributions. 
As can be seen there, the statistically optimal observables yields significant improvements, which indirectly impact the reach of the different Higgs couplings too (see~\autoref{sec:wwscale}).
This improvement should be encouraging enough for the future collider collaborations to perform dedicated and more realistic $\eeww$ analyses
with this method, using the full EFT parametrization.

%%%%%%%%%%%%%%%%%%%%%%%%%%%
\subsection{Differential Higgsstrahlung analysis}
\label{sec:hzlloo}
%%%%%%%%%%%%%%%%%%%%%%%%%%%

The rate and differential information in Higgsstrahlung production are treated separately, as in diboson production.  
Kinematic distributions have previously been exploited through the definition of angular asymmetries~\cite{Beneke:2014sba, Craig:2015wwr, Durieux:2017rsg}.
We explore here the use of optimal observables defined on the normalized fully differential distribution (see also ref.~\cite{Hagiwara:2000tk}).
Their statistical power is in principle superior.
In practice, we observe only a marginal increase in constraining power.
We exploit the $b\bar{b}e^+e^-$ and $b\bar{b}\mu^+\mu^-$ channels which are almost background free after selection cuts.
A universal $40\%$ signal efficiency is applied, following the CEPC study of ref.~\cite{An:2018dwb}.

While angular observables in $e^+e^-\to hZ$ can potentially be very useful in resolving degeneracies among parameters contributing to the Higgsstrahlung process, in practice their impact on the global-fit results is rather limited~\cite{Durieux:2017rsg}.
Once the data that would be available using the full physics program at the different colliders is combined, 
measurements at multiple energies or with different polarizations already effectively disentangle the various contributions.
Even at the CEPC where Higgsstrahlung production is only accessible at 240\,GeV with unpolarized beams, the measurements of the diboson process, the $WW$ fusion Higgs production as well as various Higgs decay channels already lift approximately flat directions in the global fit.

%%%%%%%%%%%%%%%%%%%%%%%%%%%
\subsection{Fitting procedures}
\label{sec:fit}
%%%%%%%%%%%%%%%%%%%%%%%%%%%

Two different statistical frameworks are used to perform our global fits. The independent implementations use the same inputs. We describe here the two frameworks and their differences.

\HEPfit~\cite{deBlas:2019okz} is a Bayesian analysis framework based on Markov Chain Monte Carlo (MCMC) procedures. For this work, we use the implementation of dimension-six SMEFT in the Warsaw basis~\cite{Grzadkowski:2010es} along with the EW and Higgs observables that are publicly available in the developer version.\footnote{\url{https://github.com/silvest/HEPfit}}
In~\HEPfit, priors are set for the model parameters which, in this case, are the coefficients of the relevant dimension-six operators in the SMEFT.
All priors are set to flat distributions with a range larger than the 5$\sigma$ limits of their posterior distributions.\footnote{The posterior distributions obtained in our fits are always unimodal and almost always Gaussian.
Rare deviations from Gaussianity are small and do not affect our results.}
As a result, there is no prior dependence in the fits performed.
Since all the couplings are linearized in the EFT parameters they have a flat prior distribution too.

The evolution in the space of operator coefficients is guided, using to the Metropolis--Hastings algorithm, with the logarithm of the likelihood function built from measurement projections assuming SM central values.
Where correlations between experimental projections are present, multivariate Gaussian likelihoods are built using the covariance matrix.\footnote{Most of this process is automated and inputs are in the form of plain text files that can be made available on request.}
Instead of presenting the posterior distributions of the Wilson coefficients that were varied in our fits, we provide the statistics of the posterior distributions of the effective Higgs couplings, aTGCs and EW couplings introduced in~\autoref{sec:eft}.
  
An independent fit procedure relies on the construction of a $\chi^2$ function from all observables.
Only leading order SMEFT contribution are retained, in the so-called Higgs basis of dimension-six operators~\cite{Falkowski:2001958}.
The $\chi^2$ is then quadratic in the EFT parameters, given by
\begin{equation}
	\chi^2 = \sum_{ij} (c-c_0)_i  \, [\sigma^{-2}]_{ij} \, (c-c_0)_j \,, \hspace{1cm} \mbox{where}  \hspace{0.5cm}   
	\sigma^{-2} \equiv \left( \boldsymbol{\delta c}^T {\boldsymbol \rho} ~ \boldsymbol{\delta c}  \right)^{-1}
	\,,  \label{eq:chipara}
\end{equation}
where $c_{i=1, ... }$ are the effective parameters listed in~\autoref{eq:parameters}.
The one-sigma precisions $\delta c_i$, and the correlation matrix $\rho_{ij}$, are derived from $[\sigma^{-2}]_{ij} =\frac 12 \left.\frac{\partial^2 \,\chi^2} {\partial c_i \partial c_j}\right|_{c=c_0}$.

We observe excellent agreement between the results of the two fit procedures, in particular at future lepton colliders where the leading order SMEFT expansion is well justified by the high measurement precision.

%%%%%%%%%%%%%%%%%%%%%%%%%%%
\section{Results}
\label{sec:results}
%%%%%%%%%%%%%%%%%%%%%%%%%%%
\begin{figure}[t]
\centering
\includegraphics[width=\textwidth]{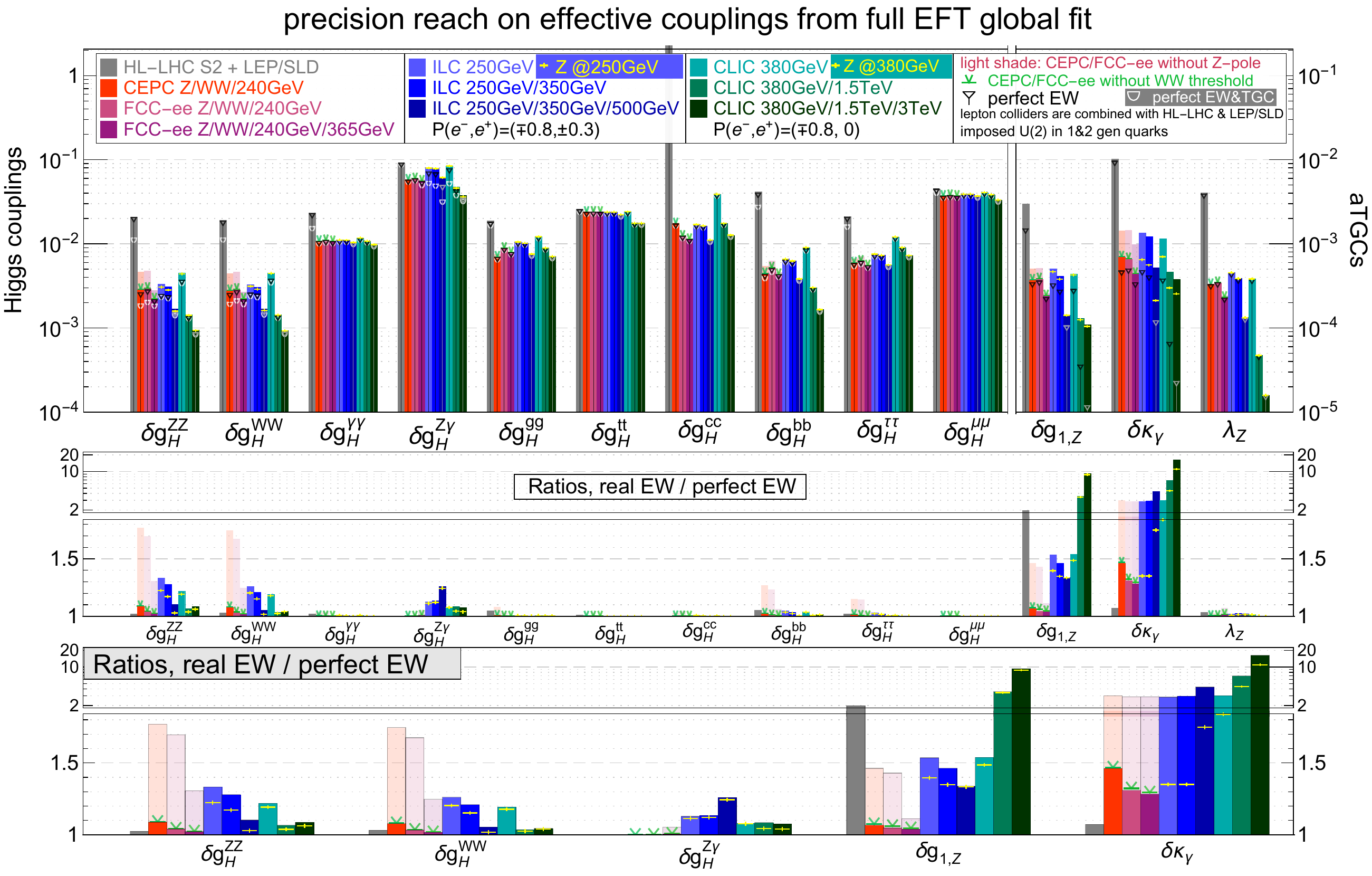}
\caption{\it
Global one-sigma reach of future lepton colliders on Higgs and triple-gauge couplings.
The run scenarios and luminosities assumed are listed in~\autoref{tab:scenarios}.
LEP and SLD electroweak measurements as well as HL-LHC prospects on Higgs and diboson processes are included in all projections.
Modifications of electroweak parameters (shown in~\autoref{fig:ew}) are marginalized over to obtain the prospects displayed as bars, and artificially set to zero to obtain those shown with triangular marks.
For the CEPC and FCC-ee, scenarios without the future $Z$-pole ($WW$ threshold) run are shown as light shaded bars (lower edges of the green marks).
For ILC, the results with the inclusion of the $A_{LR}$ measurement at 250\,GeV are shown with yellow marks.
The bottom panel highlights the couplings that are affected significantly EW uncertainties.
Numerical results are also reported in \autoref{tab:allcoll}
}
\label{fig:money}
\end{figure}

The global reach we obtain on the Higgs and triple-gauge couplings is highlighted in~\autoref{fig:money} for the four future lepton colliders considered: CEPC, FCC-ee, ILC and CLIC.
Numerical values are also provided in~\autoref{tab:allcoll}.  
The LEP and SLD measurements as well as HL-LHC prospects are included in all scenarios.
They are also combined separately, providing a reference to assess the improvement brought by future lepton collider measurements.
We also display separately the precision reached after each stage of the FCC-ee, ILC and CLIC programs accessing increasing centre-of-mass energies.
The precision reached for each Higgs and triple-gauge coupling, after marginalizing over the other parameters, is shown as a solid bar in the figure.
These results account for the finite accuracy with which EW parameters will be determined at the different future colliders (see~\autoref{fig:ew} below).
They are compared with the projections (displayed with triangular marks) obtained assuming EW coupling modifications are constrained to vanish by perfect EW measurements ($Z$-pole observables, $W$ mass, width and branching factions).
The ratio between the results assuming finite and perfect EW measurements are also displayed in the lower panel of the figure. 
Assuming both EW observables and TGCs are perfectly determined to be SM-like, one obtains the projections shown as semi-circular marks in~\autoref{fig:money}.
For the CEPC and FCC-ee, we assess the impact of the planned $Z$-pole run by excluding the latter in prospects shown with light shaded bars.
These results allow us to answer several questions raised in our introduction concerning the impact of EW parameter uncertainties on Higgs coupling determinations.
Examining in particular the middle and bottom panel of~\autoref{fig:money}, let us discuss in turn the qualitatively distinct cases of circular and linear colliders.

\begin{figure}[t]
\centering
\includegraphics[width=0.69\textwidth]{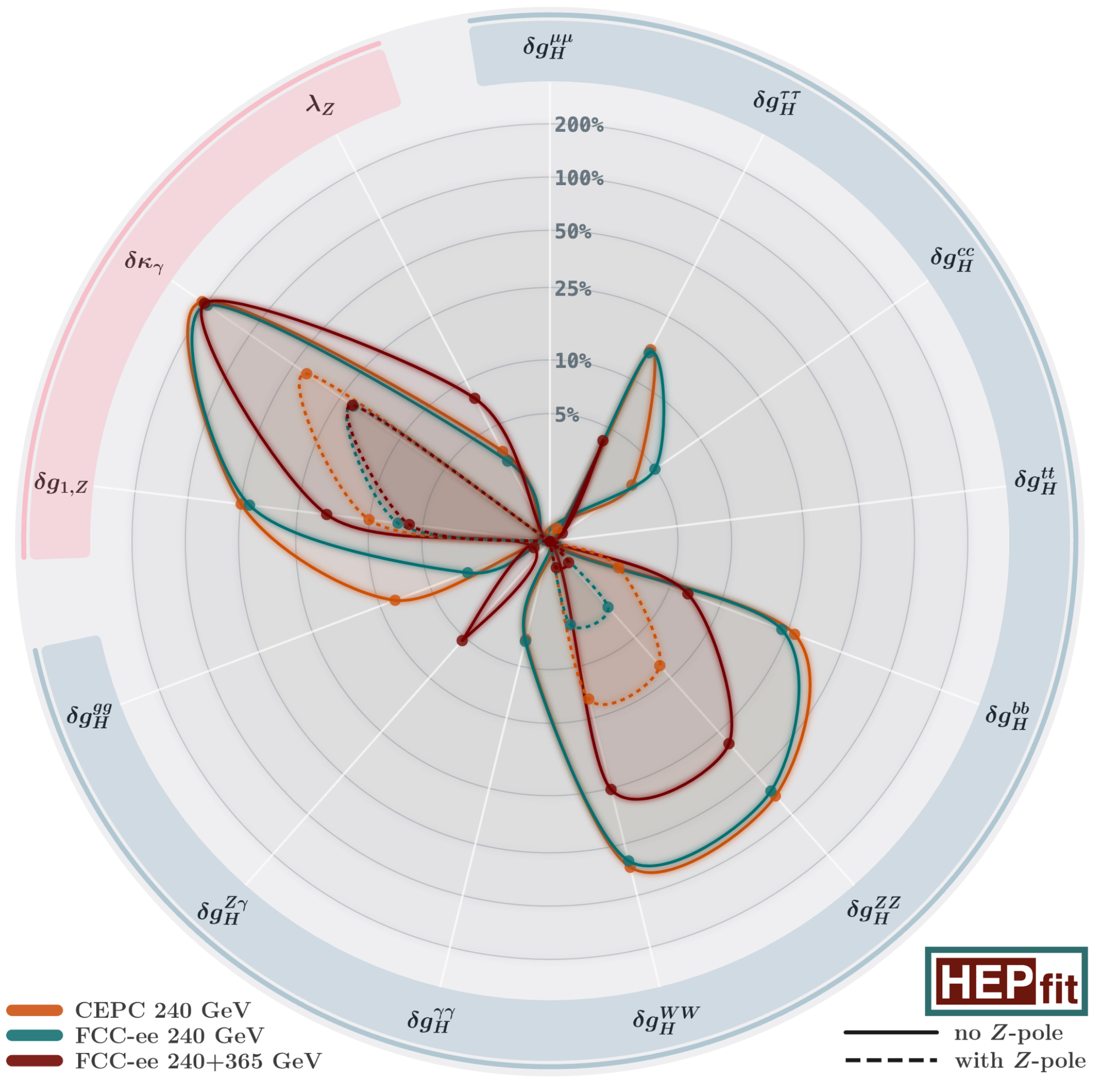}
\caption{\it
Degradation in Higgs and triple-gauge coupling determinations due to EW uncertainties at future circular colliders.
It is obtained by comparison with a perfect EW measurement scenario and quantified as ${\delta g}/{\delta g(\textrm{EW}\rightarrow 0)}-1$ expressed in percent.
The dashed and solid lines are respectively obtained with and without new $Z$-pole run.
Numerical values are also provided in \autoref{tab:allcoll}.
}
\label{fig:radarEW}
\end{figure}

%{\unskip\parfillskip 0pt \par}

\newcolumntype{g}{>{\columncolor{Gray}}c}
\begin{landscape}%
\begin{table}
\begin{center}%
\renewcommand{\arraystretch}{1.25}%
\newtabulinestyle { dd=0.25pt on 0.5pt off 2pt }
{\notsotiny%
\begin{tabu} to \textwidth {|r|g||g|[dd]g||g|[dd]g|g|[dd]g||g|g|g|g|g|g||g|g|g|}
\hline
\rowcolor{white}\multirow{4}{*}{}			&					&\multicolumn{6}{c||}{ \bf Future Circular Colliders} 												& \multicolumn{9}{c|}{\bf Future Linear Colliders}\\
				 		\hhline{|~|~||---------------|}
\rowcolor{white}					&					&\multicolumn{2}{c||}{\bf CEPC}	&\multicolumn{4}{c||}{\bf FCC-ee} 	 							&\multicolumn{6}{c||}{{\bf ILC} [P1 $\Rightarrow$ ($\mp$80\%,$\pm$30\%)] [UP $\Rightarrow$ unpolarized]} 		&\multicolumn{3}{c|}{\bf CLIC}\\
						\hhline{|~|~||---------------|}
\rowcolor{white}				&\multirow{-3}{*}{\bf HL-LHC}	&\multicolumn{2}{c||}{240\,GeV}		&\multicolumn{2}{c|}{240\,GeV} 	&\multicolumn{2}{c||}{+365\,GeV}		&\multicolumn{2}{c|}{250\,GeV}	&\multicolumn{2}{c|}{+350\,GeV}	&\multicolumn{2}{c||}{+500\,GeV}	&380\,GeV	 	&+1.5\,TeV		&+3\,TeV\\
						\hhline{|~|----------------|}
\rowcolor{white}								&S2		&\CW\xcancel{Z-pole}&\CW $Z$-pole			&\CW\xcancel{Z-pole}&\CW $Z$-pole		&\CW\xcancel{Z-pole}&\CW $Z$-pole			&\CW\;P1\; 	&\CW\;UP\;			&\CW\;P1\;	&\CW\;UP\;				&\CW\;P1\;	&\CW\;UP\;				&\multicolumn{3}{c|}{($\mp$80\%,0\%)}\\
\hline
\hline										
\rowcolor{white}\multirow{2}{*}{$\delta g^{\mu\mu}_{H}$\tp}  	&4.49	&3.68		& 3.68			&3.76		&3.75		&3.67		&3.68			&3.90	&3.93			&3.89	&3.91				&3.70	&3.72				&4.08	&3.86	&3.33	\\\tabucline[0.15pt on 0.5pt off 2pt]{2-17}
											&4.45	&\multicolumn{2}{c||}{\CG3.67}		&\multicolumn{2}{c|}{\CG3.74}	&\multicolumn{2}{c||}{\CG3.67}	&3.89	&3.92			&3.89	&3.91				&3.70	&3.72				&4.08	&3.85	&3.31	\\
\hline										
\rowcolor{white}\multirow{2}{*}{$\delta g^{\tau\tau}_{H}$\tp} 	&2.13	&~0.69\as	& 0.60			&~0.72\as	&0.63		&0.57		&0.55			&0.77	&0.89			&0.75	&0.83				&0.57	&0.58				&1.22	&0.91	&0.73	\\\tabucline[0.15pt on 0.5pt off 2pt]{2-17}
											&2.07	&\multicolumn{2}{c||}{\CG0.59}		&\multicolumn{2}{c|}{\CG0.63}	&\multicolumn{2}{c||}{\CG0.55}	&0.74	&0.82			&0.72	&0.79				&0.56	&0.58				&1.21	&0.90	&0.73	\\
\hline										
\rowcolor{white}\multirow{2}{*}{$\delta g^{cc}_{H}$\tp} 		&--		&1.80		& 1.74			&1.30		&1.25		&1.15		&1.13			&1.72	&1.87			&1.65	&1.79				&1.12	&1.14				&3.93	&1.77	&1.28	\\\tabucline[0.15pt on 0.5pt off 2pt]{2-17}
											&--		&\multicolumn{2}{c||}{\CG1.74}		&\multicolumn{2}{c|}{\CG1.24}	&\multicolumn{2}{c||}{\CG1.13}	&1.70	&1.83			&1.64	&1.76				&1.11	&1.14				&3.91	&1.76	&1.28	\\
\hline
\rowcolor{white}\multirow{2}{*}{$\delta g^{tt}_{H}$\tp}  		&2.62	&2.35		& 2.34			&2.39		&2.39		&2.38		&2.38			&2.41	&2.41			&2.41	&2.41				&2.39	&2.38				&2.44	&~2.22\as	&~2.20\as	\\\tabucline[0.15pt on 0.5pt off 2pt]{2-17}
											&2.58	&\multicolumn{2}{c||}{\CG2.35}		&\multicolumn{2}{c|}{\CG2.40}	&\multicolumn{2}{c||}{\CG2.38}	&2.40	&2.41			&2.41	&2.41				&2.37	&2.37				&2.43	&1.84	&1.83	\\
\hline										
\rowcolor{white}\multirow{2}{*}{$\delta g^{bb}_{H}$\tp} 		&4.27	&~0.57\as	& 0.46			&~0.65\as	&0.53		&0.46		&0.44			&0.68	&~0.83\as		&0.65	&0.76				&0.39	&0.40				&0.92	&0.30	&0.17	\\\tabucline[0.15pt on 0.5pt off 2pt]{2-17}
											&3.95	&\multicolumn{2}{c||}{\CG0.45} 		&\multicolumn{2}{c|}{\CG0.53}	&\multicolumn{2}{c||}{\CG0.44}	&0.65	&0.74			&0.62	&0.71				&0.39	&0.39				&0.89	&0.30	&0.16	\\
\hline										
\rowcolor{white}\multirow{2}{*}{$\delta g^{ZZ}_{H}$\tp} 		&2.08	&~0.48\as	& 0.30			&~0.50\as	&0.30		&~0.28\as	&0.22			&~0.35\as&~0.57\as		&~0.33\as&~0.49\as			&~0.18\as&~0.19\as			&~0.47\as	&0.14	&0.09	\\\tabucline[0.15pt on 0.5pt off 2pt]{2-17}
											&2.00	&\multicolumn{2}{c||}{\CG0.28}		&\multicolumn{2}{c|}{\CG0.30}	&\multicolumn{2}{c||}{\CG0.22}	&0.25	&0.42			&0.24	&0.39				&0.16	&0.17				&0.38	&0.14	&0.09	\\
\hline										
\rowcolor{white}\multirow{2}{*}{$\delta g^{WW}_{H}$\tp} 		&1.95	&~0.47\as	& 0.30			&~0.48\as	&0.30		&~0.27\as	&0.22			&~0.34\as&~0.56\as		&~0.32\as&~0.48\as			&0.18	&~0.19				&~0.47\as	&0.15	&0.09	\\\tabucline[0.15pt on 0.5pt off 2pt]{2-17}
											&1.87	&\multicolumn{2}{c||}{\CG0.27}		&\multicolumn{2}{c|}{\CG0.29}	&\multicolumn{2}{c||}{\CG0.22}	&0.26	&0.42			&0.26	&0.39				&0.17	&0.18				&0.39	&0.14	&0.09	\\
\hline								
\rowcolor{white}\multirow{2}{*}{$\delta g^{\gamma\gamma}_{H}$\tp}&2.35&1.10			& 1.07			&1.13		&1.09		&1.07		&1.06			&1.12	&1.19			&1.11	&1.16				&1.04	&1.05				&1.18	&1.09	&0.98	\\\tabucline[0.15pt on 0.5pt off 2pt]{2-17}
											&2.27	&\multicolumn{2}{c||}{\CG1.07}		&\multicolumn{2}{c|}{\CG1.09}	&\multicolumn{2}{c||}{\CG1.06}	&1.11	&1.15			&1.10	&1.14				&1.04	&1.04				&1.17	&1.08	&0.98	\\
\hline									
\rowcolor{white}\multirow{2}{*}{$\delta g^{Z\gamma}_{H}$\tp}	&9.13	&5.73		& 5.72			&5.95		&5.92		&5.78		&5.54			&~8.09\as&8.78			&~7.96\as&8.70				&~6.31\as&~7.71\as			&8.57	&4.59	&3.61	\\\tabucline[0.15pt on 0.5pt off 2pt]{2-17}
											&9.14	&\multicolumn{2}{c||}{\CG5.72}		&\multicolumn{2}{c|}{\CG5.94}	&\multicolumn{2}{c||}{\CG5.49}	&7.28	&8.71			&7.10	&8.55				&5.02	&6.73				&7.94	&4.19	&3.40	\\
\hline									
\rowcolor{white}\multirow{2}{*}{$\delta g^{gg}_{H}$\tp} 		&1.84	&0.74		& 0.69			&0.90		&0.87		&0.78		&0.77			&1.03	&1.06			&1.00	&1.04				&0.75	&0.76				&1.19	&0.87	&0.70	\\\tabucline[0.15pt on 0.5pt off 2pt]{2-17}
											&1.76	&\multicolumn{2}{c||}{\CG0.69}		&\multicolumn{2}{c|}{\CG0.87}	&\multicolumn{2}{c||}{\CG0.77}	&1.02	&1.05			&0.99	&1.03				&0.74	&0.76				&1.18	&0.87	&0.70	\\
\hline
\hline								
\rowcolor{white}\multirow{2}{*}{$\delta g_{1,Z}$\tpm} 			&~2.96\as&~0.52\as	& 0.38			&~0.53\as	&0.39		&~0.26\as	&0.24			&~0.50\as&~0.63\as		&~0.41\as&~0.52\as			&0.14	&~0.20\as			&~0.42\as	&~0.14\as	&~0.13\as	\\\tabucline[0.15pt on 0.5pt off 2pt]{2-17}
											&1.46	&\multicolumn{2}{c||}{\CG0.34}		&\multicolumn{2}{c|}{\CG0.36}	&\multicolumn{2}{c||}{\CG0.22}	&0.33	&0.50			&0.28	&0.44				&0.11	&0.17				&0.27	&0.03	&0.01	\\
\hline
\rowcolor{white}\multirow{2}{*}{$\delta\kappa_\gamma$\tpm} 		&9.93	&~1.57\as	& ~0.72\as		&~1.58\as	&~0.64\as	&~1.09\as	&~0.43\as		&~1.53\as&~1.73\as		&~1.39\as&~1.58\as			&~0.58\as&~0.75\as				&~1.29\as	&~0.47\as	&~0.32\as	\\\tabucline[0.15pt on 0.5pt off 2pt]{2-17}
											&9.49	&\multicolumn{2}{c||}{\CG0.50}		&\multicolumn{2}{c|}{\CG0.53}	&\multicolumn{2}{c||}{\CG0.36}	&0.51	&0.77			&0.44	&0.68				&0.13	&0.23				&0.39	&0.07	&0.02	\\
\hline								
\rowcolor{white}\multirow{2}{*}{$\lambda_Z$\tpm} 				&3.99	&0.34		& 0.33			&0.35		&0.34		&0.24		&0.24			&0.46	&0.51			&0.39	&0.45				&0.13	&0.14				&0.39	&0.05	&0.02	\\\tabucline[0.15pt on 0.5pt off 2pt]{2-17}
											&3.86	&\multicolumn{2}{c||}{\CG0.33}		&\multicolumn{2}{c|}{\CG0.34}	&\multicolumn{2}{c||}{\CG0.23}	&0.45	&0.50			&0.38	&0.43				&0.13	&0.14				&0.38	&0.05	&0.02	\\
\hline
\end{tabu}%
}%
\caption{\it
Global one-sigma reach of future collider measurements on Higgs and triple-gauge couplings.
All projections include current EW measurements and HL-LHC prospects.
Projections in grey rows assume perfect EW measurements.
The numbers marked with a \as~are then improved by more than 10\%.
``$Z$-pole'' and ``\xcancel{$Z$-pole}'' refer to the inclusion of new $Z$-pole runs at circular colliders.
The results are graphically presented in \autoref{fig:money}.
The effects of a $Z$-pole run at circular colliders is stressed in \autoref{fig:radarEW}.
}
\label{tab:allcoll}%
\end{center}%
\end{table}%
\end{landscape}

%\noindent
At circular colliders, as expected, the full programs (dark shaded bars) provide sufficient constraints on EW parameters to limit the impact of their uncertainties on Higgs coupling determinations.
For all couplings except $\delta\kappa_\gamma$, imperfect determinations of EW parameters impact Higgs coupling prospects by less than $10\%$.
The high luminosities collected at the $Z$ pole and the low systematics are crucial in this respect.
Removing the future $Z$-pole runs (light shaded bars), one observes significant degradations, reaching for instance factors of $1.7$ for $\delta g_H^{ZZ}$ and $\delta g_H^{WW}$, $1.4$ for $\delta g_{1,Z}$, and $1.25$ for $\delta g_H^{bb}$ at CEPC.
The inclusion of higher-energy runs ($\sqrt{s}=350,365\,\gev$) available for the FCC-ee somewhat mitigates the impact of an absence of $Z$-pole run.
On the other hand, the $WW$ threshold run has a rather limited impact on the precision reach for all Higgs and triple-gauge couplings.
It only improves the prospects for $\delta \kappa_\gamma$ by a factor of $1.05$ ($1.10$) at the CEPC (FCC-ee).
The impact of a $Z$-pole run at circular colliders is further illustrated in \autoref{fig:radarEW}.
It shows the degradation in Higgs and triple-gauge couplings due to EW uncertainties, obtained by comparison with perfect EW measurement scenarios.
The figure of merit employed is ${\delta g}/{\delta g(\textrm{EW}\rightarrow 0)}-1$ expressed in percent.
The solid and dashed lines are respectively obtained in the absence and presence of a new $Z$-pole runs.

At linear colliders, the lack of runs dedicated to electroweak coupling measurements renders substantial the contaminations from their uncertainties in Higgs coupling determinations.
Beam polarization (further discussed in~\autoref{sec:ILCPol}) does not seem to be entirely sufficient to mitigate these indirect uncertainties arising from marginalizing over electroweak parameters.
With one single energy run, the $\delta g_H^{ZZ}$, $\delta g_H^{WW}$ coupling reaches are for instance worsened by a factor of about $1.2$--$1.3$ at the ILC and CLIC.
These degradations are however reduced to factors of about $1.1$ with the inclusion of higher-energy runs which are sensitive to different combinations of parameters and can help resolving approximate degeneracies.
This is remarkable since improvements in the absolute strength of constraints at high energies also tend to increase the relative impact of EW uncertainties.
This effect is observed for $\delta g_H^{Z\gamma}$ as well as for $\delta g_{1,Z}$ and $\delta \kappa_\gamma$, indicating a striking breakdown of the TGC dominance assumption.
At the ILC, the measurement of Higgsstrahlung production with polarized beams provides the main constraint on $\delta g_H^{Z\gamma}$~\cite{Durieux:2017rsg} and is more sensitive to EW contaminations than the $h\to Z\gamma$ decay.
The relative contaminations from EW uncertainties on $\delta g_{1,Z}$ and $\delta \kappa_\gamma$ increase much faster with higher-energy runs than the absolute constraints.
This is a consequence of the fact that high centre-of-mass energies drastically improve constraints only on specific combinations of parameters including electroweak coupling modifications~\cite{Franceschini:2017xkh}.
Relative degeneracies are thus effectively enhanced.

{Besides $e^+e^-\to W^+W^-$, other electroweak measurements could help controlling electroweak uncertainties in the centre-of-mass energy range envisioned for future linear colliders.
One could for instance exploit the lower tail of the beam energy spectrum to access the $Z$ pole through \emph{radiative return}~\cite{Karliner:2015tga}, or resolved photon emission in association with a $Z$ boson ($e^+e^+\to Z\gamma$), or di-$Z$ production.
Radiative return to the $Z$ pole has for instance been considered with measurements of the left-right production asymmetry $A_{LR}$, as well as improvements in the measurements of $Z$ decays and asymmetries in final states with charged leptons, $b$- and $c$-quarks.
Preliminary prospects for the determination of $A_{LR}$ at $\sqrt{s}=250\,\gev$ claim the relative statistical error can be reduced to about 
\unskip\parfillskip 0pt \par}
% CONTINUED AFTER THE TABLE

\begin{landscape}%
\begin{table}%
\begin{center}%
\renewcommand{\arraystretch}{1.60}%
{\scriptsize
\begin{tabu}{|l|c||c|[dd]g||c|[dd]g|c|[dd]g||c|c|c|c|c|c||c|c|c|}
\hline
\multirow{3}{*}{}			&					&\multicolumn{6}{c||}{ \bf Future Circular Colliders} 												& \multicolumn{9}{c|}{\bf Future Linear Colliders}\\
						\cline{3-17}
						%\cline{3-17}
					&					&\multicolumn{2}{c||}{\bf CEPC}	&\multicolumn{4}{c||}{\bf FCC-ee} 	 							&\multicolumn{6}{c||}{{\bf ILC} [P1 $\Rightarrow$ ($\mp$80\%,$\pm$30\%)] [UP $\Rightarrow$ unpolarized]} 		&\multicolumn{3}{c|}{\bf CLIC}\\
						\cline{3-17}
				&\multirow{-3}{*}{\bf HL-LHC}	&\multicolumn{2}{c||}{240\,GeV}		&\multicolumn{2}{c|}{240\,GeV} 	&\multicolumn{2}{c||}{+365\,GeV}		&\multicolumn{2}{c|}{250\,GeV}	&\multicolumn{2}{c|}{+350\,GeV}	&\multicolumn{2}{c||}{+500\,GeV}	&380\,GeV	 	&+1.5\,TeV		&+3\,TeV\\
				        \hhline{|~|----------------|}
				% 		\tabucline{2-17}
~~(\textperthousand)	    			&S2 		&\xcancel{$Z$-pole}& $Z$-pole		&\xcancel{Z-pole}&Z-pole		&\xcancel{Z-pole}&Z-pole			&\;\;P1\;\; 	&\;\;UP\;\;			&\;\;P1\;\;	&\;\;UP\;\;				&\;\;P1\;\;	&\;\;UP\;\;				&\multicolumn{3}{c|}{($\mp$80\%,0\%)}\\
\hline
\hline										
$\delta g^{ee}_{Z,L}$\as  					&0.28	&0.21		&0.05			&0.20		&0.02		&0.14		&0.02			&0.21	&0.22			&0.19	&0.20				&0.10	&0.12				&0.19	&0.10	&0.08	\\

\hline										
$\delta g^{\mu\mu}_{Z,L}$\as 				&1.13	&0.81		&0.08			&0.77		&0.02		&0.61		&0.02			&1.06	&1.07			&1.05	&1.06				&1.03	&1.03				&1.06	&1.04	&1.04	\\

\hline										
$\delta g^{\tau\tau}_{Z,L}$ 				&0.59	&0.52		&0.10			&0.51		&0.02		&0.50		&0.02			&0.58	&0.58			&0.58	&0.58				&0.58	&0.58				&0.58	&0.58	&0.58	\\

\hline
$\delta g^{ee}_{Z,R}$\as  					&0.27	&0.22		&0.05			&0.22		&0.02		&0.18		&0.02			&0.21	&0.23			&0.20	&0.23				&0.12	&0.17				&0.22	&0.13	&0.10	\\

\hline										
$\delta g^{\mu\mu}_{Z,R}$\as 				&1.28	&1.01		&0.08			&0.99		&0.01		&0.84		&0.01			&1.16	&1.18			&1.15	&1.16				&1.11	&1.12				&1.17	&1.13	&1.13	\\

\hline										
$\delta g^{\tau\tau}_{Z,R}$ 				&0.62	&0.60		&0.11			&0.60		&0.02		&0.60		&0.02			&0.60	&0.60			&0.60	&0.60				&0.60	&0.60				&0.60	&0.60	&0.60	\\

\hline
\hline										
$\delta g^{e\nu}_{W}$\as\as 				&6.33	&0.05		&0.05			&0.05		&0.04		&0.04		&0.04			&0.07	&0.08			&0.07	&0.07				&0.05	&0.05				&0.13	&0.10	&0.10	\\

\hline								
$\delta g^{\mu\nu}_{W}$\as\as				&5.84	&0.21		&0.18			&0.18		&0.13		&0.17		&0.12			&0.37	&0.39			&0.36	&0.38				&0.28	&0.29				&0.65	&0.54	&0.50	\\

\hline									
$\delta g^{\tau\nu}_{W}$\as\as 				&7.89	&0.21		&0.18			&0.18		&0.13		&0.17		&0.12			&0.37	&0.39			&0.36	&0.38				&0.28	&0.29				&0.66	&0.54	&0.50	\\

\hline
\hline									
$\delta g^{uu}_{Z,L}$\as 					&2.84	&2.36		&0.34			&2.35		&0.59		&2.35		&0.59			&2.36	&2.36			&2.36	&2.36				&2.35	&2.36				&2.39	&2.37	&2.36	\\

\hline									
$\delta g^{uu}_{Z,R}$ 						&4.66	&4.44		&0.61			&4.45		&1.20		&4.43		&1.20			&4.44	&4.45			&4.44	&4.45				&4.42	&4.42				&4.46	&4.39	&4.39	\\

\hline									
$\delta g^{dd}_{Z,L}$\as 					&2.73	&2.35		&0.37			&2.35		&0.60		&2.35		&0.60			&2.35	&2.35			&2.36	&2.35				&2.35	&2.36				&2.36	&2.35	&2.35	\\

\hline									 
$\delta g^{dd}_{Z,R}$\as 					&14.4	&10.4		&1.79			&10.4		&3.00		&10.4		&2.99			&10.6	&10.6			&10.6	&10.6				&10.5	&10.5				&10.7	&10.5	&10.5	\\

\hline									
$\delta g^{bb}_{Z,L}$ 						&1.61	&1.55		&0.27			&1.55		&0.30		&1.51		&0.30			&1.55	&1.57			&1.54	&1.56				&1.46	&1.49				&1.56	&1.47	&1.45	\\

\hline									
$\delta g^{bb}_{Z,R}$ 						&8.22	&8.02		&1.45			&8.00		&1.62		&7.75		&1.62			&7.98	&8.08			&7.90	&8.02				&7.43	&7.61				&8.03	&7.45	&7.35	\\

\hline
\end{tabu}%
}%
\caption{\it
Global one-sigma reach on electroweak couplings, in \textperthousand.
All scenarios include HL-LHC projections as well as LEP and SLC electroweak measurements.
With respect to this baseline, the couplings marked with a \as~see their constraints improved by more than 10\% and those marked with \as\as~by more than an order of magnitude.
``$Z$-pole'' and ``\xcancel{$Z$-pole}'' refer to the inclusion of new $Z$-pole runs at circular colliders.
The results are graphically presented in \autoref{fig:ew}.}
\label{tab:allcollEW}%
\end{center}%
\end{table}%
\end{landscape}

\begin{figure}[t]
\centering
\includegraphics[width=\textwidth]{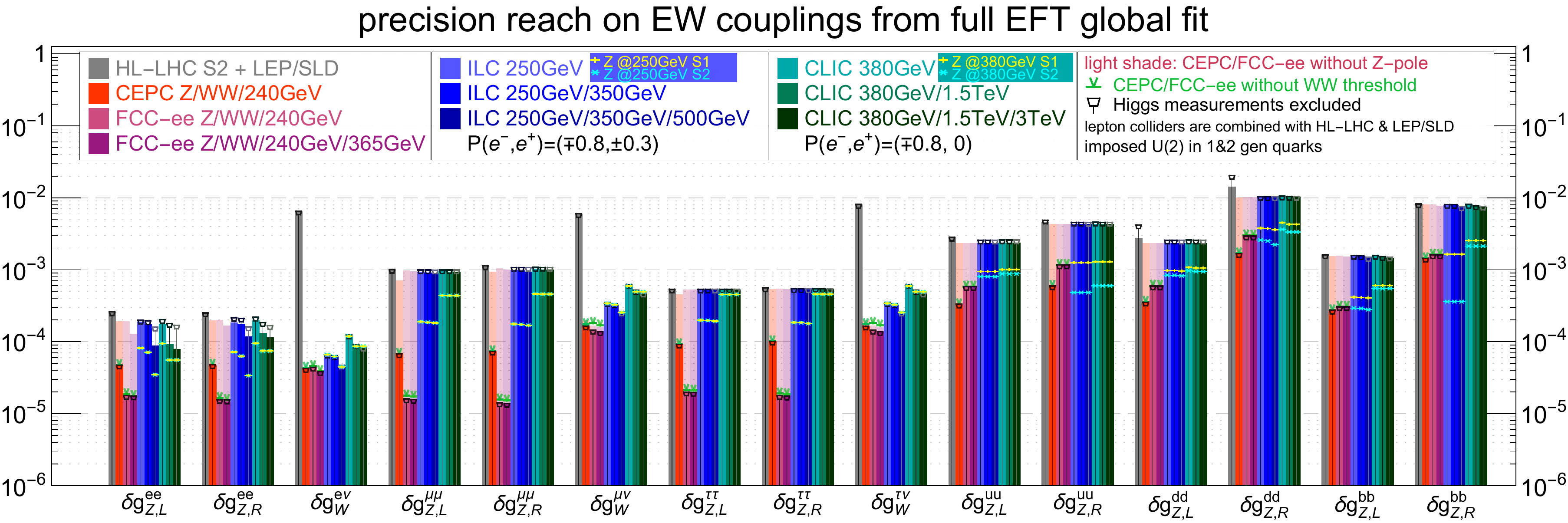}
\caption{\it
Global one-sigma reach on electroweak couplings for the same scenarios as in \autoref{fig:money}.
Higgs and triple-gauge coupling modifications are marginalized over. 
Trapezoidal and green marks respectively indicate the prospects obtained with Higgs and $WW$ threshold measurements excluded.
The numerical results are reported in \autoref{tab:allcollEW}.
}
\label{fig:ew}
\end{figure}

\noindent
$0.1\%$~\cite{Ueno:2018}, a factor of 15 improvement with respect to the $1.5\%$ one obtained by SLD~\cite{ALEPH:2005ab}.  
The dominant uncertainties associated to the knowledge of polarization are included and seem to be smaller than the statistical ones.
Still this estimate will need to be confirmed after full detector simulation, resolved photon production, and minute systematic uncertainties are fully accounted for.
As illustration, we nevertheless display the improvement that would be brought by such a measurement with yellow marks in~\autoref{fig:money}.  
It would mostly benefit the triple-gauge coupling $\delta\kappa_\gamma$.  
If additional electroweak measurements appear insufficient to control EW uncertainties contaminations to a satisfactory level, collecting some amount of luminosity at lower centre-of-mass energies might be advantageous.

Other than the prospects on the Higgs and triple-gauge couplings provided in \autoref{fig:money}, we present in \autoref{fig:ew} projections for the rest of electroweak couplings in the same run scenarios.
Numerical results are provided in \autoref{tab:allcollEW}.
Note that the only electroweak measurements included in HL-LHC projections are that of diboson production~\cite{Grojean:2018dqj} and of the $W$ mass~\cite{Azzi:2019yne}.
They are combined with LEP and SLD ones.
The latter will continue to dominate the constraints on $Z$-boson couplings to fermions until a new lepton collider is built, which naturally brings significant improvements either from direct $Z$-pole measurements or from measurements using $Z$-radiative return. 
Diboson measurements accessible to all future lepton colliders have a dramatic impact on our knowledge of the couplings of $W$-boson to the leptons.\footnote{No study of the correlations at the future colliders between the different decay modes of the $W$ is available.
They were of the order $10\%-20\%$ at LEP2, but better flavour tagging is expected at any future collider and these correlations are not anticipated to change significantly our results.}
The lower energy runs at circular colliders provide the best reaches on these couplings given the higher $\eeww$ production rates and luminosities.
Runs at the $WW$ production threshold however only play a marginal role once high luminosities are collected at centre-of-mass energies of $240\,\gev$ and above.

\begin{figure}[t !]
  \centering
  \includegraphics[width=0.9\textwidth]{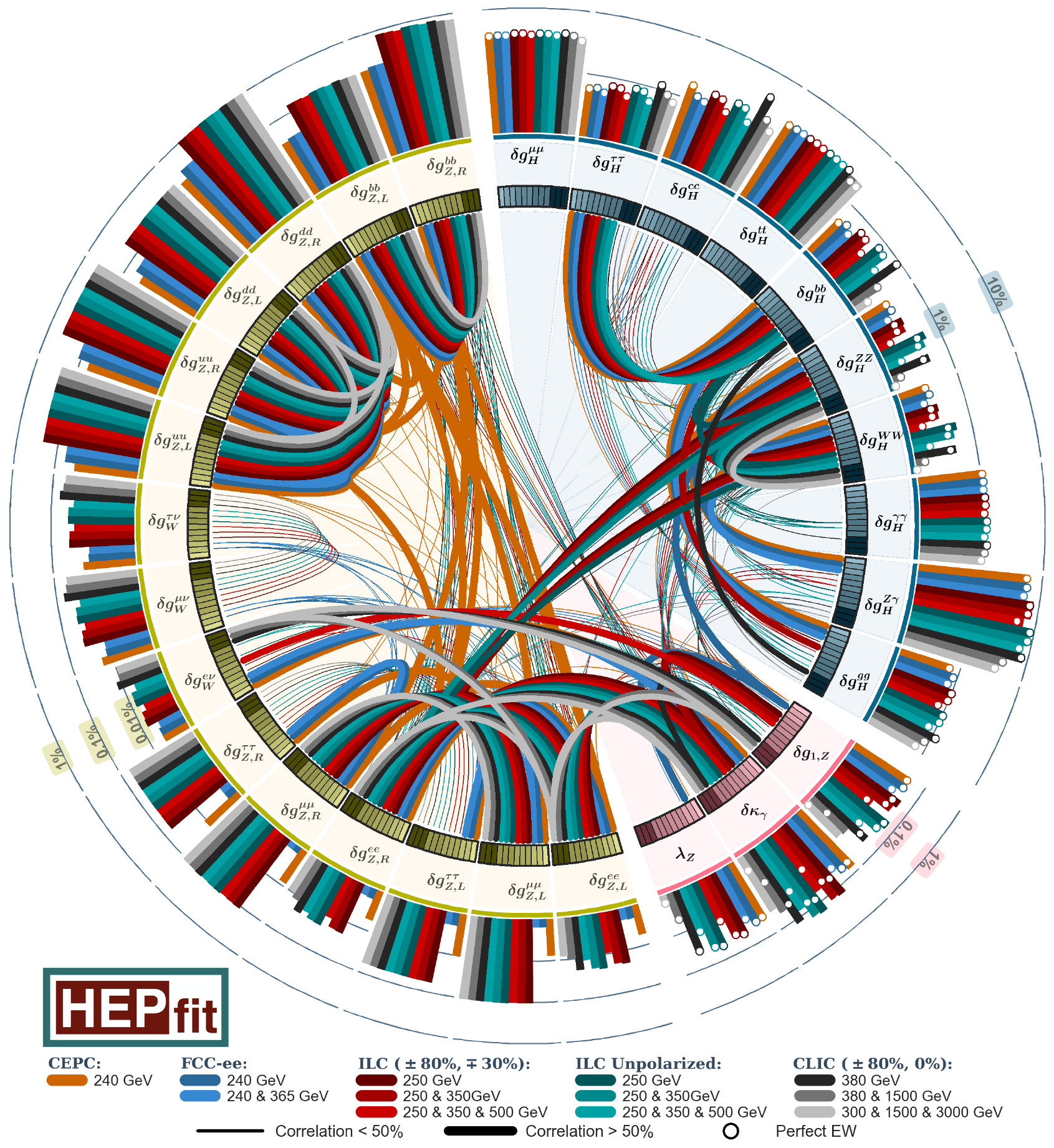}
  \caption{\it
A scheme-ball illustration of the correlations between Higgs and EW sector couplings.
The $Z$-pole runs  are included for both FCC-ee and CEPC.
Projections from HL-LHC and measurements from LEP and SLD are included in all scenarios.
The outer bars give the one-sigma precision on the individual coupling (see \hyperref[tab:allcoll]{tables~\ref{tab:allcoll}} and~\ref{tab:allcollEW}).}
  \label{fig:sb-all}
\end{figure}

The potential impact of Higgs measurements on EW parameters is assessed by comparison with the prospects obtained without Higgs measurements, shown with trapezoidal marks.
Sizeable effects are only seen, at linear colliders, on the $Z$-boson couplings to electrons.
Those would also be the most affected by an improvement of the left-right polarization asymmetry $A_{LR}$ mentioned earlier.  
At the HL-LHC, the impact of Higgs measurements on EW couplings is only visible for the gauge couplings of the light quarks, of down type in particular ($d$ and $s$), which are poorly constrained at LEP and SLD.
The $Vh$ and diboson production processes, mostly initiated by light quarks at the LHC, are sensitive to these couplings~\cite{Banerjee:2018bio}.

%\FloatBarrier

In addition to the precision reach on each coupling, the correlations among them also contain important information, and are particularly relevant for understanding the interplay of Higgs and EW measurements.
To avoid showing a large set of $28\times28$ matrices, we present a scheme-ball illustration in \autoref{fig:sb-all}, which highlights large correlations with lines connecting pairs of couplings in its inner circle.
The circular collider projections include both $Z$-pole and $WW$ threshold measurements.
At linear colliders, the EW and the Higgs sector appear clearly connected due to the absence of new $Z$-pole measurements.
Strong correlations are present between aTGCs and other electroweak couplings.
This clearly shows again that the electroweak, triple-gauge, and Higgs sectors of the effective field theory would become significantly entangled with the advent of future lepton colliders.

We further investigate the impacts of diboson measurements and beam polarizations in the rest of this section.

%%%%%%%%%%%%%%%%%%%%%%%%%%%
\subsection{Impact of \texorpdfstring{$WW$}{WW} measurements}
\label{sec:wwscale}
%%%%%%%%%%%%%%%%%%%%%%%%%%%

\begin{figure}[t]
\centering
\includegraphics[width=\textwidth]{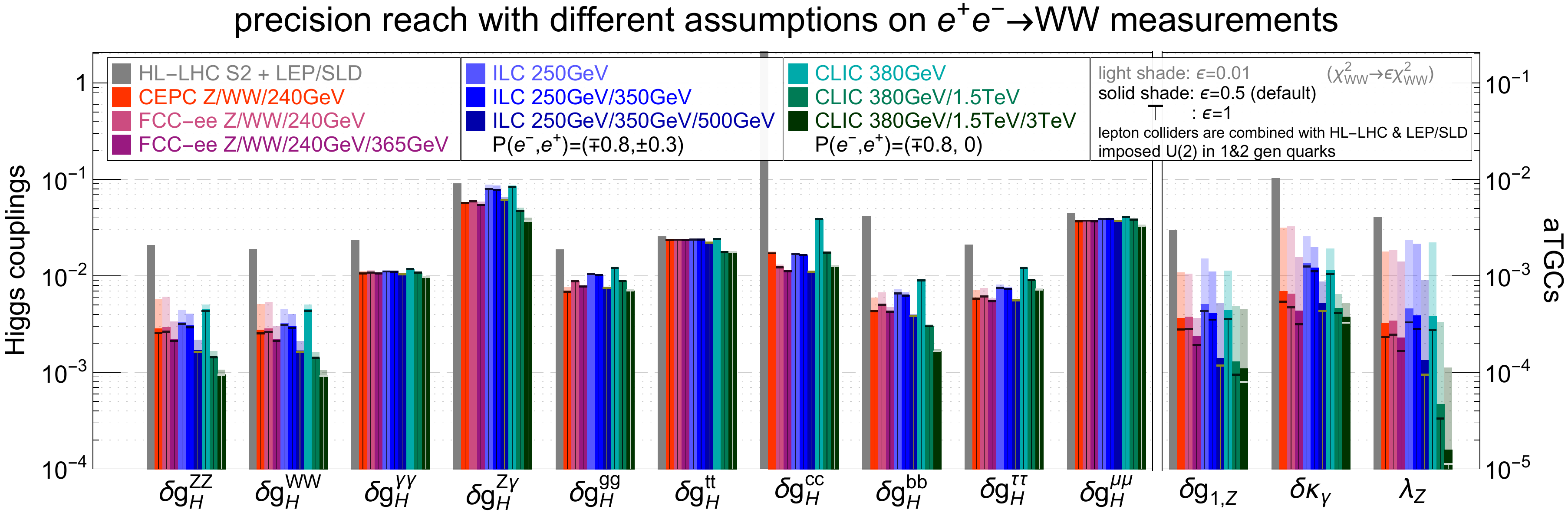}
\caption{\it
Impact of diboson measurement precision on Higgs and triple-gauge couplings.
Our default assumption, adopted in \autoref{fig:money}, is also shown here as dark-shaded bars.
It corresponds to an overall efficiency $\epsilon$ of 50\% (see \autoref{sec:wwinput}).
The results obtained with an ideal 100\% and a lower 1\% efficiency are  shown as vertical lines and light shaded bars respectively.
The run scenarios of the future lepton colliders are summarized in \autoref{tab:scenarios}.
}
\label{fig:wwscale}
\end{figure}

As explained in \autoref{sec:wwinput}, our prospects for $WW$ measurements neglect backgrounds, detector effects and systematic uncertainties but assume a conservative overall efficiency $\epsilon$ of 50\%.
We examine in \autoref{fig:wwscale} the impact of different assumptions for $\epsilon$ on Higgs and triple-gauge coupling prospects.
This exercise also more generally allows us to visualize the constraining power of diboson measurements.
In comparison with the default $\epsilon =50\%$ prospects shown as dark-shaded bars, the ideal $\epsilon =100\%$ and pessimistic $\epsilon=0.01$ ones are respectively shown with vertical lines and light shaded columns.
The results in \autoref{fig:wwscale} clearly show that $WW$ measurements dominate the reach on aTGCs.
A sizeable impact is also observed on the $\delta g^{ZZ}_H$ and $\delta g^{WW}_H$ couplings constrained by measurements in which aTGCs also enter.
It is more severe at the CEPC and FCC-ee when only Higgs measurements at 240\,GeV are included.
Reducing $\epsilon$ from 50\% to 1\% worsens diboson measurement precision by a factor of $\sqrt{0.5/0.01}\simeq 7$ and increases the uncertainties on $\delta g^{ZZ}_H$ and $\delta g^{WW}_H$ by a factor of about $2$.
This also indirectly affects $\delta g^{bb}_H$ and $\delta g^{\tau\tau}_H$.
Including higher energy runs helps reducing the impact of diboson measurements.
Higgs measurements alone are then able to disentangle the different EFT contributions.
Other Higgs couplings, mostly constrained by Higgs decays, are much less sensitive to the $WW$ measurements.

%%%%%%%%%%%%%%%%%%%%%%%%%%%
\subsection{Impact of beam polarization}
\label{sec:ILCPol}
%%%%%%%%%%%%%%%%%%%%%%%%%%%

In the conceptualization of linear colliders, one of the matters that has been discussed extensively is the impact and necessity of polarization for both the $e^+$ and $e^-$ beams.
Given that electroweak interactions have $O(1)$ parity violation, beam polarization is expected to leave a large effect on the physics case~\cite{Bambade:2019fyw}.
More specifically, the signal and background cross sections for several processes depend crucially on the beam polarization, with the signal showing a marked increase over that for unpolarized beams~\cite{Baer:2013cma} along with a simultaneous suppression of the background.
Beam polarization is also credited with having the benefit of being diagrammatically selective.
It thus allows one to probe independent parameter space directions and hence yields a more complete physics program.
Lastly, beam polarization permits a large gain in the control of systematic uncertainties.
Studies of polarization effects have been performed for both the higher-energy~\cite{Baer:2013cma} and the 250\,GeV ILC runs~\cite{Fujii:2017vwa,Fujii:2018mli}.
A comprehensive list of all the processes projected to be studied at the ILC, including new-physics searches and the importance of polarization on these channels can be found in table~1.1 of ref.~\cite{Baer:2013cma}.

The scaling with polarization of the $s$-channel $e^+e^-$ annihilation of massless fermions to a vector boson is given by~\cite{Fujii:2018mli, Bambade:2019fyw}
\begin{equation}
\sigma_{P_{e^+}P_{e^-}} = \sigma_0(1-P_{e^+}P_{e^-})\left[1-A_{LR}\frac{P_{e^-}-P_{e^+}}{1-P_{e^+}P_{e^-}}\right],
\label{eq:polscale}
\end{equation}
where $\sigma_{P_{e^+}P_{e^-}}$ is the cross section corresponding to polarizations $P_{e^+}$ and $P_{e^-}$ for the $e^+$ and $e^-$ beams respectively and $\sigma_0$ is the cross section for unpolarized beams.
$A_{LR}$ is the intrinsic left-right asymmetry of the production cross section.
For the SM, $e^+e^-\to Zh$ production channel $A_{LR}=0.151$.\footnote{Given left- and right-handed couplings of charged lepton to the $Z$ are respectively proportional to $-1+2\sin^2{\theta_w}$ and $2\sin^2{\theta_w}$, this polarization asymmetry is approximated by $(1-4\sin^2{\theta_w})/(1-4\sin^2{\theta_w}+8\sin^4{\theta_w})$ and is very sensitive to the sine of the weak mixing angle $\sin{\theta_w}$.}

The $\nu\nu h$ production proceeds through $t$-channel $W$ boson fusion which only involves left-handed fermions and right-handed anti-fermions.
The scaling of its rate with polarization is therefore simpler ($A_{LR}=1$ in \autoref{eq:polscale}):
\begin{equation}
\sigma_{P_{e^+}P_{e^-}} = \sigma_{0}(1-P_{e^-})(1+P_{e^+}).
\label{eq:polscaleLR}
\end{equation}
In this case it is clear a negative polarization for the electron and a positive one for the positron enhance the cross-section while an opposite configuration reduces it.
The process $e^+e^-\to W^+W^-$ has both $s$-channel $Z/\gamma$ and $t$-channel $\nu$ exchange contributions.
$A_{LR}$ is in practice very close to $1$.

The prescriptions we adopt for the scaling of statistical uncertainties from one polarization to the other are the following:
\begin{itemize}
\item \label{item:hz-pol-prescription} {$e^+e^-\to Zh:$} As described in ref.~\cite{Barklow:2017suo}, $A_{LR}$ being small, the rate enhancement for the $P(e^-,e^+)=(-80\%,+30\%)$ beam polarization configuration over the ($+80\%,-30\%$) is compensated by the slightly lower background in the latter.
So we assume that the statistical uncertainties will be the same for the ($\pm80\%,\mp30\%$) configurations.
For scaling to other polarization configurations, we assume no significant role is played by $A_{LR}$ in \autoref{eq:polscale} and use the following formula:
\begin{equation}
\frac{\Delta_{Zh}^2(P^a_{e^+},P^a_{e^-})}{\Delta_{Zh}^2(P^b_{e^+},P^b_{e^-})}
	= \frac{1-P_{e^+}^bP_{e^-}^b}{1-P_{e^+}^aP_{e^-}^a}
	.
\label{eq:zh_pol_scaling}
\end{equation}

\item {$e^+e^-\to \nu\nu h:$} Being driven by $W$ boson fusion, we use \autoref{eq:polscaleLR} to scale the statistical errors for the different polarizations.
\end{itemize}
\begin{figure}[t]%[htb]
\centering
\includegraphics[width=\textwidth]{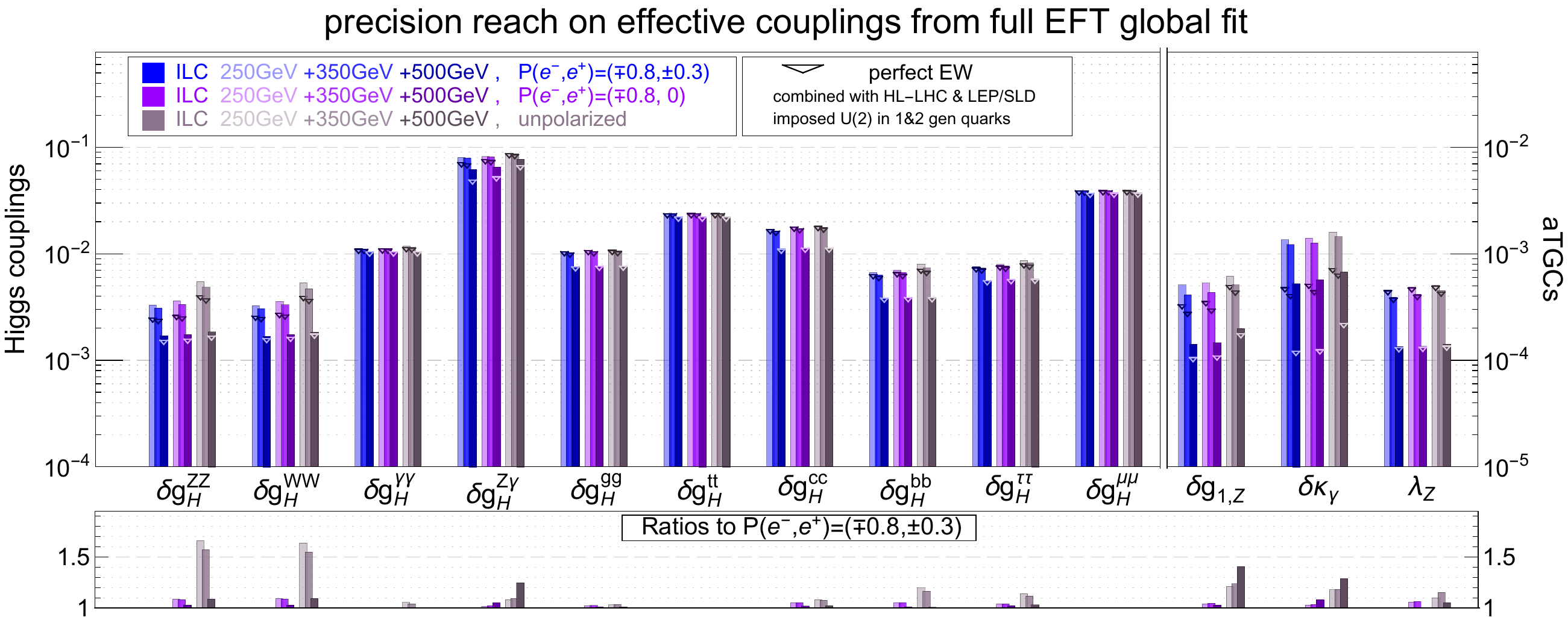}
\caption{\it
Global one-sigma reach on Higgs and triple-gauge couplings at the ILC, for three different beam polarization configurations.
Electroweak measurements from LEP and SLD as well as HL-LHC projections are included in all scenarios.
Electroweak parameters (not shown) are marginalized over.
}
\label{fig:ilcpo}
\end{figure}
On the other hand, systematic uncertainties are assumed to be polarization independent.
For unpolarized beams, no uncertainty is however associated with the determination of the polarization.

In the rest of this section, focusing for concreteness on ILC run scenarios, we briefly investigate the effects of beam polarization on Higgs and triple-gauge coupling measurements.
The reaches obtained with $P(e^-,e^+) = (\mp 80\%,\pm30\%)$, $(\mp 80\%,0\%)$ and the unpolarized configurations are shown in \autoref{fig:ilcpo}.
Electroweak parameters are marginalized over.
Compared to the reach obtained with unpolarized beams, after the first stages of ILC, sizeable improvements of about 50\% are brought by polarization on $\delta g^{ZZ}_H$ and $\delta g^{WW}_H$ coupling precisions.  
A small indirect impact is also observed on $\delta g^{bb}_H$ and $\delta g^{\tau\tau}_H$.
It arises mainly from the additional discrimination power provided by the  two different beam polarizations, each sensitive to different combinations of parameters.
The increase in $e^+e^-\to Zh$ cross-section only induces a limited improvement.
This explains the small impact of positron beam polarization: two $(\mp 80\%,0\%)$ electron beam polarization configurations already effectively resolve approximate degeneracies.
The inclusion of a 500\,GeV run provides significant discriminating powers and can largely compensate the lack of beam polarizations in the $\delta g^{ZZ}_H$ and $\delta g^{WW}_H$ coupling reach.

\begin{figure}%[ht!]
\centering
\includegraphics[width=0.7\textwidth]{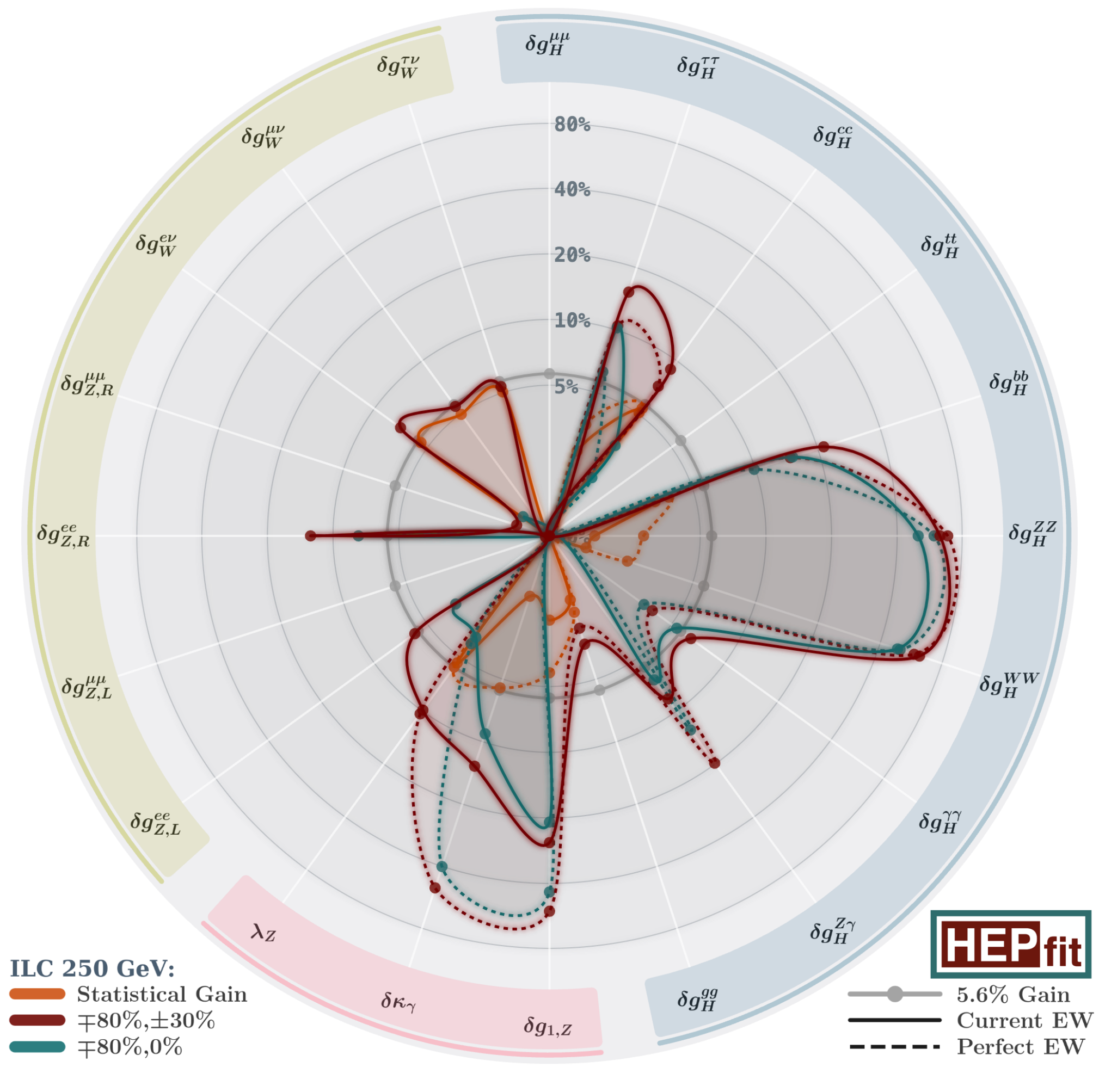}
\caption{\it
Strengthening in global constraints arising from the introduction of $P(e^-,e^+)=(\mp80\%,\pm30\%)$ and $(\mp80\%,0\%)$ beam polarizations at a centre-of-mass energy of $250\,$GeV (in red and green, respectively) quantified as ${\delta g (unpolarized)}/{\delta g(\textrm{polarized})}-1$ expressed in percent.
For comparison, the improvement of constraints brought by a factor $1.12$ increase in luminosity in shown in orange.
This factor is the purely statistical gain on $e^+e^-\to hZ$ and $e^+e^-\to \nu\nu h$ rate incurred with $(\mp80\%,\pm30\%)$ beam polarization.
The grey band is representative of a 5.6\% gain ($\sqrt{1.24 \times 0.9}-1$).
The numerical inputs for $P(e^-,e^+)=(\mp80\%,\pm30\%)$ and unpolarized beams are taken from \autoref{tab:allcoll}.
}
\label{fig:ilcpolrad}
\end{figure}

We also observe small differences (10--20\%) between polarized and unpolarized cases for the $\delta g^{Z\gamma}_H$, $\delta g_{1,Z}$ and $\delta \kappa_\gamma$ couplings.
These are enhanced by the inclusion of the higher-energy runs.
As pointed out in ref.~\cite{Durieux:2017rsg}, the sensitivity of the $hZ$ production process to $\delta g^{Z\gamma}_H$ suffers from an accidental suppression for unpolarized beams. 
The $h\to Z\gamma$ measurement at the HL-LHC however effectively constrains this coupling, so that the loss in reach incurred without beam polarization is limited.
Additional measurements of the $hZ$ process at higher energies improve the reach on $\delta g^{Z\gamma}_H$ but also make it more sensitive to the polarizations.  
For $\delta g_{1,Z}$ and $\delta \kappa_\gamma$, the discriminating power provided by the higher-energy runs is also insufficient to offset the enhanced degeneracies in the diboson process, as observed previously in \autoref{fig:money}.
Losing the handle of beam polarizations thus further enhances the degeneracies and reduces the reach.

\begin{figure}[t!]
\centering
\includegraphics[width=\textwidth]{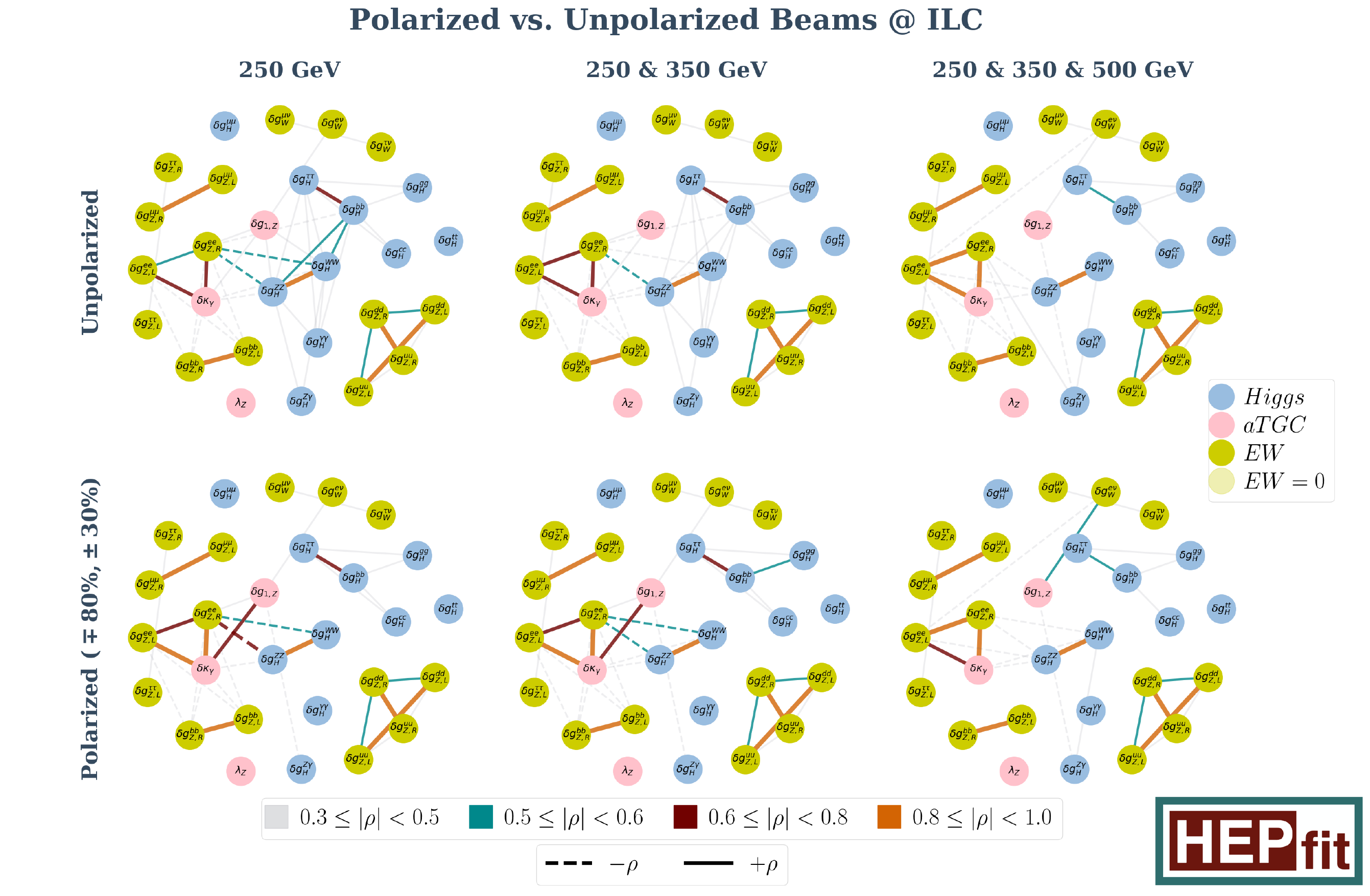}
\caption{\it A comparison of correlations between the different sectors for polarized beams ($\mp80\%,\pm30\%$) and unpolarized beams at the ILC. HL-LHC projections have been included.}
\label{fig:np-pol}
\end{figure}

Focusing on the 250\,GeV run, \autoref{fig:ilcpolrad} further highlights the complementarity of opposite beam polarization configurations for lifting approximate degeneracies.
It shows the relative improvement obtained between polarized and unpolarized scenarios.
The cases of $P(e^-,e^+)=(\mp80\%,\pm30\%)$ and $(\mp80\%,0\%)$ beam polarization configurations are respectively displayed in red and green.
For reference, the gain expected from the increase in sheer rate is displayed as orange lines.
It is obtained by artificially augmenting luminosities by a factor of  $1.24\times 0.9\simeq 1.12$ in our default unpolarized beam scenario.
The factor of $1.24$ is the statistical increase in the precision of the $hZ$ cross-section determination when adopting a $P(e^-,e^+) = (\mp 80\%,\pm30\%)$ configuration (following the prescription of \autoref{eq:zh_pol_scaling}) and the same for $\nu\nu h$.
Note that no such statistical gain is obtained in the absence of positron polarization.
The factor of $0.9$ is compensating for the $10\%$ of luminosity collected with same-sign polarization configuration and not used in our prospects.

As already noted above, polarized beams induce a sizeable improvement (up to $80\%$) in the precision achievable on several Higgs couplings, while positron beam polarization has a marginal impact.
As seen in the figure, this improvement is often much larger than the bare statistical gain in $hZ$ and $\nu\nu h$ rate due to polarization (up to $5.6\%$ shown by the grey line).
Runs  with two different polarization configurations are indeed effective in reducing approximate degeneracies.
Including higher-energy runs also reduces degeneracies and therefore limits the relative impact of beam polarization.
Imposing perfect EW measurements only affects $\delta g_{1,Z}$ and $\delta \kappa_\gamma$, increasing the improvement brought by polarization to $40$--$50\%$ level as for $\delta g_H^{ZZ}$ and $\delta g_H^{WW}$.
Considering EW couplings, the gain on $\delta g_W^{l\nu}$ coupling precisions is commensurate with the purely statistical one and small in the case of  and $\delta g_{Z,R}^{ee}$.

From \autoref{fig:np-pol} we get some insight into the difference in the correlation maps between the case of the polarized beams and the unpolarized ones. Removing positron polarization does not change the correlation map of for the polarized beams. It can be seen that $\delta\kappa_\gamma$ is always correlated with $\delta g^{ee}_{Z,L}$ and $\delta g^{ee}_{Z,R}$. The latter are progressively better constrained with the growth of energy for the case of polarized beams when compared to unpolarized as is apparent from \autoref{tab:allcollEW}. The correlation between $\delta g_{1,Z}$ and $\delta g_W^{e\nu}$ at all energies is also distinctive for the case of the polarized beams and absent for unpolarized beams.

Beam polarization also helps controlling systematic uncertainties, an aspect we have not fully accounted for here. 
The details of the different configurations of beam polarization that may be used to control systematics can be found in ref.~\cite{Bambade:2019fyw}.  
Our conclusions here are also drawn assuming only SLC precision on the polarization asymmetry $A_{LR}$, although we have seen a measurement in radiative return to the $Z$ pole has the potential of helping to reduce the contamination of electroweak uncertainties in Higgs couplings.

%%%%%%%%%%%%%%%%%%%%%%%%%%%
\section{Summary}
\label{sec:summary}
%%%%%%%%%%%%%%%%%%%%%%%%%%%

All proposed future lepton colliders, CEPC, FCC-ee, ILC and CLIC, will likely face critical decision making in the coming years. 
The determination of their physics potentials, in particular the reach on the Higgs coupling measurements, is an important and pressing matter.
In the effective-field-theory framework, the Higgs sector can not simply be isolated from the rest of the standard model.  
A successful Higgs precision program at future lepton colliders thus requires a global analysis of Higgs and electroweak measurements.
We perform such a consistent treatment for the first time in this paper.
In particular, all relevant dimension-six contributions are included in our implementation of current and future $Z$-pole measurements.
The impact of diboson ($\eeww$) production measurements is revisited with all CP-even dimension-six contributions, using the powerful method of statistically optimal observables.  

For the circular colliders, CEPC and FCC-ee, our results suggest that the future $Z$-pole runs are important not only to extract EW parameters, but also to determine triple-gauge and Higgs couplings.
The unprecedented precision of diboson measurements makes them particularly sensitive to EW parameters.
The significant cross-talk with aTGCs is only reduced by $Z$-pole measurements of matching precision.
The improvement on Higgs couplings arising from new $Z$-pole measurements comes both directly from the improved reach on the $hZee$ contact interactions, and indirectly from the improvement on the aTGCs which are related to the $hZZ$ and $hWW$ anomalous couplings.  
In particular, the reach on the $hZZ$ and $hWW$ couplings are worsened by a factor of about $50\%$ for the 240\,GeV run in the absence of a new $Z$-pole run.
On the other hand, the proposed $Z$ program at either CEPC or FCC-ee is sufficient for eliminating the uncertainties on the Higgs couplings propagated from the EW measurements, and are essentially equivalent to having perfect electroweak measurements for this purpose.

For the linear colliders, ILC and CLIC, we observe the lack of $Z$-pole run renders significant the contaminations from EW parameter uncertainties to Higgs couplings.
They are largely mitigated by higher-energy runs, which however dominantly probe given combinations of parameters and therefore also induce some relative enhancements of approximate degeneracies.  
With runs at three different centre-of-mass energies, the ILC and CLIC reach on the Higgs couplings are worsened by at most 10--15\% compared with the case of perfect EW measurements.
On the other hand, the relative impact of EW uncertainties on the $\delta g_{1,Z}$ and $\delta \kappa_\gamma$ aTGCs significantly increases at high energies.
Their absolute constraints are however very tight.
Especially during first run stages, beam polarization is effective in resolving approximate degeneracy and in improving the precision reach on $hZZ$ and $hWW$ couplings.
Besides $\eeww$ production, additional EW measurements above the $Z$-pole have the potential to help reducing the contaminations from EW-sector uncertainties to Higgs coupling determinations.
Including preliminary prospects for the measurements of $Z$ decays and asymmetries in radiative return to the $Z$ pole at the ILC 250\,GeV and CLIC 380\,GeV was for instance observed to have a significant impact.  
Further studies of such measurements, maybe complemented by dedicated GigaZ runs, are thus highly relevant for the Higgs programs at linear colliders.
We stress however that runs at higher energies, a clear option for linear colliders, can mitigate, in the extraction of the Higgs couplings, the EW parametric uncertainties inherited from LEP/SLD.

\acknowledgments
We thank Alain Blondel, Patrick Janot, Marumi Kado, Zhijun Liang, Jenny List, Zhen Liu, Michael Peskin, Roman P\"oschl, Philipp Roloff, Aidan Robson, Manqi Ruan, Roberto Tenchini, Junping Tian, Lian-Tao Wang and the members of the Higgs@Future Colliders WG for many useful discussions. 
We are also thankful the authors of ref.~\cite{Krzywinski18062009} for the access to their software which was used to make some of the figures in this article.
AP would like to thank Hyunju Kim for her help with visualizing network plots.
AP is grateful for the partial support of this research by the Munich Institute for Astro- and Particle Physics (MIAPP) of the DFG cluster of excellence ``Origin and Structure of the Universe'' during the completion of this work. 
The work of JG has been supported by the Cluster of Excellence ``Precision Physics, Fundamental Interactions, and Structure of Matter'' (PRISMA+ EXC 2118/1) funded by the German Research Foundation (DFG) within the German Excellence Strategy (Project ID 39083149). 
The work of CG and AP was in part funded by the Deutsche Forschungsgemeinschaft under Germany's Excellence Strategy -- EXC 2121 ``Quantum Universe'' -- 390833306 and by the Helmholtz Association through the recruitment initiative program.
The work of GD is supported in part at the Technion by a fellowship from the Lady Davis Foundation.
The computational resources for this work were provided in part through the Maxwell and BIRD facilities at Deutsches Elektronen-Synchrotron (DESY), Hamburg, Germany.

\appendix

\section{Comparison with Higgs@Future Colliders ECFA Working Group report}
\label{sec:ecfa}
We briefly comment on the differences between our study and the results of the recent report in ref.~\cite{deBlas:2019rxi}.
The work presented here extends and complements the results in that reference for the case of future lepton colliders.
Indeed, the results presented there are based almost exclusively on the official input provided by the different experimental collaborations.
As explained in that reference and also emphasized in this document, from the point of view of the diboson observables most future collider projects present the projections only in the case of aTGC dominance.
This prevents a precise assessment of the impact of the EW constraints on the global fit results, which we address here.
Furthermore, as explained in \autoref{app:inputs}, in some cases the status of the projections for several observables is not at the same level of completeness across the different experiments.
For instance, $H\to Z\gamma$ is presented by the CEPC project but not by FCC-ee, but there is no reason the latter cannot obtain a similar measurement.
These gaps are filled in our study using reasonable assumptions, aiming to offer a more complete and accurate picture of the global constraints achievable at each machine.
Finally, we discuss in detail other issues not covered in ref.~\cite{deBlas:2019rxi}, such as the impact of polarization in the global fit results.
On the other hand, there are some relevant topics already discussed in ref.~\cite{deBlas:2019rxi} which we therefore chose not to include in our study, such as the impact of SM theory uncertainties in the fits.
To prevent the interference of these effects with the main purpose of our study, i.e.\ to quantify the interplay between the different types of measurements at future colliders, we neglect the effects of such theory errors.
Because of all these reasons, several of the results presented here and in ref.~\cite{deBlas:2019rxi} show some differences.
Finally, the last version of ref.~\cite{deBlas:2019rxi} also includes extra scenarios for linear colliders, including a run at the $Z$ pole or an ILC run at $\sqrt{s}=1$ TeV.
Such runs are not part of the proposed baselines of ILC or CLIC and were not considered in our study.

%%%%%%%%%%%%%%%%%%%%%%%%%%%
\section{Results in other bases}
\label{app:d6basis}
%%%%%%%%%%%%%%%%%%%%%%%%%%%

%
\begin{table}[b]
\centering
\begin{tabular}{l|l} \hline\hline
$\mathcal{O}_H = \frac{1}{2} (\partial_\mu |H^2| )^2$ &  $\mathcal{O}_{GG} =  g_s^2 |H|^2 G^A_{\mu\nu} G^{A,\mu\nu}$  \\ 
$\mathcal{O}_{WW} =  g^2 |H|^2 W^a_{\mu\nu} W^{a,\mu\nu}$  & $\mathcal{O}_{y_u} = y_u |H|^2 \bar{q}_L \tilde{H} u_R \,+\, {\rm h.c.}$ \hspace{0.25cm} {\scriptsize $(u \to t, c)$}  \\
$\mathcal{O}_{BB} =  g'^2 |H|^2 B_{\mu\nu} B^{\mu\nu}$ &  $\mathcal{O}_{y_d} = y_d |H|^2 \bar{q}_L H d_R \,+\, {\rm h.c.}$ \hspace{0.3cm} {\scriptsize $(d \to b)$}  \\
$\mathcal{O}_{HW} =  ig(D^\mu H)^\dagger \sigma^a (D^\nu H) W^a_{\mu\nu}$  &  $\mathcal{O}_{y_e} = y_e |H|^2 \bar{l}_L H e_R \,+\, {\rm h.c.}$ \hspace{0.36cm} {\scriptsize $(e \to \tau, \mu)$}  \\
$\mathcal{O}_{HB} =  ig'(D^\mu H)^\dagger  (D^\nu H) B_{\mu\nu}$ &      $\mathcal{O}_{3W} = \frac{1}{3!} g \epsilon_{abc} W^{a\,\nu}_\mu W^b_{\nu \rho} W^{c\,\rho\mu} $  \\   \hline%\hline
&\\[-0.4cm]
$\mathcal{O}_{W} =  \frac{ig}{2}(H^\dagger \sigma^a \overleftrightarrow{D_\mu} H) D^\nu W^a_{\mu\nu} $  & $\mathcal{O}_{B} =  \frac{ig'}{2}(H^\dagger \overleftrightarrow{D_\mu} H) \partial^\nu B_{\mu\nu} $ \\  \hline%\hline
&\\[-0.4cm]
$\mathcal{O}_{WB} = gg' H^\dagger \sigma^a H W^a_{\mu\nu} B^{\mu\nu}$  &   $\mathcal{O}_{H\ell} = i  H^\dagger \overleftrightarrow{D_\mu} H \bar{\ell}_L \gamma^\mu \ell_L  $  \\
$\mathcal{O}_{T} = \frac{1}{2}(H^\dagger \overleftrightarrow{D_\mu} H)^2 $  &   $\mathcal{O}'_{H\ell} = i H^\dagger \sigma^a \overleftrightarrow{D_\mu} H \bar{\ell}_L \sigma^a \gamma^\mu \ell_L $  \\
$\mathcal{O}_{\ell\ell} =  (\bar{\ell}_L \gamma^\mu \ell_L) (\bar{\ell}_L \gamma_\mu \ell_L)$  &   $\mathcal{O}_{He} = i H^\dagger \overleftrightarrow{D_\mu} H \bar{e}_R \gamma^\mu e_R $  \\ \hline%\hline
&\\[-0.4cm]
$\mathcal{O}_{Hq} = i  H^\dagger \overleftrightarrow{D_\mu} H \bar{q}_L \gamma^\mu q_L  $ &  $\mathcal{O}_{Hu} = i H^\dagger \overleftrightarrow{D_\mu} H \bar{u}_R \gamma^\mu u_R $ \\
$\mathcal{O}'_{Hq} = i H^\dagger \sigma^a \overleftrightarrow{D_\mu} H \bar{q}_L \sigma^a \gamma^\mu q_L $ &  $\mathcal{O}_{Hd} = i H^\dagger \overleftrightarrow{D_\mu} H \bar{d}_R \gamma^\mu d_R $ \\  \hline\hline
\end{tabular}
\caption{A redundant set of dimension-six operators that contributes to the Higgs and EW processes in our analysis.  Flavour indices are omitted. The operators $\mathcal{O}_{WW}$, $\mathcal{O}_{WB}$, and the flavour universal components of $\mathcal{O}_{H\ell}$ and $\mathcal{O}'_{H\ell}$ are eliminated in the SILH' basis; $\mathcal{O}_{W}$, $\mathcal{O}_{B}$, $\mathcal{O}_{H\ell}$ and $\mathcal{O}'_{H\ell}$ are eliminated in the modified-SILH' basis; $\mathcal{O}_{W}$, $\mathcal{O}_{B}$, $\mathcal{O}_{HW}$ and $\mathcal{O}_{HB}$ are eliminated in the Warsaw basis.}
\label{tab:d6}
\end{table}

In this appendix, we present the results in terms of the reaches on the Wilson coefficients of dimension six operators.  
We consider several basis choices, including the Warsaw~\cite{Grzadkowski:2010es} and the SILH' bases~\cite{Elias-Miro:2013mua}.
The latter is obtained from the SILH basis~\cite{Giudice:2007fh} with the operators $\mathcal{O}_{2W} = -\frac{1}{2}(D^\mu W^a_{\mu\nu})^2$ and $\mathcal{O}_{2B} = -\frac{1}{2}(\partial^\mu B_{\mu\nu})^2$ replaced by four fermion operators, which do not contribute to $Z$-pole observables at the leading order except for the measurements of the Fermi constant.  We also considered a modified version of the SILH' basis utilized in ref.~\cite{Durieux:2017rsg} (also similar to the ones in ref.~\cite{Hagiwara:1993ck}) for better separation of the Higgs and EW measurements.  
For the sake of compactness and easy comparison, we use the same operator conventions for all three bases, which follow closely the ones in ref.~\cite{Elias-Miro:2013mua}.  
A list of redundant operators relevant for the Higgs and EW measurements are provided in \autoref{tab:d6}.  The flavour indices are also omitted, which can be trivially restored.  The three bases can be obtained by eliminating different operators via the relations from integration by parts,
\begin{align}
\mathcal{O}_B = &~ \mathcal{O}_{HB} + \frac{1}{4} \mathcal{O}_{BB} + \frac{1}{4} \mathcal{O}_{WB}  \,, \nonumber \\
\mathcal{O}_W = &~ \mathcal{O}_{HW} + \frac{1}{4} \mathcal{O}_{WW} + \frac{1}{4} \mathcal{O}_{WB}  \,,
\end{align}
and from the SM equations of motion of the gauge fields,
\begin{align}
\frac{1}{g'^2} \mathcal{O}_B =&~ -\frac{1}{2} \mathcal{O}_T + \frac{1}{2} \sum_f (Y_{f_L} \mathcal{O}_{Hf_L} + Y_{f_R} \mathcal{O}_{Hf_R} )     \,, \nonumber \\
 \frac{1}{g^2} \mathcal{O}_W =&~ -\frac{3}{2} \mathcal{O}_H + 2 \mathcal{O}_6 + \frac{1}{2} \sum_{f} \mathcal{O}_{y_f} + \frac{1}{4} \sum_f  \mathcal{O}'_{Hf_L}      \,,
\end{align}
where $f_L = \ell, \, q$ are the left-handed fermion doublets, $f_R = e,\, u,\, d$ are the right-handed fermion singlets, and $Y$ is the hyperchange ($Y_\ell = -\frac{1}{2}$, {\it etc.}).
Note that only the entries of $\mathcal{O}_{Hf_{L,R}}$ and $\mathcal{O}'_{Hf_L}$ proportional to the SM fermionic currents enter in the previous equation.
In the SILH' basis, $\mathcal{O}_{WW}$, $\mathcal{O}_{WB}$ and the above-mentioned flavour universal entries of $\mathcal{O}_{H\ell}$ and $\mathcal{O}'_{H\ell}$ are eliminated. The modified-SILH' basis is obtained from the SILH' basis trading $\mathcal{O}_{W}$ and $\mathcal{O}_{B}$ by $\mathcal{O}_{WW}$ and $\mathcal{O}_{WB}$. In the Warsaw basis, $\mathcal{O}_{W}$, $\mathcal{O}_{B}$, $\mathcal{O}_{HW}$ and $\mathcal{O}_{HB}$ are eliminated from \autoref{tab:d6}.
It should also be noted that our definition of $\mathcal{O}_{WB}$ differs from the one in ref.~\cite{Barklow:2017awn} by a factor of two.

\begin{figure}[t!]
\centering
\includegraphics[width=\textwidth]{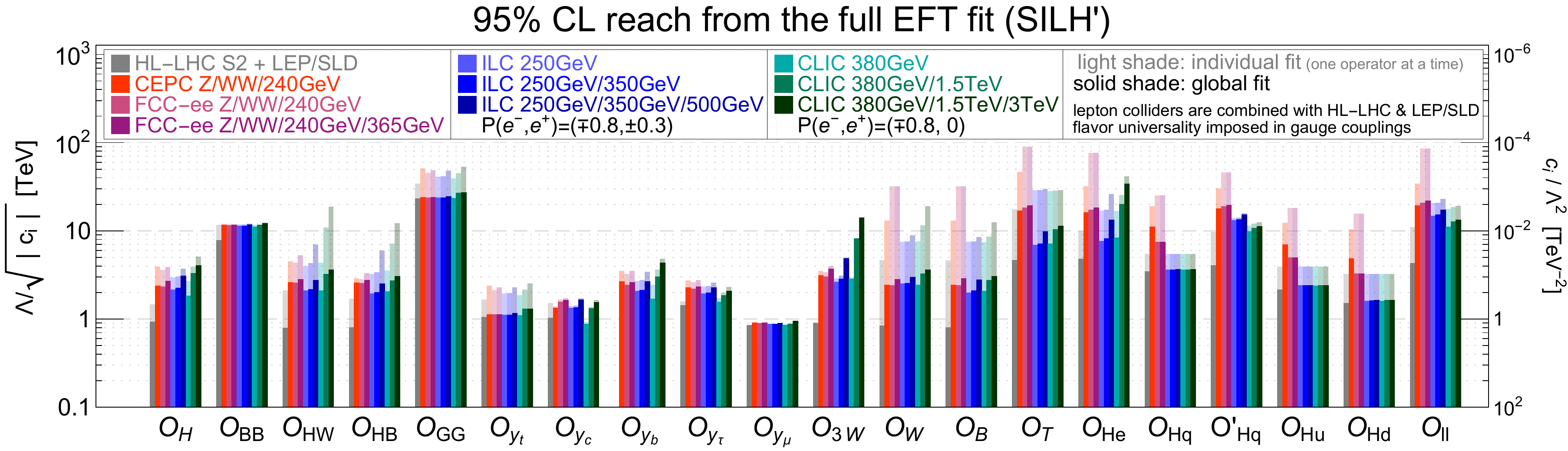}  \\ \vspace{0.5cm}
\includegraphics[width=\textwidth]{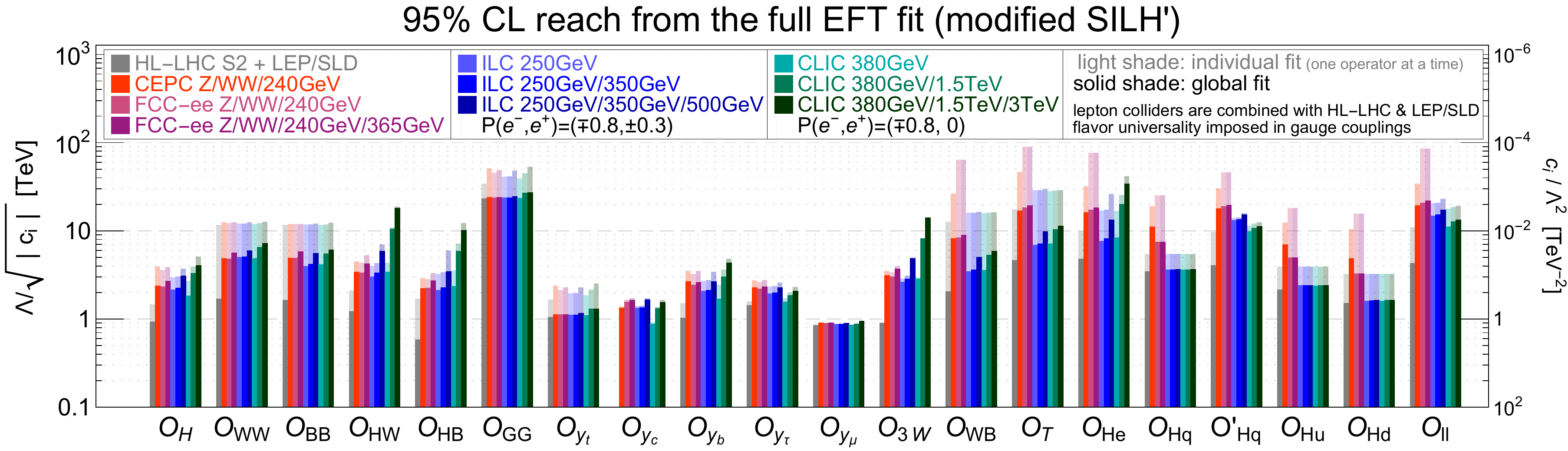}  \\ \vspace{0.5cm}
\includegraphics[width=\textwidth]{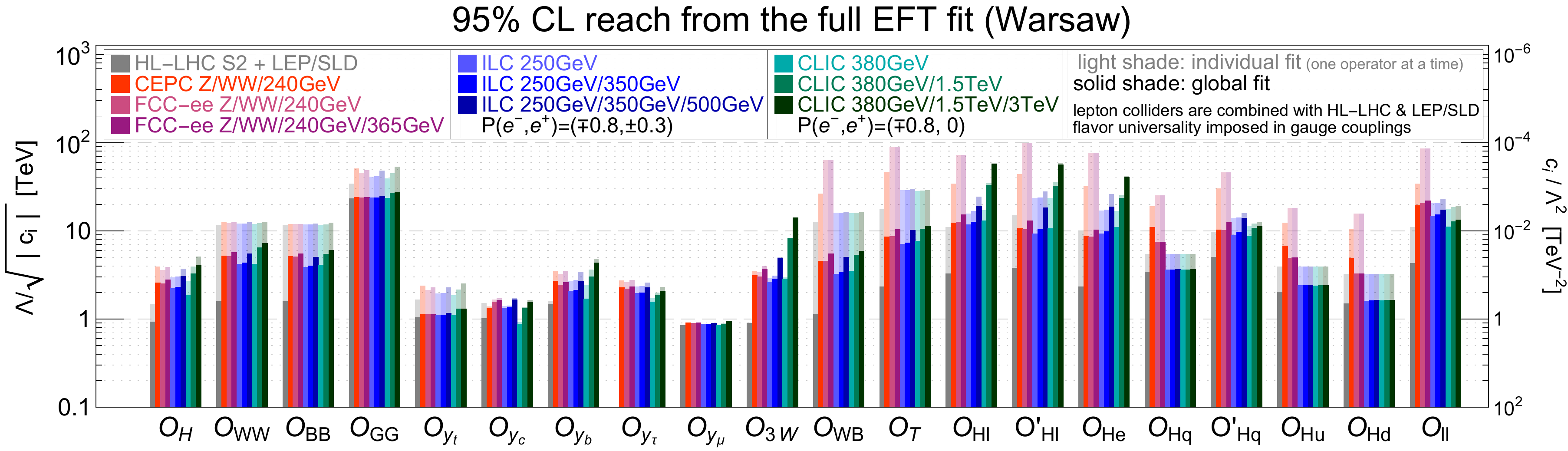}
\caption{\it
The 95\% CL reaches for $\Lambda/ \sqrt{|c_i|}$ for HL-LHC and the future lepton colliders in the SILH' (top), modified-SILH' (middle) and Warsaw (bottom) bases.
The corresponding values for $c_i/\Lambda^2$ are shown on the right-hand side in units of $[{\rm TeV}^{-2}]$.
The columns with solid shades shows the results from a global fit, and the ones with light shades are obtained by switching on one operator at a time.
}
\label{fig:d6}
\end{figure}

Our results for the three bases are presented in \autoref{fig:d6} in terms of the 95\% confidence level (CL) ($\Delta \chi^2=4$) reach for $\Lambda/ \sqrt{|c_i|}$, with $c_i$ and $\Lambda$ defined in \autoref{eq:EFT_Lops}.
This is particularly convenient for comparing $\Lambda$ with the bounds on new particle masses from direct searches (which are usually in terms of 95\% CL).
However, it should be emphasized that in an EFT analysis one always constrain the combination $\Lambda/ \sqrt{c_i}$ (or $c_i/\Lambda^2$) rather than $\Lambda$ itself~\cite{Contino:2016jqw}, and such comparisons are only valid if the sizes of $c_i$ are known, for instance, from assumptions of the UV theory.
Unlike in the rest of the paper, in the results presented here we will impose,  for simplicity, the flavour universality condition in the operators modifying the gauge-fermion couplings.
This reduces the total number of new physics fit parameters to twenty.
For each operator, we show both the reach from a global fit (solid shade) and the individual one with all other operator coefficients set to zero (light shade).
The corresponding values for $c_i/\Lambda^2$ are also shown on the right-hand side of the plots.

With the naive assumption of $c_i\sim 1$, we observe a global reach on $\Lambda$ in the range from $\sim 1\,$TeV to more than 10\,TeV for various operators at the future lepton colliders.
The combination of Higgs and EW measurements indeed provide robust constraints on the relevant operator coefficients and leaves no unresolved exact degeneracy/flat direction in any of three bases.
Nevertheless, the gaps between the global reaches and individual ones are still sizeable for some of the operators, with a difference of up to one order of magnitude.
For instance, the measurement of the decay $h\to \gamma\gamma$ provides the best sensitivity to the operators $\mathcal{O}_{WW}$ and $\mathcal{O}_{BB}$, but would only constrain one combination of them with $\mathcal{O}_{WB}$, while additional measurements are required to further resolve them.
This results in a significant gap of the global and individual reach for these two operators, except in the SILH' basis where $\mathcal{O}_{WW}$ and $\mathcal{O}_{WB}$ are eliminated in favour of $\mathcal{O}_{W}$ and $\mathcal{O}_{B}$, leaving $\mathcal{O}_{BB}$ the only contribution to the decay $h\to \gamma\gamma$.
On the other hand, the inclusion of $\mathcal{O}_{W}$ and $\mathcal{O}_{B}$ in the SILH' basis induces a large correlation between them and $\mathcal{O}_{HW}$, $\mathcal{O}_{HB}$, especially for the high energy runs at linear colliders where the diboson measurements provide very strong sensitivities on them but could not resolve these operators individually.
Interestingly, for those high energy runs the correlations in the diboson measurements can be removed by appropriate basis choices.
This is the case for both the modified-SILH' and the Warsaw bases, as the former eliminates the operators that modifies the left-handed lepton gauge couplings ($\mathcal{O}_{H\ell}$ and $\mathcal{O}'_{H\ell}$) and the latter eliminates $\mathcal{O}_{HW}$, $\mathcal{O}_{HB}$ which directly contribute to the aTGCs $\delta g_{1,Z}$ and $\delta \kappa_\gamma$.
The degeneracy between operators $\mathcal{O}_{GG}$ and $\mathcal{O}_{y_t}$, both of which contribute to the decay $h\to gg$, is lifted by the $t\bar{t}h$ measurements at the LHC.
Sizeable leftover correlations are also observed among the EW operators for both circular and linear colliders, as the future $Z$-pole measurements or the high energy diboson/Higgs measurements are not able to completely resolve them.

%%%%%%%%%%%%%%%%%%%%%%%%%%%
\section{Higgs-electroweak correlations at circular colliders}
\label{app:z-pole-run}
%%%%%%%%%%%%%%%%%%%%%%%%%%%
\begin{figure}%[]
  \centering
  \includegraphics[width=0.9\textwidth]{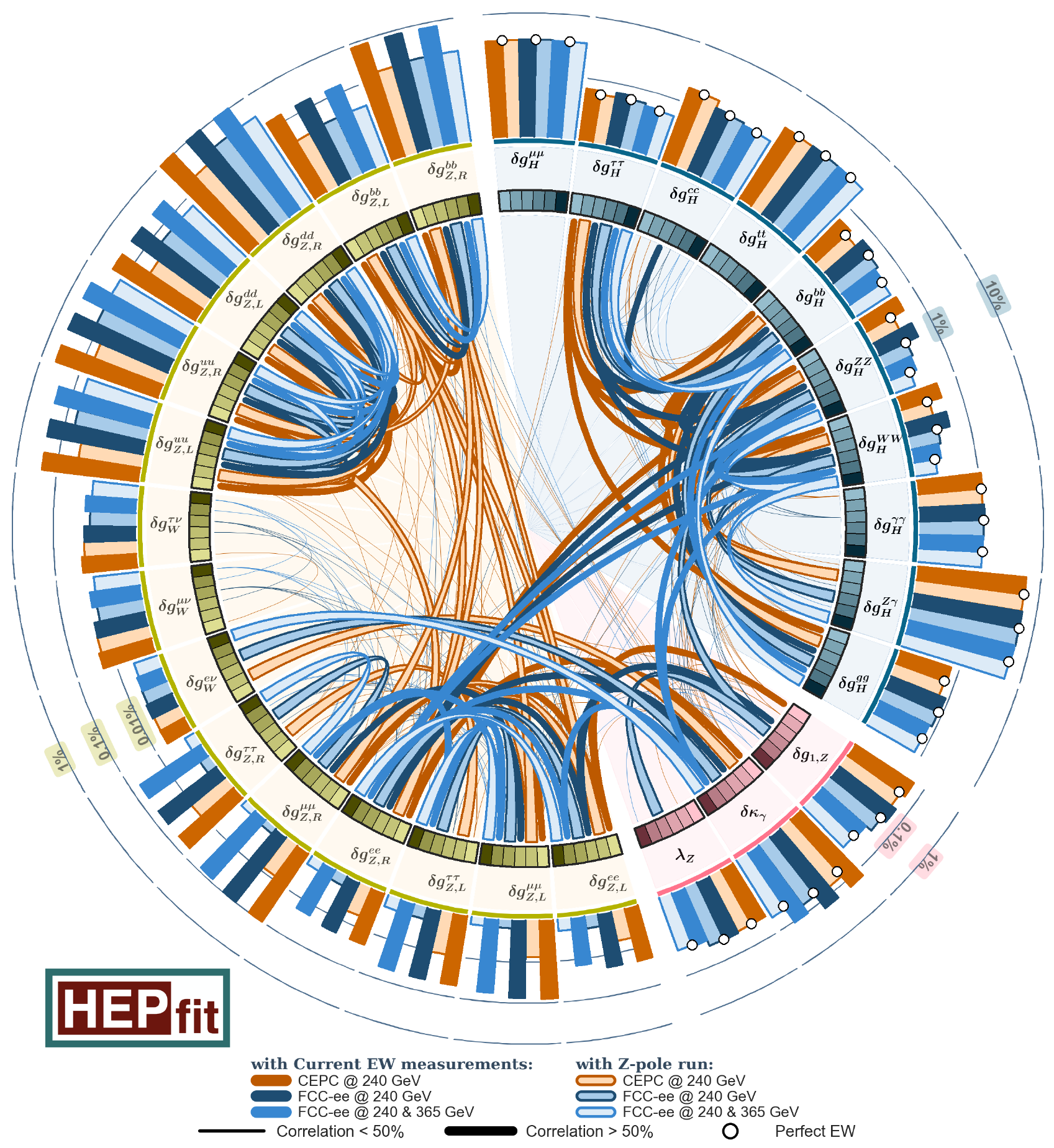}
  \caption{\it A scheme-ball illustration of the constraints on and correlations between all the effective couplings with and without a $Z$-pole run at CEPC and FCC-ee.}
  \label{fig:sb-Zpole}
 \end{figure}
 
 \begin{figure}%[t!]
  \centering
  \includegraphics[width=\textwidth]{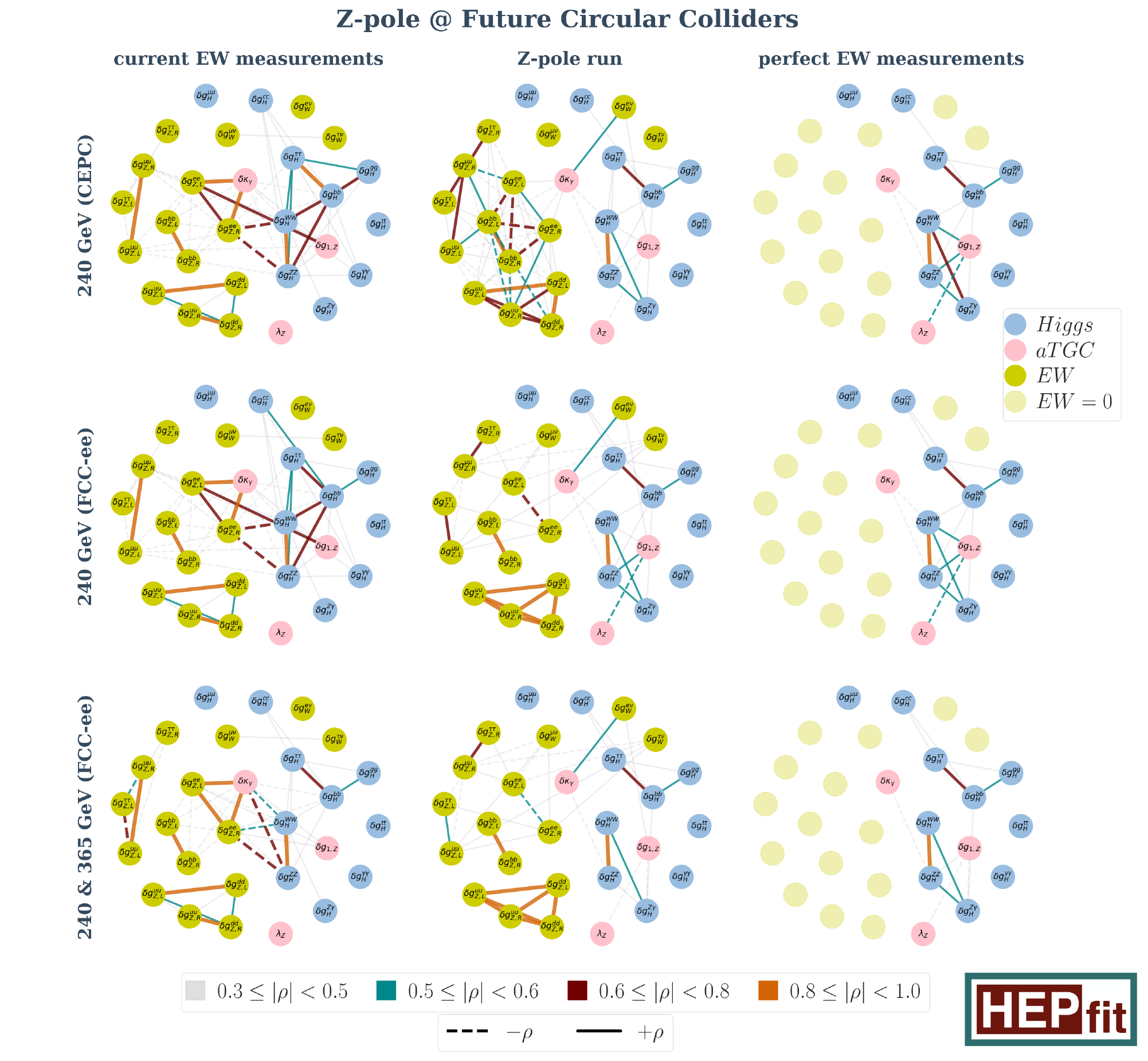}
  \caption{\it Changes in correlations between couplings depending on the precision of EW measurements assumed. The top row is for CEPC and the bottom two rows are for FCC-ee. HL-LHC projections are included for all scenarios.}
  \label{fig:np-Zpole}
\end{figure}

To further understand the interplay of the EW and the Higgs sectors, we study here the evolution of the correlations between the different parameters as one increases the precision of the $Z$-pole measurements from the current uncertainties to those that would be possible at the CEPC and FCC-ee.
We compare three different scenarios for EW measurements in \autoref{fig:sb-Zpole}.
The links and bar plots marked in red, blue and dark grey refer to the case where we use the current EW measurements and combine them with the projections for other observables at CEPC (240\,GeV) and FCC-ee (240\,GeV and 240+365\,GeV), respectively.
It can be clearly seen that there are significant correlations ($>50\%$) between $\delta g_{Z,R}^{ee}$ and the effective Higgs couplings $\delta g_H^{ZZ}$ and $\delta g_H^{WW}$. Large correlations also exist between $\delta g_{1,Z}$ and $\delta g_{Z,L}^{ee}$ and between $\delta\kappa_\gamma$ and $\delta g_{Z,R}^{ee}$ for 240\,GeV at both colliders. For the higer energy run at FCC-ee $\delta\kappa_\gamma$ is also correlated with $\delta g_{Z,L}^{ee}$.
Therefore, when one assumes perfect EW measurements shown with the white dots on the on the left side of the scheme-ball, the bounds on the these couplings in the Higgs sector are significantly stronger as they are  affected by the assumption we make about the EW measurements.

The lighter colours, orange, green and light grey, mark the bar plots and correlations for the case where we include the $Z$-pole runs for CEPC (240\,GeV) and FCC-ee (240\,GeV and 240+365\,GeV), respectively.
All of the large correlations between the effective Higgs couplings and the EW couplings drop off leaving only correlations between $\delta\kappa_\gamma$ and $\delta g^{e\nu}_W$ for all energies. Correlations between $\delta g_H^{ZZ}$ and $\delta g_{1,Z}$ remain as significant correlations between the effective Higgs couplings and the aTGCs for the 240\,GeV runs at both CEPC and FCC-ee .

The change in the correlations from one EW scenario to another for both CEPC and FCC-ee can also be seen from \autoref{fig:np-Zpole}.
For both the colliders at 240\,GeV, meshes of significant correlations can be identified between the Higgs and the EW sectors.
With the inclusion of the $Z$-pole projections these two sectors get decoupled.
While we see from \autoref{tab:allcoll} that the assumption of perfect EW measurements and the case for the inclusion of a $Z$-pole run give numerically similar bounds for both the colliders, from \autoref{fig:np-Zpole} we see that the correlation maps are different.
It can then be understand from these variations of the correlation map why $\delta\kappa_\gamma$ is still affected by the EW assumptions made even after the inclusion of EW measurements from a $Z$-pole run at the lepton colliders since the bound on it is diluted by it correlation with $\delta g_W^{e\nu}$ when EW couplings are not measured with perfect precision. 

The differences in the correlation maps between FCC-ee and CEPC at 240\,GeV when including their $Z$-pole runs are due to the differences in the input that can be found in \autoref{tab:Z-poleInput}. While the projections are given in terms of $A_{FB}^{0,(l,q)}$ for CEPC, they are given in terms of the asymmetries, $A_{l,q}$, for FCC-ee. $A_{FB}^{0,(l,q)}$ induces asymmetries from both the production and the decay at a lepton collider making the asymmetry parameters extracted from $A_{FB}^{0,(l,q)}$ highly correlated. However these correlations are not given for FCC-ee and hence the correlation maps in \autoref{fig:np-Zpole} are different for the two colliders.

%%%%%%%%%%%%%%%%%%%%%%%%%%%
\section{Statistically optimal observables}
\label{sec:oo}
%%%%%%%%%%%%%%%%%%%%%%%%%%%

Statistically optimal observables~\cite{Atwood:1991ka, Diehl:1993br} have been employed for studying the prospective sensitivity of the $e^+e^-\to W^+W^-$ and $\eehz, Z\to \ell^+ \ell^-$ processes to new physics effects in the EFT approach.

\paragraph{Definitions}
Let us briefly review general definitions, following ref.\,\cite{Diehl:1993br}.
Given a $S(\Phi) = S_0(\Phi) + C_i S_i(\Phi)$ phase-space distribution linear in the $C_i$ coefficients, the
\begin{equation}
O_i \equiv \sum_{k\text{ events}} \frac{S_i(\Phi_k)}{S_0(\Phi_k)}
\end{equation}
observables have been shown to maximally exploit the rate and differential information to provide the most precise $C_i$ measurements around the $\{C_i=0,\:\forall i\}$ point~\cite{Diehl:1993br}.
Technically, they form a set of so-called \emph{joint efficient estimators} which saturate the Cramér-Rao bound for unbiased estimators.
It is assumed that the number of collected events $n$ follows a Poisson distribution with expected value $E[n] = \mathcal{L}\sigma$ where $\mathcal{L}$ is the total integrated luminosity and $\sigma=\int\text{d}\Phi S(\Phi)$ is the total rate,
while the phase-space distribution has a probability distribution function given by $S(\Phi)\text{d}\Phi/\sigma$.
Since
\begin{equation}
\begin{aligned}
E[O_i O_j] &= E\bigg[
	\sum_k \frac{S_i(\Phi_k)}{S_0(\Phi_k)}
	\sum_l \frac{S_j(\Phi_l)}{S_0(\Phi_l)}
	\bigg]
	\\&= E\bigg[\sum_{k=l} 1\bigg]
	E\bigg[
	\frac{S_i(\Phi)}{S_0(\Phi)}
	\frac{S_j(\Phi)}{S_0(\Phi)}
	\bigg]
	+ E\bigg[\sum_{k\ne l} 1\bigg]
	E\bigg[\frac{S_i(\Phi)}{S_0(\Phi)}\bigg]
	E\bigg[\frac{S_j(\Phi)}{S_0(\Phi)}\bigg],
\end{aligned}
\end{equation}
where $E[\sum_{k\ne l} 1] = E[n^2-n] = E[n]^2$,
the asymptotic formula for the covariance matrix between the $O_i$ observables is
\begin{equation}
\cov(O_i,O_j) 
	= E[O_i O_j] - E[O_i] E[O_j]
	= \mathcal{L} \int\text{d}\Phi \frac{S_i(\Phi)S_j(\Phi)}{S_0(\Phi)}
	+ \mathcal{O}(C_k).
\end{equation}
The same symmetric $M_{ij}\equiv\mathcal{L} \int\text{d}\Phi \frac{S_i(\Phi)S_j(\Phi)}{S_0(\Phi)}$ matrix also appears in the expected value for $O_i$ in terms of $C_j$,
\begin{equation}
E[O_i] = \mathcal{L} \int\text{d}\Phi S_i(\Phi) +  C_j \mathcal{L}\int\text{d}\Phi \frac{S_i(\Phi)S_j(\Phi)}{S_0(\Phi)},
\end{equation}
so that the covariance matrix between the $C_i$ measurements is actually $M^{-1}$ at zeroth order in $C_k$:
\begin{equation}
\cov(C_i,C_j) = [M^T M^{-1} M]^{-1}_{ij} + \mathcal{O}(C_k)
	%= M^{-1}_{ij}
	 = \left(
	 \mathcal{L}\int\text{d}\Phi \frac{S_i(\Phi)S_j(\Phi)}{S_0(\Phi)}
	 \right)^{-1}
	  + \mathcal{O}(C_k).
\end{equation}
Assuming ideal experimental conditions, the statistical power of optimal observable measurements can therefore simply be estimated through this phase-space integral.
A finite efficiency can be introduced by trivially rescaling the integrated luminosity $\mathcal{L}\to \epsilon\mathcal{L}$.

To avoid using total rate information, one can also define statistically optimal observables as 
\begin{equation}
\tilde{O}_i\equiv \frac{1}{n}\sum_{k\text{ events}} \frac{S_i(\Phi_k)}{S_0(\Phi_k)}  \label{eq:oonorate1}
\end{equation}
on normalized differential distributions. Reproducing the derivation above, one easily obtains the following expression for the asymptotic covariance matrix between operator coefficient determination around the standard-model point $\{C_i=0,\:\forall i\}$ (see Eq.(40,43) of ref.~\cite{Diehl:1993br}):
\begin{equation} \cov(C_i,C_j)^{-1}/\mathcal{L} = 
\int\text{d}\Phi \frac{S_i(\Phi)S_j(\Phi)}{S_0(\Phi)}
-\frac{\int\text{d}\Phi S_i(\Phi)\: \int\text{d}\Phi S_j(\Phi) }
	{\int\text{d}\Phi S_0(\Phi)}
	  + \mathcal{O}(C_k).   \label{eq:oonorate2}
\end{equation}

\paragraph{Treatment of diboson production}
We apply the optimal observable technique to the effective-field-theory dependence of the $e^+e^-\to W^+W^-$ production process. For simplicity, optimal observables are defined on the normalized distributions, and the information about the total rate in different channels is included separately.

Unlike in LEP analyses~\cite{Abbiendi:2003mk, Achard:2004ji, Schael:2004tq, Abdallah:2010zj}, we restrict ourselves to linear effective field theory dependences.
Simple checks we performed indicate that higher-order dependences on the three CP-conserving anomalous triple gauge couplings are subleading.
In our approximation of vanishing masses for quarks and leptons, the dipole interactions as well as the right-handed couplings of the $W$ boson only appear at the quadratic level.

The semileptonic final state is considered and only doubly resonant contributions are included in the narrow-width approximation.
The differential distribution is symmetrized under the exchange of the two quarks which are experimentally indistinguishable.
Cuts, beam structure, efficiencies, detector effects and systematic uncertainties are not included.
The neutrino momentum is assumed to be perfectly reconstructed.

An analytical and a numerical computation are compared.
The latter relies on tree-level amplitudes computed with \mg~\cite{Alwall:2014hca} and uses a simple Monte-Carlo phase-space integration to compute covariance matrices.
Results for the standard-model and linear effective-field-theory rates are in good agreement with default \mg\ computation chain.
The \texttt{BSMC} implementation~\cite{Falkowski:2015wza}\footnote{It is available at \url{https://feynrules.irmp.ucl.ac.be/wiki/BSMCharacterisation}.} of the standard-model dimension-six effective field theory in the so-called Higgs basis~\cite{Falkowski:2001958} is employed.
In the analytical analysis we use for final results, the effective-field-theory dependence of the $W$ mass is included in addition to vertex corrections.

\begin{figure}%[t]
\centering
\includegraphics[width=0.7\textwidth]{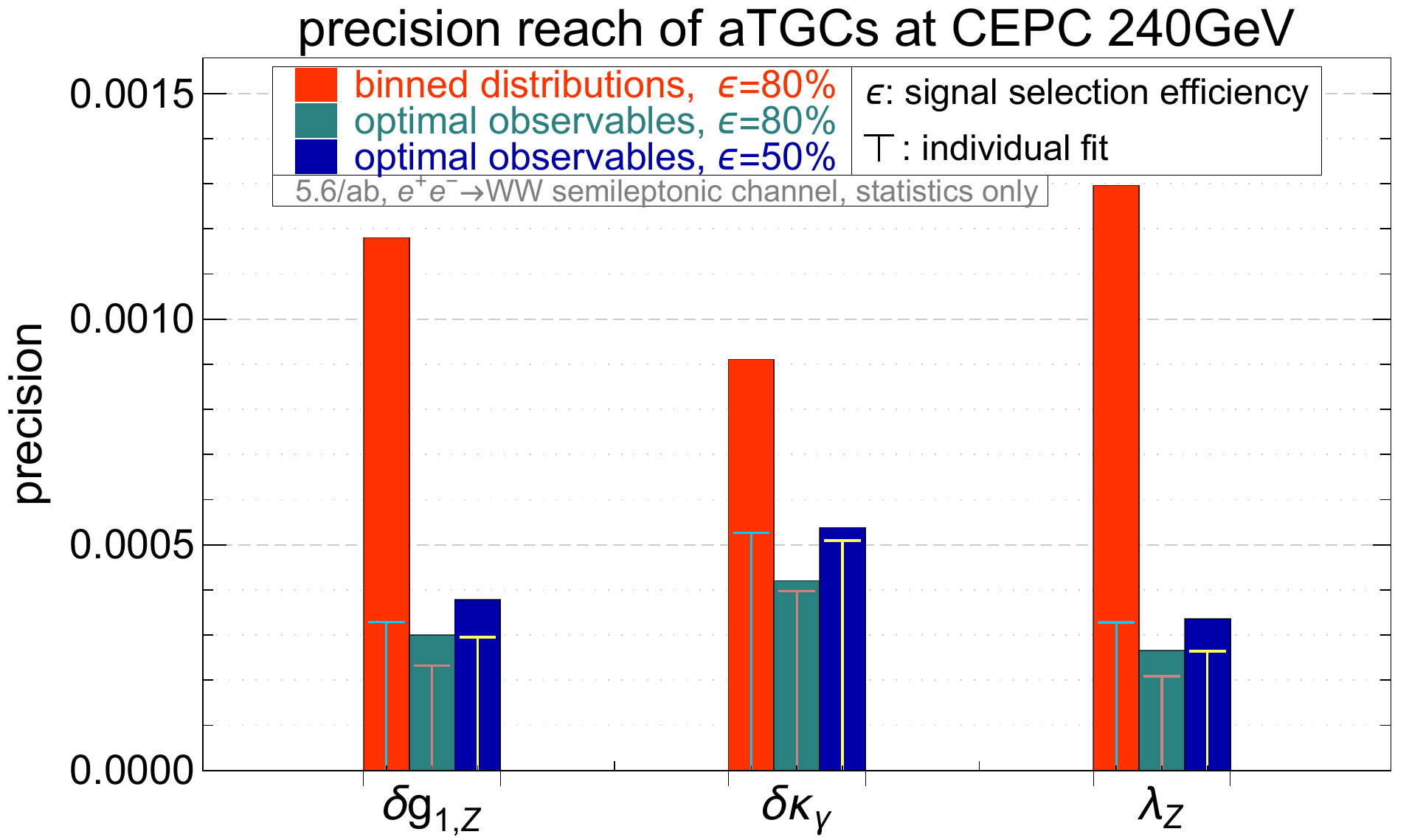}
\caption{\it
A comparison of the reach on aTGCs from the binned method used in ref.~\cite{CEPCStudyGroup:2018ghi} and the optimal observables for the diboson measurement at CEPC 240\,GeV.
To match ref.~\cite{CEPCStudyGroup:2018ghi}, we use both the total rate and the normalized distributions of the semileptonic channel, and impose the TGC dominance assumption.
A $80\%$ signal selection efficiency is assumed in ref.~\cite{CEPCStudyGroup:2018ghi}.}
\label{fig:cepctgc1}
\end{figure}

As an illustration of the power of the optimal observables, we show in \autoref{fig:cepctgc1} a comparison with the conventional binned distribution method used in ref.~\cite{CEPCStudyGroup:2018ghi} for CEPC 240\,GeV.
To match the inputs and assumptions of ref.~\cite{CEPCStudyGroup:2018ghi}, we use both the total rate and the normalized distributions of the semileptonic channel of $\eeww$, make the TGC dominance assumption and perform a global fit among the three aTGCs.
If a 80\% signal selection efficiency is assumed as in ref.~\cite{CEPCStudyGroup:2018ghi}, we observe a factor of 4-5 improvement in $\delta g_{1,Z}$ and $\lambda_Z$ with the use of optimal observables, and a some what smaller improvement (by a factor of $\sim 2$) for $\delta \kappa_\gamma$.
In particular, a better discrimination between $\delta g_{1,Z}$ and $\lambda_Z$ is achieved using optimal observables, which reduced the strong correlation between them from $-0.9$ (of the binned distribution method) to $-0.6$.
The improvement is still outstanding even with the conservative 50\% efficiency used in our analysis.
Note however that they remain degeneracies between Higgs and EW parameters that cannot be resolved with $WW$ measurements alone, even with optimal use of the available differential information.

\paragraph{Treatment of Higgsstrahlung production}

The three relevant angles in the process $\eehz, Z\to \ell^+ \ell^-$ are the production polar angle and the $Z$ decay polar and azimuthal angles.
In refs.~\cite{Beneke:2014sba, Craig:2015wwr}, the information contained in angular distributions was extracted using asymmetries.
While this approach captures all the essential information, the correlations among the asymmetry observables are omitted, which results in a reduction in the sensitivity.
We instead construct statistically optimal observables from these three angles using \autoref{eq:oonorate1} and \eqref{eq:oonorate2}, keeping only the linear CP-even EFT dependences.
We use only the $h\to b\bar{b}$ and $Z\to e^+e^-/\mu^+\mu^-$ channel, which is almost background free after the selection cuts.
The $\chi^2$ is computed analytically, including only statistical uncertainties with a universal $40\%$ signal efficiency.
Note that the $b\bar{b}$ pair is only used for tagging the Higgs and reducing backgrounds.
The flat distribution of scalar decay product does not contain useful information.

%%%%%%%%%%%%%%%%%%%%%%%%%%%
\section{Input for the global fits}
\label{app:inputs}
%%%%%%%%%%%%%%%%%%%%%%%%%%%

In this section, we  give a list of inputs that we used in the fits for the various colliders.
The same inputs can also be provided as configuration files for \HEPfit on request which can be used for reproducing our results.
While we  try to give a complete list of inputs in this section of the appendix, large correlation matrices like those that appear for the Higgs signal strength projection for HL-LHC and those available in ref.~\cite{ALEPH:2005ab} for the EW inputs will not be presented here.
The inputs can be broadly divided into four categories, those at the $Z$ pole, those pertaining to $W$ mass, width and branching fractions, those pertaining to Higgs signal strengths at various energies and configurations of the colliders and those for the optimal observables. 

Of these inputs, we will provide the input for the optimal observables as covariance matrices from which likelihoods can be built. The rest will be given as variations around the SM input values. In addition we vary the SM parameters in \HEPfit which we list below. The variances of these parameters can be dependent on the collider being studied.

%%%%%%%%%%%%%%%%%%%%%%%%%%%
\subsection{SM inputs}
\label{app:SM-Input}
%%%%%%%%%%%%%%%%%%%%%%%%%%%

\begin{table}%[h!]
\centering
\renewcommand{\arraystretch}{1.35}
{\footnotesize
	\begin{tabu}{|l|l|l|l|l|l|}
	\hline
	\multicolumn{6}{|c|}{\bf SM input parameters at future colliders}\\
	\hhline{:======:}
	{\bf parameter}				&\multicolumn{5}{c|}{\bf uncertainties}\\
%	{\bf parameter}				&\multicolumn{5}{c|}{\bf root mean square\\
	\hhline{:======:}
								&HL-LHC				&CEPC					&FCC-ee 				&ILC &CLIC\\
	\hhline{:======:}
	$\alpha_s(m_Z)$				&$2\cdot10^{-4}$	&$2\cdot10^{-4}$		&$2\cdot10^{-4}$		&$2\cdot10^{-4}$		&$2\cdot10^{-4}$	\\\tabucline[0.15pt on 0.5pt off 2pt]{1-6}
	$\Delta\alpha_{\mathrm{had}}^{(5)}(m_Z)$		&$5\cdot10^{-5}$	&$5\cdot10^{-5}$		&$3\cdot10^{-5}$		&$5\cdot10^{-5}$		&$5\cdot10^{-5}$	\\\tabucline[0.15pt on 0.5pt off 2pt]{1-6}
	$m_t$ (GeV)					&0.4				&0.4\da					&0.02					&0.05					&0.05				\\\tabucline[0.15pt on 0.5pt off 2pt]{1-6}
	$m_h$ (GeV)					&$2\cdot10^{-2}$	&$5.9\cdot10^{-3}$		&$5.9\cdot10^{-3}$\as	&$1.5\cdot10^{-2}$		&$2\cdot10^{-2}$\da\\\tabucline[0.15pt on 0.5pt off 2pt]{1-6}
	\hhline{|------|}
	\end{tabu}
	\caption{\it Gaussian errors on the SM input parameters varied in the fits (see \autoref{app:Zpole} for $m_Z$). Unless the units are specified, all uncertainties are relative to the respective central value. \da HL-LHC projection. \as CEPC projection.}
	\label{tab:SM-Input}
	}
\end{table}

In \autoref{tab:SM-Input} we list the uncertainties for the different SM parameters that were varied in the fits, with the exception of the $Z$ mass, which is reported in \autoref{app:Zpole}. 
For the strong coupling constant we use the projected uncertainty from future lattice calculations. For the uncertainty on the hadronic contribution to the running of the electromagnetic constant, $\Delta\alpha_{\mathrm{had}}^{(5)}(m_Z)$, we use the projection from future experiments measuring the $e^+ e^- \to \mbox{hadrons}$ cross section. For FCC-ee we use the expected precision on $\alpha_s(m_Z)$ from the measurement of $A_{FB}^{\mu\mu}$ below and above the $Z$ pole~\cite{Abada:2019lih,Abada:2019zxq}.
The projection for $m_h$ at CEPC can be found in ref.~\cite{CEPCStudyGroup:2018ghi}. Due to the absence of an official projection for the Higgs mass measurement from FCC-ee, we use the CEPC value. For all CLIC stages we use the same 20 MeV precision for the Higgs mass envisioned at the end of the HL-LHC era. This is comparable to the precision that would be available at the end of the CLIC 3 TeV physics program~\cite{Robson:2018zje}. 
For the uncertainty on the top mass at CEPC we use the HL-LHC projection, since no run at the $t\bar{t}$ threshold in currently proposed for CEPC.
The projections for $m_t$ at FCC-ee can be found in refs.~\cite{Abada:2019lih,Abada:2019zxq}; the ones for CLIC are in ref.~\cite{Charles:2018vfv}; and the projections for ILC are taken from ref.~\cite{Aihara:2019gcq}.

%%%%%%%%%%%%%%%%%%%%%%%%%%%
\subsection{Electroweak measurements}
\label{app:Zpole}
%%%%%%%%%%%%%%%%%%%%%%%%%%%

\begin{table}%[h!]
\centering
\renewcommand{\arraystretch}{1.35}
{\notsotiny
	\begin{tabu}{|l|l|l|l|l|l||l|l|l|l|l|l|l}
	\hline
	\multicolumn{12}{|c|}{\bf Current and future EWPO uncertainties}\\
	\hhline{:======:t:======:}
	$(10^{-3})$				&L/S	&CEPC	&FCC-ee		&ILC &CLIC &$(10^{-3})$	&L/S	&CEPC	&FCC-ee &ILC &CLIC\\
	\hhline{:======::======:}
	$M_Z$ (GeV)				&2.1		&0.5		&0.1		&--		&--		&$A_e$\as\as			&14.3		&--			&0.11		&1.00				&4.20			\\\tabucline[0.15pt on 0.5pt off 2pt]{1-6}\hhline{~~~~~~||~~~~~~|}\tabucline[0.15pt on 0.5pt off 2pt]{6-10}
	$\Gamma_Z$ (GeV)		&2.3		&0.5		&0.1		&--		&--		&$A_\mu$\as\as			&102.0		&--			&0.15		&5.41				&26.5			\\\tabucline[0.15pt on 0.5pt off 2pt]{1-12}
	$\sigma_{had}$ (nb)		&37.0		&\CG5.0		&5.0		&--		&--		&$A_\tau$\as\as		&102.0		&--			&0.3		&5.71				&37.4			\\\tabucline[0.15pt on 0.5pt off 2pt]{1-12}
	$R_e$					&2.41		&\CG0.6		&0.3		&1.14	&2.70	&$R_b$				&3.06		&0.2		&0.3		&1.06				&1.76			\\\tabucline[0.15pt on 0.5pt off 2pt]{1-12}
	$R_\mu$					&1.59		&0.1		&0.05		&1.14	&2.70	&$R_c$				&17.4		&\CG1.13	&1.5		&5.03				&5.56			\\\tabucline[0.15pt on 0.5pt off 2pt]{1-12}
	$R_\tau$				&2.17		&\CG0.2		&0.1		&1.15	&5.99	&$A_{\rm FB}^{0,b}$&15.5		&1.0		&--			&--					&--				\\\tabucline[0.15pt on 0.5pt off 2pt]{1-12}
	$A_{\rm FB}^{0,e}$		&154.0		&5.0		&--			&--		&--		&$A_{\rm FB}^{0,c}$&47.5		&\CG3.08	&--			&--					&--				\\\tabucline[0.15pt on 0.5pt off 2pt]{1-12}
	$A_{\rm FB}^{0,\mu}$	&80.1		&3.0		&--			&--		&--		&$A_b$				&21.4		&--			&3.0		&$^{0.64}_{3.05}$\da&$_{4.88}^{4.03}$\da\\\tabucline[0.15pt on 0.5pt off 2pt]{1-12}
	$A_{\rm FB}^{0,\tau}$	&104.8		&\CG5.0		&--			&--		&--		&$A_c$				&40.4		&--			&8.0		&$^{2.12}_{8.27}$\da&$_{8.49}^{3.01}$\da\\\tabucline[0.15pt on 0.5pt off 2pt]{1-12}
	$A_e$\as				&33.3		&--			&--			&--		&--		&$A_s$				&97.3 		&--			&--			&--					&--				\\\tabucline[0.15pt on 0.5pt off 2pt]{1-12}
	$A_\tau$\as				&29.2		&--			&--			&--		&--		&source:&\cite{ALEPH:2005ab}\cite{Abe:2000uc}&\cite{Abada:2019zxq}&\cite{CEPCStudyGroup:2018ghi}&&\\
	\hhline{|------||------|}
	\end{tabu}
	\caption{\it Comparison of current precision of EW observables (L/S) vs. the projections for CEPC, FCC-ee, ILC and CLIC. Projections correspond to 8 ab$^{-1}$ of integrated luminosity at CEPC~\cite{CEPCStudyGroup:2018ghi} and 150 ab$^{-1}$ at FCC-ee~\cite{Abada:2019zxq}. For ILC and CLIC the numbers correspond to the projections for measurements using radiative returns at 250\,GeV~\cite{Fujii:2019zll} and 380\,GeV~\cite{deBlas:2019rxi} respectively. The cells marked in grey denote the projections that we estimated. Unless the units are specified all numbers are relative to the respective SM central value. All numbers should be multiplied by $10^{-3}$.\as From $\tau$ polarization measurements at LEP-I. \as\as From lepton polarization and LR asymmetry measurements at SLC. LEP/SLD data is collected from ref.~\cite{ALEPH:2005ab} with $A_s$ from ref.~\cite{Abe:2000uc}.\da Two different assumptions for systematic errors have been used for $A_b$ and $A_c$ at ILC and CLIC, the upper numbers (S2) being with their estimate of systematics and the lower (S1) being with the systematics estimated for FCC-ee.}
	\label{tab:Z-poleInput}
	}
\end{table}

Only the future circular colliders, namely CEPC and FCC-ee, have runs proposed at the $Z$ pole. 
So for these colliders we include their projected sensitivities for the EW measurements.
There are Giga-$Z$ proposals for both ILC and CLIC.
However, the actual EW projections for these scenarios are still under discussion, see e.g.\ ref.~\cite{Fujii:2019zll}, and we do not include them in our work at this stage.
For CEPC, the numbers are available from table~11.9 of the CDR~\cite{CEPCStudyGroup:2018ghi}.
Some of the numbers listed in \autoref{tab:Z-poleInput} are our estimates and we clearly mark them.
For the FCC-ee, the numbers are taken from ref.~\cite{Abada:2019zxq}.
For comparison, we also list the current LEP/SLD precision for the EW observables that we use in our fits converted to numbers relative to the SM central value with the exception of $M_Z$, $\Gamma_Z$ and $\sigma_{had}$.
The former two are in units of $10^{-3}$\,GeV and the latter in $10^{-3}\,$nb.
All LEP/SLD data given in \autoref{tab:Z-poleInput} can be found in ref.~\cite{ALEPH:2005ab} except $A_s$ which can be found in ref.~\cite{Abe:2000uc}.
For the linear colliders, we use the projections for the measurements of the EW observables using radiative return to the $Z$ pole.
This is proposed at 250\,GeV for ILC and at 380\,GeV for CLIC.
We study two different scenarios for the systematic uncertainties affecting $A_b$ and $A_c$ (S1 and S2 in \autoref{fig:ew}).
S2 is using the systematics estimated by ILC and CLIC collaborations.
S1 assumes universal systematic uncertainties for ILC, CLIC, and FCC-ee and adopts the estimates of the latter collaboration.
While this does not make a large difference for the error estimated for CLIC, with just a somewhat larger error in $A_c$, for ILC this significantly enlarges the error in $A_b$ and to a lesser extent in $A_c$. For CLIC we also use an uncertainty in $R_\nu$ of $9.43\cdot10^{-3}$.

For LEP/SLD, there are small correlations between various observables which are mostly negligible.
The only significant correlations are between $\Gamma_Z$ and $\sigma_{had}$ ($-0.297$) and between $R_e$ and $A_{\rm FB}^{0,e}$ ($-0.371$).
The full correlation matrices can be found in ref.~\cite{ALEPH:2005ab}.
No correlations are assumed for CEPC and FCC-ee.

%%%%%%%%%%%%%%%%%%%%%%%%%%%
\subsection{\texorpdfstring{$W$}{W} mass, width and branching fraction measurements}
\label{sec:WW-Input}
%%%%%%%%%%%%%%%%%%%%%%%%%%%
\begin{table}%[h]
\renewcommand{\arraystretch}{1.3}
\newtabulinestyle { dd=0.25pt on 0.5pt off 2pt }
	\centering
	{\scriptsize
	\begin{tabu}{|m{0.1cm}R{1.2cm}|   |m{0.1cm}R{1.2cm}|   |m{0.1cm}R{1.2cm}|[dd]R{1.55cm}|   |m{0.1cm}R{1.2cm}|   |m{0.1cm}R{1.2cm}|[dd]R{1.55cm}|}
	\hhline{--::--::---::--::---}
	\multicolumn{2}{|c||}{\bf CEPC}&\multicolumn{2}{c||}{\bf FCC-ee}&\multicolumn{3}{c||}{\bf ILC ($\mp80\%,\pm30\%$)}&\multicolumn{2}{c||}{\bf ILC (UP)}&\multicolumn{3}{c|}{\bf CLIC ($\mp80\%,0\%$)}\\
	\hhline{==::==::===::==::===}
	\multicolumn{2}{|r||}{\bf $10^{-4}$}&\multicolumn{2}{r||}{\bf $10^{-4}$}&\multicolumn{2}{r|[dd]}{\bf {\tiny (--,+)} $10^{-4}$}&{\bf {\tiny (+,--)} $10^{-3}$}&\multicolumn{2}{r||}{\bf$10^{-4}$}&\multicolumn{2}{r|[dd]}{\bf {\tiny (--,+)} $10^{-4}$}&{\bf {\tiny (+,--)} $10^{-3}$}\\
	\hhline{==::==::===::==::===}
	\multirow{2}{*}{\rb{ \bf ~~161~~}}		&3.76	&\multirow{2}{*}{\rb{\bf ~~161~~}}	&1.92	&\multirow{2}{*}{\rb{\bf ~~250~~}}	&2.72	&1.03	&\multirow{2}{*}{\rb{\bf ~~250~~}}	&2.78	&\multirow{2}{*}{\rb{\bf ~~380~~}}	&5.28	&1.54\\
											&\CG7.52&									&\CG3.84&									&\CG5.44&\CG2.06&									&\CG5.56&									&\CG10.6&\CG3.09\\
	\hhline{==::==::===::==::===}
	\multirow{2}{*}{\rb{ \bf ~~240~~}}		&1.63	&\multirow{2}{*}{\rb{\bf ~~240~~}}	&1.72	&\multirow{2}{*}{\rb{\bf ~~350~~}}	&8.48	&5.70	&\multirow{2}{*}{\rb{\bf ~~350~~}}	&10.6	&\multirow{2}{*}{\rb{\bf ~~1500~~}}	&7.04	&4.19\\
											&\CG3.26&									&\CG3.45&									&\CG17.0&\CG11.4&									&\CG21.3&									&\CG14.1&\CG8.39\\
	\hhline{==::==::===::==::===}
	\multirow{2}{*}{\rb{ \bf ~~\gev~~}}      &\bf $\sigma_{ee\to WW}$	&\multirow{2}{*}{\rb{\bf ~~365~~}}		&3.98	&\multirow{2}{*}{\rb{\bf ~~500~~}}	&3.09	&1.22	&\multirow{2}{*}{\rb{\bf ~~500~~}}	&2.98	&\multirow{2}{*}{\rb{\bf ~~3000~~}}	&8.75	&5.22\\
	                                        &\CG \boldmath $\BR_{W\to\ell\nu}$&							&\CG7.96&									&\CG6.18&\CG2.44&									&\CG5.96&									&\CG17.5&\CG10.4\\
	\hhline{==::==::===::==::===}
	{\bf $\delta m_W:$}				&1.0	&									&0.5&\multicolumn{3}{r||}{2.4}														&	&2.4								&\multicolumn{3}{r|}{2.4}\\
	{\bf $\delta\Gamma_W\,\,:$}		&2.8	&									&1.2&\multicolumn{3}{r||}{--}														&	&--								&\multicolumn{3}{r|}{--}\\
	\hhline{|--||--||---||--||---|}	
	\end{tabu}
	\caption{\it Mass, width and leptonic branching fractions of the $W$ boson for all the future colliders at different energies.
Deviations to $m_W$ and $\Gamma_W$ are quoted in MeV. $\BR(W^\pm\to\ell^\pm\nu)$ is marked in grey and $\sigma_{ee\to WW}$ is marked in white and both are normalized with their SM values. The values for the latter for FCC-ee 350\,GeV run are numerically insignificant and can be found in the text.}
	\label{tab:WInput}
	}
\end{table}%

The mass, width and branching fraction measurements of the $W$ boson that we use in the fits are listed in \autoref{tab:WInput}. For the circular colliders which have proposed threshold scans around the $WW$ threshold, we list the projections for 161\,GeV. Both the cross section, $\sigma_{ee\to WW}$ and the leptonic branching fraction $\BR(W^\pm\to\ell^\pm\nu)$, with $\ell=e,\mu\tau$, are normalized with their SM values. The projections for the leptonic branching fractions are the same for all the leptons. The corresponding numbers in the table are {\em not} summed over three lepton families. There are small correlations of -0.125 between the branching fraction of each of the three families of leptons. While we use the correlations in our fits, they are numerically insignificant. For the 350\,GeV run at FCC-ee, the integrated luminosity is supposed to be quite low yielding very loose constraints on both the cross section ($10.6\times10^{-4}$) and the branching fractions ($21.3\times10^{-4}$) relative to the SM values, which are about a factor of 5 worse that those from 240\,GeV and a factor of 3 worse than those from 365\,GeV. The derivation of $\sigma_{ee\to WW}$ and $\BR(W^\pm\to\ell^\pm\nu)$ for the future colliders based on the number of $W$ bosons produced is briefly described in \autoref{sec:wwinput}.

In the absence of threshold scans, there are no additional projections for $\Gamma_W$ for the linear colliders since the bounds can then only be derived from the branching fraction measurements by kinematic reconstruction of the $W$.
The ILC projections for $m_W$ can be found in ref.~\cite{Fujii:2017vwa}.
In the absence of a projection for CLIC, we use the same number as ILC. 
For FCC-ee the projections for mass and width measurements can be found in the CDR~\cite{Abada:2019zxq}.
The CEPC projections can be found in table 11.12 of the CEPC CDR~\cite{CEPCStudyGroup:2018ghi}.

%%%%%%%%%%%%%%%%%%%%%%%%%%%%
\subsection{Higgs measurements}
\label{sec:Hinput}
%%%%%%%%%%%%%%%%%%%%%%%%%%%%

We tabulate the Higgs signal strengths, $\mu$, that we used in \autoref{tab:HLLHCinput} for HL-LHC and \autoref{tab:HiggsInputs} for all the future lepton colliders. Where available we also used the correlations between the projected measurements. Please note the following details for each collider:

\begin{itemize}
    \item {\bf HL-LHC} We focus only on the S2 configuration for systematics. While ATLAS provides the bounds for $\mu_{Z\gamma}$, there are no official numbers for CMS. Assuming similar bounds from the two experiments we use the same numbers and correlations for ATLAS and CMS for the $Z\gamma$ final state. The same holds true for $\mu_{ggh}^{bb}$ where only CMS provides the extrapolation and not ATLAS. Only CMS provides projections for $\mu_{Wh}^{WW}$ and $\mu_{Zh}^{WW}$ and we do not use this for ATLAS. The extrapolations for the future measurements for both the experiments can be found in table 35 of ref.~\cite{Cepeda:2019klc} as cross-sections per production channel per decay mode and not as signal strengths. However, the correlations are not given there. The correlations are only significant ($>0.4$) between the pair $\mu_{Zh}^{ZZ}$ and $\mu_{Wh}^{ZZ}$ (-0.76) and the pair $\mu_{tth}^{WW}$ and  $\mu_{tth}^{\tau\tau}$ (-0.61) for ATLAS which are combined in table 35 of ref.~\cite{Cepeda:2019klc}. The large negative correlation between the latter pair is accompanied by relatively weaker bounds on the two signal strengths for ATLAS when compared with CMS. For CMS, several decay modes in the $ggh$ productions channel are correlated as are the pair $\mu_{Zh}^{ZZ}$ and $\mu_{Wh}^{ZZ}$ (-0.68) and the pair $\mu_{VBF}^{\mu\mu}$ and $\mu_{ggh}^{\mu\mu}$ (-0.53). The full correlations matrices is provided as ancillary files while the central values for the signal strengths are reported in \autoref{tab:HLLHCinput}. The numbers we use for the fit (including correlations) are from~\cite{CMS-HLLHC} for CMS and received through private communications for ATLAS.
    \begin{table}[h!]
        \centering
        \renewcommand{\arraystretch}{1.2}
        {\footnotesize
        \caption*{\bf CMS HL-LHC}\vspace{-0.1in}
        \begin{tabular}{l|rrrr}
                                    &$\mu_{ggh}^{\gamma\gamma}$&$\mu_{ggh}^{ZZ}$&$\mu_{ggh}^{WW}$&$\mu_{gh}^{\tau\tau}$\\
                                    \addlinespace[0.1cm]
                                    \hline
                                    \addlinespace[0.1cm]
            $\mu_{ggh}^{\gamma\gamma}$&1.00	&0.48	&0.54	&0.46   \\
            $\mu_{ggh}^{ZZ}$         &0.48	&1.00	&0.52	&0.37   \\
            $\mu_{ggh}^{WW}$         &0.54	&0.52	&1.00	&0.42   \\
            $\mu_{ggh}^{\tau\tau}$   &0.46	&0.37	&0.42	&1.00   \\
        \end{tabular}
        }\vspace{-0.1in}
        \label{tab:HL-LHCCorr}
    \end{table}

    \item {\bf CEPC:} While the correlations for 7 of the decay channels are given, only a few show significant correlation, the most important being that between $\mu_{Zh}^{bb}$ and $\mu_{\nu\nu h}^{bb}$ at -0.48 while the others can be considered as being small. The projections for CEPC can be found in refs.~\cite{CEPCStudyGroup:2018ghi,An:2018dwb}. The correlation matrix is given by:
    \begin{table}[h!]
        \centering
        \renewcommand{\arraystretch}{1.2}
        {\footnotesize
        \caption*{\bf CEPC 240\,GeV}\vspace{-0.1in}
        \begin{tabular}{l|rrrrrrr}
                                    &$\mu_{Zh}^{bb}$&$\mu_{Zh}^{cc}$&$\mu_{Zh}^{gg}$&$\mu_{Zh}^{\tau\tau}$&$\mu_{Zh}^{WW}$&$\mu_{Zh}^{ZZ}$&$\mu_{\nu\nu h}^{bb}$\\
                                    \addlinespace[0.1cm]
                                    \hline
                                    \addlinespace[0.1cm]
            $\mu_{Zh}^{bb}$         &1.000 &-0.055	&-0.130	&0.001  &0.010	&-0.022 &-0.482 \\
            $\mu_{Zh}^{cc}$         &-0.055	&1.000  &-0.237	&0.008	&-0.034	&-0.013 &0.027  \\
            $\mu_{Zh}^{gg}$         &-0.130	&-0.237	&1.000  &0.010	&-0.039	&-0.021	&0.063  \\
            $\mu_{Zh}^{\tau\tau}$   &0.001	&0.008	&0.010	&1.000  &-0.150	&-0.078 &-0.000 \\
            $\mu_{Zh}^{WW}$         &0.010	&-0.034	&-0.039	&-0.150	&1.000  &-0.245	&-0.005 \\
            $\mu_{Zh}^{ZZ}$         &-0.022	&-0.013	&-0.022	&-0.078	&-0.245	&1.000  &0.011  \\
            $\mu_{\nu\nu h}^{bb}$   &-0.482 &0.027	&0.063	&0.000	&-0.005	&0.011	&1.000  \\
        \end{tabular}
        }\vspace{-0.1in}
        \label{tab:CEPCCorr}
    \end{table}
    
    \item {\bf FCC-ee:} No correlations are provided for either of the energies, although, in principle, there should be significant correlation between $\mu_{Zh}^{bb}$ and $\mu_{\nu\nu h}^{bb}$ at 240\,GeV. There are no official estimates for $\mu_{Zh}^{Z\gamma}$ for FCC-ee. So we rescale the estimate from the CEPC with the projected luminosity for FCC-ee at 240\,GeV given in \autoref{tab:scenarios}. It should also be noted that we use the combined Higgs signal strength bounds for 350 and 365\,GeV as is officially reported instead of using them separately. This is labelled as 365\,GeV throughout our work. The projections for FCC-ee can be found in ref.~\cite{Abada:2019zxq}.
    
    \item {\bf ILC:} The official projections of statistical errors for Higgs boson measurements at ILC are available only for the baseline polarization of ($-80\%, +30\%$) in table~XI of ref.~\cite{Bambade:2019fyw}. The projected precision for the other polarizations are computed from this set using the method delineated in \autoref{sec:ILCPol}\footnote{Estimates for beam polarization of ($+80\%, -30\%$) is also available from~\cite{Barklow:2017suo}}. To these projections we add the recommended 0.1\% systematic error each for luminosity and polarization measurements (except for unpolarized beams). In addition there is a systematic error associated with flavour tagging which appears for the $h\to bb$ decay mode and is estimated to scale with the luminosity as $0.3\%\sqrt{0.250/L}$. Most importantly, no official numbers are available as yet for $\mu_{Z\gamma}$ for ILC. Furthermore, the only correlation available for ILC is that between  $\mu_{Zh}^{bb}$ and $\mu_{\nu\nu h}^{bb}$ at 250\,GeV of -0.34. These two channel are uncorrelated at higher energies as it is much easier to separate these two modes with the recoil mass distribution.
    
    \item {\bf CLIC:} 
We use the most recent update of the Higgs studies at CLIC from ref.~\cite{Robson:2018zje}. The projections available in that reference for the different Higgs rates were obtained for centre-of-mass energies of 350/1400/3000\,GeV and assuming unpolarized beams. These run energies are slightly different than those in the current CLIC baseline, namely 380/1500/3000\,GeV. We checked that these small differences are not so relevant for the final results and hence we present all results for the latter configuration. Furthermore, the current CLIC baseline foresees the use of electron polarization, see table~\autoref{tab:scenarios}. We therefore scale the projected precisions for the different Higgs observables to account for the changes in the cross sections in the scenario with polarized beams, under the assumption that signal and backgrounds scale similarly. The correlation between the different channels involving hadronic final states are also taken into account in our analysis. These are assumed to remain unchanged in the scaling with the polarization. The correlation matrices at the different energies are given by:
    \begin{table}[h!]
    {\footnotesize
    \renewcommand{\arraystretch}{1.2}
        \begin{subtable}{.6\linewidth}
        \centering
        \subcaption*{\bf CLIC 380\,GeV}\vspace{-0.1in}
        \begin{tabular}{l|rrrrrr}
                                    &$\mu_{\nu\nu h}^{bb}$&$\mu_{\nu\nu h}^{cc}$&$\mu_{\nu\nu h}^{gg}$&$\mu_{Zh}^{bb}$&$\mu_{Zh}^{cc}$&$\mu_{Zh}^{gg}$\\
                                    \addlinespace[0.1cm]
                                    \hline
                                    \addlinespace[0.1cm]
            $\mu_{\nu\nu h}^{bb}$   &1.000  &-0.072 &-0.040 &-0.244 &0.022  &0.022  \\
            $\mu_{\nu\nu h}^{cc}$   &-0.072 &1.000  &-0.203 &0.025  &-0.311 &0.059  \\
            $\mu_{\nu\nu h}^{gg}$   &-0.040 &-0.203 &1.000  &0.029  &0.069  &-0.323 \\
            $\mu_{Zh}^{bb}$         &-0.244 &0.025  &0.029  &1.000  &-0.080 &-0.086 \\
            $\mu_{Zh}^{cc}$         &0.022  &-0.311 &0.069  &-0.080 &1.000  &-0.202 \\
            $\mu_{Zh}^{gg}$         &0.022  &0.059  &-0.323 &-0.086 &-0.202 &1.000  \\
        \end{tabular}
        \end{subtable}
        \begin{subtable}{.35\linewidth}
        \centering
        \caption*{\bf CLIC 1.5\,TeV}\vspace{-0.1in}
        \begin{tabular}{l|rrr}
                                    &$\mu_{\nu\nu h}^{bb}$&$\mu_{\nu\nu h}^{cc}$&$\mu_{\nu\nu h}^{gg}$\\
                                    \addlinespace[0.1cm]
                                    \hline
                                    \addlinespace[0.1cm]
            $\mu_{\nu\nu h}^{bb}$   &1.000  &-0.095 &-0.066    \\
            $\mu_{\nu\nu h}^{cc}$   &-0.095 &1.000  &-0.192    \\
            $\mu_{\nu\nu h}^{gg}$   &-0.066 &-0.192 &1.000     \\
        \end{tabular}\vspace{0.1in}
        \caption*{\bf CLIC 3\,TeV}\vspace{-0.1in}
        \begin{tabular}{l|rrr}
                                    &$\mu_{\nu\nu h}^{bb}$&$\mu_{\nu\nu h}^{cc}$&$\mu_{\nu\nu h}^{gg}$\\
                                    \addlinespace[0.1cm]
                                    \hline
                                    \addlinespace[0.1cm]
            $\mu_{\nu\nu h}^{bb}$   &1.000  &-0.104 &-0.053    \\
            $\mu_{\nu\nu h}^{cc}$   &-0.104 &1.000  &-0.178    \\
            $\mu_{\nu\nu h}^{gg}$   &-0.053 &-0.178 &1.000     \\
        \end{tabular}
        \end{subtable}
        }\vspace{-0.1in}
        \label{tab:CLICCorr}
    \end{table}
    
Unlike the other colliders, the projections for the production cross section for the $Zh$ channel for CLIC is split into the $ll$ and $qq$ final states for the decay of the $Z$ boson.
Scaling the precisions in ref.~\cite{Robson:2018zje} to the case of ($-80\%, 0\%$) polarization we get 0.036 and 0.017 for $\sigma_{Zh}(Z\to ll)$ and $\sigma_{Zh}(Z\to qq)$, respectively.
For the ($+80\%, 0\%$) polarization we find 0.041 and 0.020 for $\sigma_{Zh}(Z\to ll)$ and $\sigma_{Zh}(Z\to qq)$, respectively. 
    
\end{itemize}

\addtocounter{table}{-1}
\begin{table}[h]
        \centering
        \renewcommand{\arraystretch}{1.2}
        {\scriptsize
        \begin{tabu}{|l|[dd]c|l|[dd]c|l|[dd]c|l|[dd]c|l|[dd]c|}
        \hline
        \multicolumn{10}{|c|}{\bf HL-LHC CMS \& ATLAS (3 ab$^{-1}$) in \%}\\
        \hline
        \hline
            \multirow{2}{*}{$\mu_{ggh}^{\gamma\gamma}$}	&~4.20~	&\multirow{2}{*}{$\mu_{VBF}^{\gamma\gamma}$}	&~12.8~	&\multirow{2}{*}{$\mu_{Wh}^{\gamma\gamma}$}	&~13.9~	&\multirow{2}{*}{$\mu_{Zh}^{\gamma\gamma}$}	&~23.3~	&\multirow{2}{*}{$\mu_{tth}^{\gamma\gamma}$}	&~9.40~	\\
            										&\CG4.52	&										&\CG8.93	&										&\CG14.1	&										&\CG16.5	&										&\CG8.92	\\
	   \tabucline[0.25pt on 4pt off 5pt]{1-10}
            \multirow{2}{*}{$\mu_{ggh}^{ZZ}$}				&4.00	&\multirow{2}{*}{$\mu_{VBF}^{ZZ}$}				&13.4	&\multirow{2}{*}{$\mu_{Wh}^{ZZ}$}				&47.8	&\multirow{2}{*}{$\mu_{Zh}^{ZZ}$}				&78.6	&\multirow{2}{*}{$\mu_{tth}^{ZZ}$}				&24.6	\\
            										&\CG4.64	&										&\CG11.8	&										&\CG43.8	&										&\CG83.3	&										&\CG19.7	\\
	    \tabucline[0.25pt on 4pt off 5pt]{1-10}
            \multirow{2}{*}{$\mu_{ggh}^{WW}$}       			&3.70	&\multirow{2}{*}{$\mu_{VBF}^{WW}$}			&7.30	&\multirow{2}{*}{$\mu_{Wh}^{WW}$}				&13.8	&\multirow{2}{*}{$\mu_{Zh}^{WW}$}				&18.4	&\multirow{2}{*}{$\mu_{tth}^{WW}$}				&9.70	\\
            										&\CG6.16	&										&\CG6.68	&										&\CG--	&										&\CG--	&										&\CG114	\\
	   \tabucline[0.25pt on 4pt off 5pt]{1-4}\cline{5-8}\tabucline[0.25pt on 4pt off 5pt]{9-10}
            \multirow{2}{*}{$\mu_{ggh}^{\tau\tau}$}  			&5.50	&\multirow{2}{*}{$\mu_{VBF}^{\tau\tau}$}			&4.40	&\multicolumn{4}{c|}{\multirow{6}{*}{}}																	&\multirow{2}{*}{$\mu_{tth}^{\tau\tau}$}			&14.9	\\
            										&\CG8.79	&										&\CG8.06	&\multicolumn{4}{c|}{	}																				&										&\CG73.3	\\
	   \tabucline[0.25pt on 4pt off 5pt]{1-4}\cline{9-10}									
            \multirow{2}{*}{$\mu_{ggh}^{\mu\mu}$}  			&13.8	&\multirow{2}{*}{$\mu_{VBF}^{\mu\mu}$}			&54.0	&\multicolumn{4}{c}{	}																				&\multicolumn{2}{c}{\multirow{4}{*}{}	}					\\
            										&\CG18.5	&										&\CG36.1	&\multicolumn{4}{c}{	}																				&\multicolumn{2}{c}{}									\\
	   \tabucline[0.25pt on 4pt off 5pt]{1-4}									
            \multirow{2}{*}{$\mu_{ggh}^{Z\gamma}$} 		&--		&\multirow{2}{*}{$\mu_{VBF}^{Z\gamma}$}		&--		&\multicolumn{4}{c}{	}																				&\multicolumn{2}{c}{}									\\
            										&\CG33.3	&										&\CG68.2	&\multicolumn{4}{c}{	}																				&\multicolumn{2}{c}{}									\\
	   \tabucline[0.25pt on 4pt off 5pt]{1-2}\cline{3-10}
            \multirow{2}{*}{$\mu_{ggh}^{bb}$}		 		&24.7	&\multicolumn{2}{c|}{\multirow{2}{*}{}}			&\multirow{2}{*}{$\mu_{Wh}^{bb}$}						&9.40	&\multirow{2}{*}{$\mu_{Zh}^{bb}$}				&6.5		&\multirow{2}{*}{$\mu_{tth}^{bb}$}				&11.6	\\
            										&\CG--	&\multicolumn{2}{c|}{}						&												&\CG10.1	&										&\CG5.85	&										&\CG14.8	\\
            \cline{1-2}\cline{5-10}
        \end{tabu}
        \caption{\it HL-LHC projections for CMS and ATLAS for luminosities of 3ab$^{-1}$ individually. The white cells are for CMS and the grey ones are for ATLAS. The details of correlations and the exact use of these numbers in the fits can be found in the text.}
        \label{tab:HLLHCinput}
        }
\end{table}

\begin{landscape}
\begin{table}
\begin{center}
\renewcommand{\arraystretch}{1.75}
{\tiny
\begin{tabular}{|l|c||c|c||g|c|g|g|c|g|g|c|g||c|c|c|c|c|c|}
\hline
\multirow{4}{*}{$\mu$ (\%)}					&\multicolumn{3}{c||}{ \bf Future Circular Colliders} 												& \multicolumn{15}{c|}{\bf Future Linear Colliders}\\
						\cline{2-19}
							&{\bf CEPC}	&\multicolumn{2}{c||}{\bf FCC-ee} 	 &\multicolumn{9}{c||}{{\bf ILC} [PM $\Rightarrow$ (+80\%,-30\%)] [MP $\Rightarrow$ (-80\%,+30\%)] [UP $\Rightarrow$ unpolarized]\da\da\da} 	&\multicolumn{6}{c|}{{\bf CLIC} [P\texttt{0} $\Rightarrow$ (+80\%,0\%)] [M\texttt{0} $\Rightarrow$ (-80\%,0\%)]}\\
						\cline{2-19}
							&240\,GeV		&240\,GeV 	&365\,GeV			&\multicolumn{3}{c|}{250\,GeV}		&\multicolumn{3}{c|}{350\,GeV}		&\multicolumn{3}{c||}{500\,GeV}							&\multicolumn{2}{c|}{380\,GeV}	 	&\multicolumn{2}{c|}{1.5\,TeV}		&\multicolumn{2}{c|}{3\,TeV}\\
						\hhline{|~|---|---|---|---|--|--|--|}
							&\multicolumn{3}{c||}{unpolarized}				&\;\;PM\;\; &\;\;MP\;\;	&\;\;UP\;\;		&\;\;PM\;\;	&\;\;MP\;\; &\;\;UP\;\;		&\;\;PM\;\;	&\;\;MP\;\; &\;\;UP\;\;							&\;\;P\texttt{0}\;\;	&\;\;M\texttt{0}\;\;	&\;\;P\texttt{0}\;\;	&\;\;M\texttt{0}\;\;	&\;\;P\texttt{0}\;\;	&\;\;M\texttt{0}\;\;	\\
\hline
\hline										
$\sigma_{Zh}$	  			&0.005		&0.005		&0.009				&0.011	&0.011	&0.008		&0.025     	&0.042     	&0.022     		&0.017   	&0.017   	&0.012   							&\da\da			&\da\da			&--				&--				&--				&--			\\

\hline
\hline										
$\mu_{Zh}^{bb}$  			&~0.31\da		&0.30		&0.50				&~0.72\da	&~0.72\da	&~0.53\da		&3.69     	&2.09     	&1.90     		&1.01   	&1.01   	&0.71   							&~0.92\da			&~0.81\da			&6.20			&2.75			&10.3			&4.54		\\

\hline										
$\mu_{Zh}^{cc}$ 			&~3.26\da		&2.20		&6.50				&4.38	&4.38	&3.27		&25.9	&15.0	&13.7 		&7.12	&7.12	&5.01							&~15.1\da			&~13.4\da			&--				&--				&--				&--			\\

\hline										
$\mu_{Zh}^{\tau\tau}$ 		&~0.82\da		&0.90		&1.80				&1.69	&1.69	&1.28		&9.43	&5.45	&4.98 		&2.42	&2.42	&1.70							&6.64			&5.88			&--				&--				&--				&--			\\

\hline
$\mu_{Zh}^{\mu\mu}$  		&17.1		&19.0		&40.0				&37.9	&37.9	&28.3		&205		&118		&108 		&47.4	&47.4	&33.4							&--				&--				&--				&--				&--				&--			\\

\hline										
$\mu_{Zh}^{WW}$ 			&~0.98\da		&1.20		&2.60				&2.43	&2.43	&1.81		&13.4	&7.62	&6.97 		&3.05	&3.05	&2.15							&5.73			&5.08			&--				&--				&--				&--			\\

\hline										
$\mu_{Zh}^{ZZ}$ 			&~5.09\da		&4.40		&12.0				&9.49	&9.49	&7.09		&58.9	&34.0	&31.1 		&13.8	&13.8	&9.74							&--				&--				&--				&--				&--				&--			\\

\hline										
$\mu_{Zh}^{Z\gamma}$ 		&15.0		&15.9		&--					&--		&--		&--			&--		&--		&--	 		&--		&--		&--								&--				&--				&--				&--				&--				&--			\\
	
\hline										
$\mu_{Zh}^{\gamma\gamma}$ 	&6.84		&9.00		&18.0				&17.9	&17.9	&13.4		&91.9	&53.1	&48.6 		&18.6	&18.6	&13.1							&--				&--				&--				&--				&--				&--			\\

\hline										
$\mu_{Zh}^{gg}$ 			&~1.27\da		&1.90		&3.50				&3.69	&3.69	&2.76		&19.8	&11.4	&10.5 		&5.93	&5.93	&4.18							&~6.49\da			&~5.74\da			&--				&--				&--				&--			\\

\hline
\hline		
$\mu_{\nu\nu h}^{bb}$  		&~3.00\da		&3.10		&0.90				&~17.4\da	&~4.27\da	&~4.37\da		&17.7	    &2.48     &3.10  		&1.51     	&0.41     	&0.38     							&~4.11\da			&~1.37\da			&~1.50\da			&~0.25\da			&~1.00\da			&~0.17\da		\\

\hline										
$\mu_{\nu\nu h}^{cc}$ 		&--			&--			&10.0				&--		&--		&--			&186		&25.9	&32.5 		&14.2	&3.48	&3.37							&~56.9\da			&~19.0\da			&~23.5\da			&~3.92\da			&~22.0\da			&~3.67\da		\\

\hline										
$\mu_{\nu\nu h}^{\tau\tau}$ 	&--			&--			&8.00				&--		&--		&--			&156		&21.8	&27.4 		&15.8	&3.88	&3.75							&--				&--				&16.5			&2.75			&14.0			&2.33		\\

\hline
$\mu_{\nu\nu h}^{\mu\mu}$  	&--			&--			&--					&--		&--		&--			&1579	&218		&274 		&166		&39.5	&38.2							&--				&--				&145				&24.17			&80.0			&13.3		\\

\hline										
$\mu_{\nu\nu h}^{WW}$ 		&--			&--			&3.0				&--		&--		&--			&56.6	&7.76	&9.75 		&5.54	&1.35	&1.30							&--				&--				&4.00			&0.67			&2.00			&0.33		\\

\hline										
$\mu_{\nu\nu h}^{ZZ}$ 		&--			&--			&10.0				&--		&--		&--			&191		&27.2	&34.2 		&19.0	&4.75	&4.59							&--				&--				&21.5			&3.58			&12.5			&2.08		\\

\hline										
$\mu_{\nu\nu h}^{Z\gamma}$ 	&--			&--			&--					&--		&--		&--			&--		&--		&--	 		&--		&--		&--								&--				&--				&165				&27.5			&95.0			&15.8		\\

\hline										
$\mu_{\nu\nu h}^{\gamma\gamma}$ 	&--		&--			&22.0				&--		&--		&--			&424		&61.2	&77.0 		&43.5	&10.7	&10.3							&--				&--				&60.0			&10.0			&30.0			&5.00		\\

\hline										
$\mu_{\nu\nu h}^{gg}$ 		&--			&--			&4.50				&--		&--		&--			&75.4	&10.5	&13.2 		&9.49	&2.30	&2.22							&~22.8\da			&~7.59\da			&~19.5\da			&~3.25\da			&~13.5\da			&~2.25\da		\\
\hline
\hline
$\mu_{eeH}^{bb}$ 			&--			&--			&--					&-- 		&--		&--			&--		&--		&--	 		&--		&--		&--								&--				&--				&3.34			&1.48			&3.58			&1.58		\\

\hline
$\mu_{ttH}^{bb}$ 			&--			&--			&--					&-- 		&--		&--			&--		&--		&--	 		&--		&--		&--								&--				&--				&15.01			&5.64			&--				&--			\\

\hline
\end{tabular}
\caption{\it The Higgs signal strength inputs used for the fits in this work. All the numbers are in \%. The details of some derivations made for some of the colliders are given in the text. The couplings marked with \da are correlated. The correlation matrices are given separately in the text and in the ancillary files. \da\da For the projections of $\sigma_{Zh}$ for CLIC see text.\da\da\da The details of how we derived the projections for different beam polarization (marked in grey) at the ILC can be found in \autoref{sec:ILCPol}.}
\label{tab:HiggsInputs}
}
\end{center}
\end{table}
\end{landscape}

\bibliographystyle{apsrev4-1_title}
\bibliography{EW-HIGGS}

\end{document}